\documentclass[twocolumn,secnumarabic,amssymb, nobibnotes, aps, prb]{revtex4-2}
\usepackage{amsmath}
\usepackage{graphicx}

\begin{document}

\title{Theory of perturbatively nonlinear quantum transport I: general formulation and structure of the retarded correlator}%

\author{Varga Bonbien$^1$}%
\email[]{bonbien.varga@kaust.edu.sa}
\author{Aur\'elien Manchon$^{2}$}
\affiliation{$^1$Physical Science and Engineering Division (PSE), King Abdullah University of Science and Technology (KAUST), Thuwal 23955-6900, Saudi Arabia\\
$^2$Aix-Marseille Univ, CNRS, CINaM, Marseille, France}

\begin{abstract}
This article is the first of a trilogy that addresses various aspects of the perturbative response of general quantum systems, with possibly nontrivial ground state geometry, beyond linear order. Here, we use group theoretical considerations to investigate the structure of retarded correlators, and demonstrate how they decompose according to irreducible representations of a `time-reversal group' and relevant permutation groups, with the former probing dissipative and time-reversal properties, and the latter discerning configurational properties---longitudinal, transverse and their generalizations. We establish second order fluctuation-dissipation and fluctuation-reaction theorems, and connect them to well-known second order transport effects such as the shift and injection currents. Exploiting the Schur-Weyl duality between irreducible representations of general linear groups and irreducible representations of permutation groups, we show how to decide which terms in the decomposition based on the latter can be supported by different crystals described via the 32 point groups, and perform the full point group classification for rank 3 and 4 polar and axial tensors. Our results provide a formal basis for the extraction of uniquely differing physical effects from retarded correlators. Applications to second order charge current responses in selected cases are given in \cite{Bonbien2021c}.
\end{abstract}

\maketitle
\tableofcontents

\section{Introduction}

The accurate calculation of non-equilibrium transport coefficients without foregoing the mystique of quantum effects is a major area of research in condensed matter physics. In his seminal work \cite{Kubo1957}, Kubo put down a general framework for performing such computations. His fundamental innovation was the realization that the transport coefficients, considered as quantifying the perturbative responses to an externally applied field, could be expressed as correlation functions of quantum operators over the equilibrium configuration. The formalism was applied to lowest order in the external driving field and led to the development of, now standard, linear response theory \cite{KuboHashitsume,Rammer1991,RammerQT,RammerQFT,RadiJishi}. A great triumph of the Kubo formalism was its key role in the realization that the quantized Hall conductance, a transport coefficient, is directly related to a topological invariant \cite{TKNN}, and this cemented its place as a major actor in the quest for unraveling the secrets of the quantum realm.\\

While applications of Kubo's formalism to the calculation of linear responses has matured over the decades, with more recent applications of it including the anomalous Hall \cite{NagaosaAHErev} and spin Hall effects \cite{SinovaSHErev}, potential insights lurking within the vast labyrinth of nonlinear responses and transport effects are ripe for the picking. The pioneering steps in this endeavour were taken by Kubo himself \cite{Kubo1957}, but, with the focus being directed primarily towards the more accessible linear responses, the nonlinear formalism was not developed in considerable detail by neither him, nor the quantum transport community at large. On the other hand, the laser revolution of the 1960s brought about a surge of activity in nonlinear optics, and led to important developments in the microscopic theory of nonlinear optical susceptibilities of atomic and molecular systems \cite{BloembergenNLO,Rabin1975}. Over the next few decades, the quantum mechanical aspects of nonlinear response theory continued to be developed almost entirely by workers in nonlinear optics, and was expanded to include a description of nonlinear optical effects in semiconductors \cite{KrautBaltz1981,AversaSipeSumRules}, culminating in a general formalism put forth by Sipe and Shkrebtii \cite{SipeShkrebtii2000}. More recently, the revolutionary insights into the structure of crystalline materials gained through the lenses of geometry and topology over the last two decades \cite{HasanKane}, have broadened the scope of investigations concerning nonlinear responses \cite{MorimotoNagaosa2016,QCPGE} and the field is experiencing a renaissance \cite{Orenstein2021,Ma2021}.\\

Building on the Kubo philosophy, we address and resolve certain issues, both conceptual and technical, and introduce novel perspectives to the study of nonlinear response theory, nonlinear optics and quantum transport. In this paper we delve into the heart of response theory, the retarded correlator, and present a near-exhaustive analysis of the latter's general structures through the lense of group theory.\\

\subsection{Time-reversal, magnetic properties, and dissipation}

The question regarding how the time-reversal operation should be prescribed for linear transport coefficients of magnetically ordered crystals was a subject of significant controversy for decades. The source of the controversy was the following \cite{Grimmer1993}. Suppose we look at a static, linear transport coefficient $\sigma_{ik}$ that satisfies a constitutive relation $J_i=\sigma_{ik}E_k$ (summation over $k$ implied), where $\textbf{J}$ is a charge current and $\textbf{E}$ is an electric field. $\textbf{J}$ is odd under time-reversal `$\mathcal{T}$', whereas $\textbf{E}$ is even, hence we should have $\text{`}\mathcal{T}\text{'}\sigma_{ik}=-\sigma_{ik}$. On the other hand, for a dia or paramagnet, $\text{`}\mathcal{T}\text{'}$ would be an element of the material's magnetic symmetry group and Neumann's principle \cite{BurnsGlazer} would imply $\text{`}\mathcal{T}\text{'}\sigma_{ik}=\sigma_{ik}$, in other words, $\sigma_{ik}=0$, a conclusion not supported by experiment. The accepted resolution to this conundrum was to simply not apply $\text{`}\mathcal{T}\text{'}$ to the constitutive relations, but define the behaviour of $\sigma_{ik}$ under $\text{`}\mathcal{T}\text{'}$ according to Onsager's relations \cite{OnsagerII}, which, for a magnetic material, read $\sigma_{ik}(\textbf{m})=\sigma_{ki}(-\textbf{m}) \to \text{`}\mathcal{T}\text{'}\sigma_{ik}(\textbf{m})=\sigma_{ki}(\textbf{m})=\sigma_{ik}(-\textbf{m})$, where $\textbf{m}$ labels the magnetic structure of the material \cite{Grimmer1993,Grimmer2017}. Two immediate sources of unease start taking shape. Onsager's relations are only valid for linear transport effects. What about the nonlinear case? How should we define the general behaviour of nonlinear transport coefficients under time-reversal? With the above prescription, time-reversal $\text{`}\mathcal{T}\text{'}$ is relegated to simply represent a reversal of the magnetic structure, and thereby fails to provide a probe of the temporal properties of the system in a `complete' manner. More specifically, under `complete', we mean an operation that satisfies our intuitive expectation of what time-reversal should represent; if a coefficient is odd under it, then that coefficient is sensitive to the arrow of time and is therefore dissipative, conversely, if a coefficient is even under it, then that coefficient ignores the arrow of time and hence is reactive. How should we define time-reversal for our expectations to be unambiguously satisfied? Both sources of unease can be whisked away by realizing that there is an underlying structure in which several \textit{different} notions of `time-reversal' all play an important role. Roughly speaking, from the microscopic point of view, it is clear that the macroscopic, linear constitutive relation represents a perturbative expansion of an expectation value to leading order in the driving field, i.e., $\textbf{J}\propto\langle\hat{\textbf{J}}\rangle\propto\textbf{E}$. Since the dependence on $\textbf{m}$ arises through the equilibrium Hamiltonian, it is clear that the macroscopic average of the observable itself, which is on the left hand side of the constitutive relation, should be dependent on $\textbf{m}$, i.e., we should have $j_i(\textbf{m})=\sigma_{ik}(\textbf{m})E_k$ \cite{BonbienRev2021}. At this point we can distinguish between two operations; magnetic-inversion $\mathcal{K}^{\textbf{m}}$ that acts by only reversing $\textbf{m}$, i.e.,  $\mathcal{K}^{\textbf{m}}j_i(\textbf{m})=j_i(-\textbf{m})\to\mathcal{K}^{\textbf{m}}\sigma_{ik}(\textbf{m})=\sigma_{ik}(-\textbf{m})$, and full time-reversal $\mathcal{T}^{\text{full}}$ that acts by reversing not only $\textbf{m}$, but also the flow of current, i.e., $\mathcal{T}^{\text{full}}j_i(\textbf{m})=-j_i(-\textbf{m})\to\mathcal{T}^{\text{full}}\sigma_{ik}(\textbf{m})=-\sigma_{ik}(-\textbf{m})$. It is clear that this is consistent: the magnetic symmetry group is only concerned with the magnetic structure, hence the `time-reversal' element of it is identified as $\mathcal{K}^{\textbf{m}}$, and, for a nonmagnetic material described by one of the grey groups, we can take $\textbf{m}=0$ and we have $\mathcal{K}^{\textbf{m}}\sigma_{ik}=\sigma_{ik}$, in accordance with Neumann's principle. This is, on the other hand, not in contradiction with $\mathcal{T}^{\text{full}}\sigma_{ik}=-\sigma_{ik}$, since $\mathcal{T}^{\text{full}}$ represents a \textit{different} operation. In fact, $\mathcal{T}^{\text{full}}$ satisfies our expectations for complete time-reversal, since $\sigma_{ik}(\textbf{m})\pm\mathcal{T}^{\text{full}}\sigma_{ik}(\textbf{m})=\sigma_{ik}(\textbf{m})\mp\sigma_{ik}(-\textbf{m})$, and it is clear that upon taking $\textbf{m}=0$ only the dissipative part that is odd under $\mathcal{T}^{\text{full}}$ survives, as it should \footnote{Note that in the presence of an external magnetic field $\textbf{B}$, $\mathcal{K}^{\textbf{m}}$ would act by inverting both $\textbf{m}$ and $\textbf{B}$, i.e., $\mathcal{K}^{\textbf{m}}\sigma_{ik}(\textbf{m};\textbf{B})=\sigma_{ik}(-\textbf{m};-\textbf{B})$, meaning that it would not be the `time-reversal' element of the magnetic group, rather, this role would be fulfilled by a new operation $\mathcal{K}^{\textbf{m}}_{\mathcal{G}}$ that only inverts $\textbf{m}$, i.e., $\mathcal{K}^{\textbf{m}}_{\mathcal{G}}\sigma_{ik}(\textbf{m};\textbf{B})=\sigma_{ik}(-\textbf{m};\textbf{B})$. It is then $\mathcal{K}^{\textbf{m}}_{\mathcal{G}}$ that should be used when applying Neumann's principle for magnetic groups, and $\mathcal{K}^{\textbf{m}}$ when looking at the `time-reversal breaking' behaviour of the system as a whole.}. This point of view, that does \textit{not} rely on Onsager's relations, can be extended to arbitrary orders and allows us to identify an underlying structure. Namely, we can collect the introduced operations---together with another one to be discussed---and show that they form the representation of a group we dub the `time-reversal group'. We show that the operation $\mathcal{T}^{\text{full}}$ then emerges naturally as an interplay between different representations of the time-reversal group connected to the retarded correlation functions giving the transport coefficients, and all concerns regarding dissipative and reactive responses, and time-reversal symmetry and breaking can then be consistently and unambiguously resolved for responses of arbitrary order by studying the representation theory of this group.  An important contribution of this paper is the complete resolution of such questions via our group theoretical formulation.\\

\subsection{Fluctuation-response relations}

The remarkable link between the fluctuations and dissipative linear responses of physical systems is a well-established fact that has contributed extensively to the study of linear transport phenomena \cite{Kubo1966}. Apart from being conceptually intriguing, the latter also leads to a number of important sum rules \cite{Kubo1957,Souza2008}, and, when applied to crystalline materials, offers insights into their localization behaviour and quantum geometry \cite{Kudinov1991,Souza2000,Ozawa2019}. The link between fluctuations and responses, now reactive responses, was also extended to second order; first in a classical context \cite{Golden1972}, and later by the means of a fully quantum-mechanical treatment \cite{Kalman1987}. Moreover, as expected, it also led to sum rules \cite{Tao1990,Rommel1996}. However, the lack of attention this important relation has received, perhaps due to the rather convoluted way in which it was derived, is entirely unwarranted. In fact, it seems that its use has not left the confines of plasma physics.  Building on our general framework, we remedy this by presenting a rather simple derivation exactly analogous to the linear case, and by showing that the relation underlies an important nonlinear optical response; the circular injection current (see e.g. ref. \cite{SipeShkrebtii2000} for a description of this effect). However, we need not stop at this point. If we think about the two discussed relations; the linear fluctuation-dissipation and the quadratic fluctuation-reaction relations; there is an important principle to recognize. They connect a \textit{response} that, by causality, is a combination of fully time-ordered quantities, with a `\textit{fluctuation}' that lacks time-ordering and, consequently, has unconstrained time arguments. In the linear case, we are set; however, in the quadratic case, we have some remaining freedom since we can assign a partial time-ordering to the unordered `fluctuations' and continue to relate them to the fully time-ordered responses. These `weak' fluctuations will still have an unconstrained time argument, and upon applying the partial time-ordering to the quadratic fluctuation-reaction relation proper, we obtain a corresponding weak fluctuation-\textit{dissipation} relation. We show that this weak relation underlies a large number of second order transport and optical effects including the linear shift current, and dissipative second harmonic responses. Moreover, they lead to a new class of sum rules, special cases of which have appeared in the recent literature \cite{Patankar2018,QRSR}, albeit it was not realized that they, in fact, arise from an underlying---in this case weak---quadratic fluctuation-response relation. An important contribution of this paper is the establishment of these fundamental links in a highly general manner, purely from the properties of the correlation functions describing the responses.\\

\subsection{Permutations and responses}
In his seminal work on linear transport phenomena \cite{OnsagerI,OnsagerII}, Onsager discovered an intimate relation between the symmetry of the tensors describing the process at hand and the time-reversal conditions of the system supporting the process. While Onsager's analysis was fundamentally classical, Kubo showed that the result also follows from quantum mechanical considerations in terms of correlators \cite{Kubo1957}. An example of an Onsager relation is provided by the simple case of the static linear charge conductivity; we have $\sigma_{ik}(\textbf{m})=\sigma_{ki}(-\textbf{m})$, where $\textbf{m}$ represents a time-reversal breaking field, such as a magnetic structure and/or an external magnetic field \cite{NagaosaAHErev}. It follows that if $\textbf{m}=0$, then $\sigma_{ik}=\sigma_{ki}$, meaning that $\sigma_{ik}$ is symmetric and a non-vanishing $\textbf{m}$ is required for the antisymmetric part, describing the (anomalous) Hall effect, to contribute. In fact, the relation goes beyond simply connecting the time-reversal properties of the system to the symmetry of the tensor; in a clean crystal it uniquely separates the Fermi surface from the Fermi sea contributions to the linear conductivity \cite{Bonbien2020}. A natural question we can ask is whether any of this generalizes beyond linear order. In order to answer this question, we have two tasks: describe the behaviour of the nonlinear response under time-reversal, and find some unique way to think of tensor symmetry for higher order responses. The first task was already introduced and is accomplished by means of a group representation. This provides a hint that perhaps the second task can also be approached along these lines. Indeed, we can look at the action of permutation groups on the responses and naturally decompose the relevant correlators into terms transforming according to the irreducible representations of the relevant permutation group. In the linear case, the permutation group is $P(2)$, the group of permutations of two objects, and has two irreducible representations corresponding to the symmetric and antisymmetric parts \cite{DresselhausSymm}. Thus, we can formulate Onsager's relations as an intimate connection between representations of a time-reversal group and a permutation group. We show that this intimate relation does not generalize to higher order, in the sense that, in general, time-reversal does not restrict the permutation group irreducible representation the response correlator can transform in.  However, the decomposition into the irreducible representations of permutation groups is always beneficial, since the resulting terms are directly connected to the driving field and observer configuration. As a simple example, let us look at second order charge current conductivity $\sigma_{i_0i_1i_2}(\omega_1,\omega_2)$, where $i_0,i_1,i_2\in\{x,y,z\}$. This satisfies intrinsic permutation symmetry \cite{BoydNLO}, i.e., $\sigma_{i_0i_1i_2}(\omega_1,\omega_2)=\sigma_{i_0i_2i_1}(\omega_2,\omega_1)$. The second harmonic charge conductivity $\sigma_{i_0i_1i_2}(\omega,\omega)$, is then clearly symmetric in $i_1$ and $i_2$. Decomposing it into irreducible representation of $P(3)$, the group of permutations of three objects, then provides us with the means of uniquely separating two physically distinct responses; the `longitudinal' response given by the totally symmetric representation, in which case even if $i_1=i_2=i_0$ the response \textit{can be} non-vanishing, and the `transverse' response, which requires at least two out of the three indices to be distinct. This can clearly be generalized to all orders of practical importance, since the irreducible representations of permutation groups are well-known \cite{DresselhausSymm}. We present a detailed analysis of the retarded correlator's permutation structure and perform this decomposition explicitly.\\

A question that arises then, is how would the symmetry of the crystal structure restrict the supported responses transforming in a given permutation group irreducible representation? This analysis can be performed by utilizing, for a given tensor, the one-to-one correspondence between the irreducible representations of the general linear group of invertible matrices and those of the permutation group, often referred to as Schur-Weyl duality \cite{TungGroups,SternbergPhysics,FultonHarris}. Since the crystal point groups are subgroups of the full orthogonal group, which is in turn a subgroup of the general linear group, we can associate to each permutation group irreducible representation a collection of orthogonal group irreducible representations which can then be decomposed into crystal group irreducible representations. In this way, we can associate physically distinct responses to the different point groups in a very general way. We have performed this classification for rank 3 and rank 4 polar and axial tensors.\\

\subsection{Spectral representations and structure of the paper}

Some sections of the paper require the explicit form of the correlators. In this series of papers, we choose to use the spectral representation in terms of Green's functions and derive this for the 2, 3 and 4-point retarded correlators. While this representation of the retarded 2-point correlator has been in use for decades \cite{Shiba1971,Bastin1971} it has only recently been gaining traction for higher order, primarily 3-point correlators \cite{Bergman2013,Parker2019,Joao2020,Michishita2021,Du2021}. However, most derivations are based on either Keldysh \cite{Joao2020} or imaginary-time \cite{Bergman2013,Parker2019,Michishita2021,Du2021} techniques in the frequency domain and fall short of providing any further insight. To remedy this, we dedicate appendix \ref{SpectRep} to a different perspective of spectral representations with Green's functions, and show that they can be arrived at, still within the time-domain, by simply attaching the step functions of the retarded correlator to the interaction picture evolution factors. This allows the possibility of finding different equivalent spectral representations depending on how we attach the step functions and will be of particular importance in the case of the 3-point correlator, since, as we discuss in paper III \cite{Bonbien2021c}, the alternate expression lends us a useful separation of the second order conductivity.\\

The paper is sructured as follows. In Section \ref{densEvol} we obtain the non-equilibrium density matrix via the standard iterative solution to the von Neumann equation to arbitrary order, and then consider the expectation value of an observable dependent on the applied field and extract the corresponding response function to arbitrary order. In Section \ref{tstruct} we elucidate the temporal structure of the retarded correlator via group theoretical analysis. This is followed by Section \ref{FlucDissReact}, in which we lay bare the links between second order fluctuations and responses, and apply our results to second order charge conductivities. Next, we move on to Section \ref{pstruct} and perform a detailed group theoretical analysis of the retarded correlator's permutation structure before arriving at Section \ref{GLstruct}, in which we use the connection between permutation groups and the general linear group to perform the point group classification of nonlinear response tensors according to the irreducible representations of the relevant permutation groups. We wrap up with a short discussion on some generalizations and speculate on the connection of certain nonlinear responses to elliptic curves. Among the appendices, we highlight Appendix \ref{RetPlain} and Appendix \ref{SpectRep} which contain more than technical details of calculations. In the former, we show that it is most conducive to exploit the intrinsic permutation symmetry of the retarded correlator already within the time-domain, since the properties of step functions allow us to reduce the number of terms in a simple manner. In the latter, we offer a derivation of the retarded correlator's spectral representation using a perspective differing from the standard one, which, as discussed in the paragraph above, we believe is instructive.\\

\section{Density matrix evolution and response functions}
\label{densEvol}

We commence with a standard analysis of the density matrix evolution in the presence of an external field, and go on to use the obtained density matrix to find the expectation value of an observable whose evolution is the result of the external field coupling to the system.\\

Consider a many-particle system in equilibrium as an element of a grand canonical ensemble with Hamiltonian $\mathcal{H}_0=H_0-\mu N$, where $\mu$ is the chemical potential and $N$ is the number operator. Application of an external field, modifies the Hamiltonian to $\mathcal{H} = \mathcal{H}_0+H'(t)$. Suppose the equilibrium density matrix is $\rho_0$. Then, the application of an external field changes it to $\rho(t)=\rho_0+\rho'(t)$ and, under the condition that the external field is not strong enough to alter the probability of microstate occupation within the ensemble, its evolution is described by the von Neumann equation

\begin{equation}
\label{eq:NE}
i\hbar\frac{d\rho}{d t} = [\mathcal{H},\rho].
\end{equation}

As usual, should we move to the interaction picture, a simple equation for $\rho'(t)$, emphasizing the perturbation $H'(t)$, can be arrived at. Indeed, define

\begin{equation}
\rho_{\mathcal{H}_0}(t) = e^{\frac{i}{\hbar}\mathcal{H}_0t}\rho(t)e^{-\frac{i}{\hbar}\mathcal{H}_0t} = \rho_0+\rho_{\mathcal{H}_0}'(t).
\end{equation}

Using this in \eqref{eq:NE} leads to

\begin{equation}
\label{eq:NEint}
i\hbar\frac{d\rho_{\mathcal{H}_0}'}{d t} = [H_{\mathcal{H}_0}'(t),\rho_0+\rho_{\mathcal{H}_0}'(t)].
\end{equation}

Now suppose that the system was in equilibrium at $t=t_0\to -\infty$, and we have an initial condition $\lim_{t_0\to -\infty}\rho_{\mathcal{H}_0}'(t) = 0$. Integrating \eqref{eq:NEint} gives us

\begin{equation}
\label{eq:NEfinal}
\rho_{\mathcal{H}_0}'(t) = \frac{1}{i\hbar}\int_{-\infty}^tdt'\,\left([H_{\mathcal{H}_0}'(t'),\rho_0] +[H_{\mathcal{H}_0}'(t'),\rho_{\mathcal{H}_0}'(t')]\right).
\end{equation}

Next, we consider an observable $\mathcal{O}$. The expectation value of the observable can be expressed with the density matrix as 

\begin{equation}
\label{eq:Oexp1}
\begin{split}
\langle \mathcal{O}(t)\rangle &= \text{tr}(\rho(t)\mathcal{O}) =\text{tr}\left(\rho_0\mathcal{O}\right)+\text{tr}\left(\rho_{\mathcal{H}_0}'(t)\mathcal{O}_{\mathcal{H}_0}(t)\right).
\end{split}
\end{equation}

We thus need to solve the integral equation \eqref{eq:NEfinal}. Using the standard iterative technique, the solutions is \cite{Kubo1957}:

\begin{widetext}
\begin{equation}
\label{eq:rhoeq}
\begin{split}
\rho_{\mathcal{H}_0}'(t)=\sum_{n=1}^{\infty}\left(\frac{1}{i\hbar}\right)^n\int_{-\infty}^tdt_1\int_{-\infty}^{t_1}dt_2\dots\int_{-\infty}^{t_{n-1}}dt_n\,[H_{\mathcal{H}_0}'(t_1),[H_{\mathcal{H}_0}'(t_2),\dots,[H_{\mathcal{H}_0}'(t_n),\rho_0]\dots]]
\end{split}
\end{equation}

Putting \eqref{eq:rhoeq} into \eqref{eq:Oexp1} and using the cyclicity of the trace together with symmetrizing, we find

\begin{equation}
\label{eq:Oexp}
\begin{split}
\langle \mathcal{O}(t)\rangle =&\sum_{n=0}^{\infty}\frac{1}{n!}\left(\frac{1}{i\hbar}\right)^n\int dt_1\dots\int dt_n\,\times
\\
&\times\sum_{\pi\in S_n}\theta(t-t_{\pi(1)})\theta(t_{\pi(1)}-t_{\pi(2)})\dots\theta(t_{\pi(n-1)}-t_{\pi(n)})\text{tr}\left(\rho_0[[\dots[\mathcal{O}_{\mathcal{H}_0}(t),H_{\mathcal{H}_0}'(t_{\pi(1)})],\dots],H_{\mathcal{H}_0}'(t_{\pi(n)})]\right)
\end{split}
\end{equation}
\end{widetext}

where $S_n$, the symmetric group, denotes the group of permutations of the set $\{1,\dots,n\}$ and $\pi\in S_n$ refers to a specific permutation

\begin{equation}
\pi=
\begin{pmatrix}
1&2&\dots&n\\
\pi(1)&\pi(2)&\dots&\pi(n),
\end{pmatrix}
\end{equation}

in standard permutation notation. Furthermore we introduced Heaviside's step function $\theta(t)$.\\

Let $\textbf{F}(t)$ be a spatially uniform \textit{classical} field that we consider as the external perturbation and $\mathcal{M}_j^{(0)},\mathcal{M}_{jk}^{(1)},\dots$ denote the components in an array of Hermitian operators that the field couples to. The interaction Hamiltonian becomes 

\begin{equation}
\label{eq:HextOp}
\begin{split}
H'(t) =& \sum_{n=1}^{\infty}\sum_{j_1}\dots\sum_{j_n}\mathcal{M}^{(n-1)}_{j_1\dots j_n}F^{j_1}(t)\dots F^{j_n}(t)
\\
=&\sum_{j_1}\mathcal{M}^{(0)}_{j_1}F^{j_1}(t)+\sum_{j_1,j_2}\mathcal{M}^{(1)}_{j_1j_2}F^{j_1}(t)F^{j_2}(t)+\dots
\end{split}
\end{equation}

The arrays $\mathcal{M}^{n-1}_{j_1\dots j_n}$ are defined to be completely symmetric in $j_1,\dots,j_n$. No information is lost in this way, since we are summing over these indices and the arrays are multiplied by $F^{j_1}(t)\dots F^{j_n}(t)$, a totally symmetric expression.\\
 As an example, suppose we apply an electric field in the $x$ direction and a thermal gradient also in the $x$ direction. Then we have $F^1(t) = -A_x(t), F^2(t) = - A^T_x(t)$ and $\mathcal{M}^{(0)}_1=J_x, \ \mathcal{M}^{(0)}_2 = J^Q_x$, where $A_x,A^T_x$ are the $x$ components of the electromagnetic and thermal \cite{Tatara2015} vector potentials respectively, whereas $J_x,J^Q_x$ are the $x$ components of the total electric and heat currents respectively. Thus the indices labelling a 'component' do not always refer to a coordinate component of a vector operator, rather to a component in a vector of operators.\\

Similarly to the above, let $\mathcal{O}_i$ be the $i$-th component of a vector of observable operators whose expectation values we are looking for. Keeping in mind that these might depend on the applied field, we have

\begin{equation}
\label{eq:ObsOp}
\begin{split}
\mathcal{O}_{i_0}(\textbf{F}(t)) =\mathcal{O}^{(0)}_{i_0}+\sum_{n=1}^{\infty}\sum_{i_1,\dots,i_n}\mathcal{O}^{(n)}_{i_0i_1\dots i_n}F^{i_1}(t)\dots F^{i_n}(t).
\end{split}
\end{equation}

Just as for the couplings, we take $\mathcal{O}^{(n)}_{i_0i_1\dots i_n}$ to be completely symmetric in $i_1,\dots,i_n$. Plugging \eqref{eq:HextOp} and \eqref{eq:ObsOp} in \eqref{eq:Oexp} we can express the total expectation value as the sum of $n$-th order responses

\begin{equation}
\label{eq:ObsTotal}
\langle\mathcal{O}_{i_0}(t)\rangle = \sum_{n=0}^{\infty}\langle\mathcal{O}_{i_0}(t)\rangle_n,
\end{equation}

where

\begin{widetext}
\begin{equation}
\label{eq:Obsn}
\langle\mathcal{O}_{i_0}(t_0)\rangle_n=\sum_{i_1}\dots\sum_{i_n}\int dt_1\dots\int dt_n P_{i_0i_1\dots i_n}(t_0,t_1,\dots,t_n)F^{i_1}(t_1)\dots F^{i_n}(t_n),
\end{equation}

and the $n$-th order response function is

\begin{equation}
\label{eq:respn}
P_{i_0i_1\dots i_n}(t_0,t_1,\dots,t_n) =\frac{1}{n!}\sum_{\pi\in S_n}R^{(n)}_{i_0i_{\pi(1)}\dots i_{\pi(n)}}(t_0,t_{\pi(1)},\dots,t_{\pi(n)}),
\end{equation}

with

\begin{equation}
\label{eq:respnR}
\begin{split}
R^{(n)}_{i_0i_{1}\dots i_{n}}(t_0,t_{1},\dots,t_{n})=\sum_{l=0}^{n}\sum_{\substack{k_0,\dots,k_l\\ 
								0\leq k_0\leq n\\
								1\leq k_1,\dots,k_l\leq n\\
								k_0+\dots+k_l=n}}&C^r_{\mathcal{O}^{(k_0)}_{i_0\dots i_{k_0}}\mathcal{M}^{(k_1-1)}_{i_{k_0+1}\dots i_{k_0+k_1}}\dots \mathcal{M}^{(k_l-1)}_{i_{k_0+\dots+k_{l-1}+1}\dots i_{k_0+\dots+k_l}}}(t_{k_0},t_{k_0+k_1},\dots,t_{k_0+\dots+k_l})
\\
&\times\delta(t_0-t_1)\cdots\delta(t_0-t_{k_0})
\\
&\times\delta(t_{k_0+1}-t_{k_0+2})\cdots\delta(t_{k_0+1}-t_{k_0+k_1})
\\
&\times\cdots
\\
&\times\delta(t_{k_0+\dots+k_{l-1}+1}-t_{k_0+\dots+k_{l-1}+2})\cdots\delta(t_{k_0+\dots+k_{l-1}+1}-t_{k_0+\dots+k_{l}}),
\end{split}		
\end{equation}

where we defined the $n+1$-point retarded correlator

\begin{equation}
\label{eq:Retn}
\begin{split}
C^r_{A^0A^1\dots A^n}(t_0,t_1,\dots,t_n)=
\frac{(-i)^{n}}{\hbar^n n!}\sum_{\pi\in S_n}&\theta(t_0-t_{\pi(1)})\theta(t_{\pi(1)}-t_{\pi(2)})\cdots\theta(t_{\pi(n-1)}-t_{\pi(n)})\times
\\
&\times C_{\left[A^0A^{\pi(1)}\dots A^{\pi(n)}\right]}(t_0,t_{\pi(1)},\dots,t_{\pi(n)}),
\end{split}
\end{equation}

and the stripped correlator

\begin{equation}
\label{eq:StrN}
\begin{split}
C_{[A^0A^1\dots A^n]}(t_0,t_1,\dots,t_n)= \text{tr}\left(\rho_0\left[\left[\dots\left[A^0_{\mathcal{H}_0}(t_0),A^1_{\mathcal{H}_0}(t_1)\right],\dots\right],A^n_{\mathcal{H}_0}(t_n)\right]\right),
\end{split}
\end{equation}

for Hermitian operators $A^0,\dots,A^n$. Keeping completeness and further use in sight, we also define the $n+1$-point advanced correlator

\begin{equation}
\label{eq:Advn}
\begin{split}
C^a_{A^0A^1\dots A^n}(t_0,t_1,\dots,t_n)=
\frac{i^{n}}{\hbar^n n!}\sum_{\pi\in S_n}&\theta(t_{\pi(n)}-t_{\pi(n-1)})\theta(t_{\pi(n-1)}-t_{\pi(n-2)})\cdots\theta(t_{\pi(1)}-t_{0})\times
\\
&\times C_{\left[A^0A^{\pi(1)}\dots A^{\pi(n)}\right]}(t_0,t_{\pi(1)},\dots,t_{\pi(n)}).
\end{split}
\end{equation}

In order to retain consistency in the notation, we also define $C^r_{A}=C^r_{A}(t)=C_{[A]}(t)=\text{tr}\left(\rho_0 A_{\mathcal{H}_0}(t)\right)=\text{tr}(\rho_0 A)$ and refer to it as the `1-point correlator'. We have thus expressed the complete response of the system in terms of correlation functions in closed form.\par

Due to the apparent length of certain expressions, we introduce the common shorthand notations $\delta_{t_it_j}\equiv\delta(t_i-t_j)$ and $\theta_{t_it_j}\equiv \theta(t_i-t_j)$ that will be used throughout the paper.\par

Our primary focus will be directed towards the first, second and third order responses, so we give the explicit expressions for these response functions from the general formulae above

\begin{align}
\label{eq:resp1}
&P_{i_0 i_1}(t_0,t_1)=C^r_{\mathcal{O}^{(1)}_{i_0 i_1}}\delta_{t_0t_1}+C^r_{\mathcal{O}^{(0)}_{i_0}\mathcal{M}^{(0)}_{i_1}}(t_0,t_1),
\\\nonumber
\\\nonumber
\label{eq:resp2}
&P_{i_0i_1i_2}(t_0,t_1,t_2)=C^r_{\mathcal{O}^{(2)}_{i_0i_1i_2}}\delta_{t_0t_1}\delta_{t_0t_2}
+C^r_{\mathcal{O}^{(0)}_{i_0}\mathcal{M}^{(1)}_{i_1 i_2}}(t_0,t_2)\delta_{t_1t_2}
\\\nonumber
&\qquad\qquad\qquad\qquad+\frac{1}{2}\sum_{\pi\in S_2}C^r_{\mathcal{O}^{(1)}_{i_0i_{\pi(1)}}\mathcal{M}^{(0)}_{i_{\pi(2)}}}(t_{\pi(1)},t_{\pi(2)})\delta_{t_0t_{\pi(1)}}
\\
&\qquad\qquad\qquad\qquad+C^r_{\mathcal{O}^{(0)}_{i_0}\mathcal{M}^{(0)}_{i_1}\mathcal{M}^{(0)}_{i_2}}(t_0,t_1,t_2),
\\\nonumber
&\text{and}
\\\nonumber
\\\nonumber
\label{eq:resp3}
&P_{i_0i_1i_2i_3}(t_0,t_1,t_2,t_3)=C^r_{\mathcal{O}^{(2)}_{i_0i_1i_2i_3}}\delta_{t_0t_1}\delta_{t_0t_2}+C^r_{\mathcal{O}^{(0)}_{i_0}\mathcal{M}^{(2)}_{i_1 i_2 i_3}}(t_0,t_3)\delta_{t_1t_2}\delta_{t_1t_3}
\\\nonumber
&+\frac{1}{3!}\sum_{\pi\in S_3}\bigg(
C^r_{\mathcal{O}^{(1)}_{i_0i_{\pi(1)}}\mathcal{M}^{(1)}_{i_{\pi(2)}i_{\pi(3)}}}(t_{\pi(1)},t_{\pi(3)})\delta_{t_0t_{\pi(1)}}\delta_{t_{\pi(2)}t_{\pi(3)}}
+C^r_{\mathcal{O}^{(2)}_{i_0i_{\pi(1)}i_{\pi(2)}}\mathcal{M}^{(0)}_{i_{\pi(3)}}}(t_{\pi(2)},t_{\pi(3)})\delta_{t_0t_{\pi(1)}}\delta_{t_{0}t_{\pi(2)}}
\bigg)
\\
&+\frac{1}{3!}\sum_{\pi\in S_3}\bigg(
C^r_{\mathcal{O}^{(0)}_{i_0}\mathcal{M}^{(0)}_{i_{\pi(1)}}\mathcal{M}^{(1)}_{i_{\pi(2)}i_{\pi(3)}}}(t_0,t_{\pi(1)},t_{\pi(3)})\delta_{t_{\pi(2)}t_{\pi(3)}}
+C^r_{\mathcal{O}^{(0)}_{i_0}\mathcal{M}^{(1)}_{i_{\pi(1)}i_{\pi(2)}}\mathcal{M}^{(0)}_{i_{\pi(3)}}}(t_0,t_{\pi(2)},t_{\pi(3)})\delta_{t_{\pi(1)}t_{\pi(2)}}
\bigg)
\\\nonumber
&+\frac{1}{3!}\sum_{\pi\in S_3}
C^r_{\mathcal{O}^{(1)}_{i_0i_{\pi(1)}}\mathcal{M}^{(0)}_{i_{\pi(2)}}\mathcal{M}^{(0)}_{i_{\pi(3)}}}(t_{\pi(1)},t_{\pi(2)},t_{\pi(3)})\delta_{t_{0}t_{\pi(1)}}
\\\nonumber
&+C^r_{\mathcal{O}^{(0)}_{i_0}\mathcal{M}^{(0)}_{i_1}\mathcal{M}^{(0)}_{i_2}\mathcal{M}^{(0)}_{i_3}}(t_0,t_1,t_2,t_3).
\end{align}
\end{widetext}

For the special case of nonlinear conductivities, this agrees with the expressions up to third order in \cite{Haruki2020}. We see that, for $n=1$, \eqref{eq:Obsn} describes the linear response with response function $P_{i_0i_1}(t_0,t_1)$ given in \eqref{eq:resp1} whereas for $n=2,\,3$ it yields the second and third order responses with respective response functions $P_{i_0i_1i_2}(t_0,t_1,t_2)$ and $P_{i_0i_1i_2i_3}(t_0,t_1,t_2,t_3)$ given by \eqref{eq:resp2} and \eqref{eq:resp3}. The linear response function is expressed using 1 and 2-point retarded  correlators, the second order response with 1, 2 and 3-point retarded correlators whereas the third order response as a combination of 1, 2, 3 and 4-point retarded correlators. This is a common theme for higher order response functions; as can be observed from \eqref{eq:respnR} the $n$-th order response can be expressed using $1,2,\dots,n+1$-point retarded correlators and the number of terms gets out of hand relatively quickly.\\

\section{Temporal structure of the retarded correlator}
\label{tstruct}

We now turn our focus towards the heart of Kubo's formalism, the correlation functions. There are several structures carried by the correlators, out of which the temporal structure will be the focus of this section. This structure results from the time-dependence, both explicit and implicit, of the correlators and a detailed analysis of the latter's behaviour under the reversal of the flow of time leads to important constraints. In this section, we show that the heralded behaviour can be entirely captured using representations of the Klein four-group $\mathsf{Z}_2\times\mathsf{Z}_2$ and that physically relevant aspects, such as the dissipative and reactive parts of the correlators, correspond to the group's irreducible representations.\\

\subsection{The time-reversal group and its representations}

The cyclic group $\mathsf{Z}_2=\{E,A\}$ containing two elements, the identity $E$ and an element $A$ that is its own inverse $A*A=E$, where `$*$' is the group operation, can be realized in several ways. For example, $\{0,1\}$ with `$*$' being addition modulo 2; $\{1,-1\}$ with `$*$' being multiplication; $\{E,\Pi\}$, where $\Pi$ exchanges, yielding the permutation group $P(2)$ of two elements; the crystallographic groups  $\{E,C_2\}$ and $\{E,i\}$, where $C_2$ is a $180^{\circ}$ rotation whereas $i$ is spatial inversion; and so on. We can also consider another element $B$ that is its own inverse, $B*B=E$, and commutes with $A$ so that $A*B=B*A$. The pair $\{E,B\}$ with operation `$*$' also forms $\mathsf{Z}_2$, but we can combine the two operations into a larger group $\mathsf{Z}_2\times\mathsf{Z}_2=\{E,A,B,A*B\}$, the Klein four-group, some realizations of which are the crystallographic groups $C_{2v},\,C_{2h}$, and $D_2$ \cite{DresselhausSymm}. Now assume $A$ and $B$ to be realized as operations that reverse the flow of time and their action is represented on functions as follows. The action of $A$ corresponds to a \textit{time-inversion}, flipping \textit{only} the \textit{explicit} time-dependence of the object it acts on, whereas $B$ corresponds to a \textit{magnetic-inversion}, flipping any \textit{implicitly} time-dependent field \textbf{m}, such as a magnetic field or magnetic texture that the object it acts on might depend upon. The composition of a time-inversion and a magnetic-inversion yields a \textit{time-reversal}. The objects we are interested in applying these operations to are the correlators that depend on the time-independent equilibrium Hamiltonian $\mathcal{H}_0(\textbf{m})$ and the time-evolution governed by it. These three operations---time-inversion, magnetic-inversion, and their combination that is time-reversal---separately form three realizations of $\mathsf{Z}_2$, denoted; $\mathcal{G}_{\mathfrak{K}}=\{\mathfrak{I},\mathfrak{K}\}$, where $\mathfrak{I}$ is the identity operation and $\mathfrak{K}$ refers to time-inversion; $\mathcal{G}_{\mathfrak{K}^{\textbf{m}}}=\{\mathfrak{I},\mathfrak{K}^m\}$ where $\mathfrak{K}^m$ refers to magnetic-inversion; and $\mathcal{G}_{\mathfrak{T}}=\{\mathfrak{I},\mathfrak{T}\}$ where $\mathfrak{T}=\mathfrak{K}\mathfrak{K}^{\textbf{m}}$ is time-reversal. We can also take the product $\mathcal{G}_{\mathfrak{K}}\times\mathcal{G}_{\mathfrak{K}^{\textbf{m}}}=\{I,\mathfrak{K},\mathfrak{K}^{\textbf{m}},\mathfrak{T}\}$ to get a single group containing all three operations. This is just a realization of the Klein four-group $\mathsf{Z}_2\times\mathsf{Z}_2$ and the three groups $\mathcal{G}_{\mathfrak{K}},\,\mathcal{G}_{\mathfrak{K}^{\textbf{m}}}$ and $\mathcal{G}_{\mathfrak{T}}$ are all its subgroups. We consider all the groups $\mathcal{G}_{\mathfrak{K}},\,\mathcal{G}_{\mathfrak{K}^{\textbf{m}}},\, \mathcal{G}_{\mathfrak{T}}$ and $\mathcal{G}_{\mathfrak{K}}\times\mathcal{G}_{\mathfrak{K}^{\textbf{m}}}$ as abstract groups that are represented on the correlator in several ways depending on how we look at the correlator as a function.\\

The $n+1$-point stripped correlator $C_{[A^0A^1\dots A^n]}(t_0,t_1,\dots,t_n;\textbf{m})$ defined in \eqref{eq:StrN}, where we explicitly show the implicitly time-dependent field \textbf{m}, determining the retarded and advanced correlators can first be considered as a scalar-valued function of $t_0,\dots,t_n$ and \textbf{m}---formally, $C_{[A^0\dots A^n]}:\mathbb{R}^{n+1+d}\to \mathbb{C}$, where we take \textbf{m} to be a a collection of $d$ parameters. Denote the space of such functions as $\mathsf{C}$ and the space of operators on $\mathsf{C}$ as $\mathcal{A}(\mathsf{C})$. A representation of $\mathcal{G}_{\mathfrak{K}}\times \mathcal{G}_{\mathfrak{K}^{\textbf{m}}}$ on these functions can be defined as $\pi_{\mathsf{C}}:\mathcal{G}_{\mathfrak{K}}\times \mathcal{G}_{\mathfrak{K}^{\textbf{m}}}\to \mathcal{A}(\mathsf{C})$, with the action of the group elements being 
\begin{equation}
\label{eq:TRrep1}
\begin{split}
&(\pi_{\mathsf{C}}(\mathfrak{I})C_{[A^0\dots A^n]})(t_0,\dots,t_n;\textbf{m})
=C_{[A^0\dots A^n]}(t_0,\dots,t_n;\textbf{m}),
\\
&(\pi_{\mathsf{C}}(\mathfrak{K})C_{[A^0\dots A^n]})(t_0,\dots,t_n;\textbf{m})
\\
&\qquad\qquad\qquad=C_{[A^0\dots A^n]}(-t_0,\dots,-t_n;\textbf{m}),
\\
&(\pi_{\mathsf{C}}(\mathfrak{K}^{\textbf{m}})C_{[A^0\dots A^n]})(t_0,\dots,t_n;\textbf{m})
\\
&\qquad\qquad\qquad
=C_{[A^0\dots A^n]}(t_0,\dots,t_n;-\textbf{m}),
\\
&(\pi_{\mathsf{C}}(\mathfrak{T})C_{[A^0\dots A^n]})(t_0,\dots,t_n;\textbf{m})
\\
&\qquad\qquad\qquad=C_{[A^0\dots A^n]}(-t_0,\dots,-t_n;-\textbf{m}).
\end{split}
\end{equation}

Naturally, this representation of $\mathcal{G}_{\mathfrak{K}}\times \mathcal{G}_{\mathfrak{K}^{\textbf{m}}}$ is also a representation of the subgroups $\mathcal{G}_{\mathfrak{K}},\,\mathcal{G}_{\mathfrak{K}^{\textbf{m}}}$ and $\mathcal{G}_{\mathfrak{T}}$. Henceforward we will denote this particular representation of the non-trivial elements as $\mathcal{K}\equiv\pi_{\mathsf{C}}(\mathfrak{K})$, $\mathcal{K}^{\textbf{m}}\equiv\pi_{\mathsf{C}}(\mathfrak{K}^{\textbf{m}})$ and $\mathcal{T}\equiv\pi_{\mathsf{C}}(\mathfrak{T})$. This representation, however, does not fully probe the temporal properties of the \textit{observables} making up the correlator. In order to resolve this, we need another perspective. In fact, we already have a different way of looking at the correlators: as scalar-valued functions on the space of operators acting on the Hilbert space of states. Indeed, writing our $n+1$-point stripped correlator as

\begin{equation}
\begin{split}
&C_{[A^0\dots A^n]}(t_0,\dots,t_n;\textbf{m})
\\
&=C(\rho_0[[\dots[A^0_{\mathcal{H}_0}(t_0),A^1_{\mathcal{H}_0}(t_1)],\dots],A^n_{\mathcal{H}_0}(t_n)])
\\
&=\text{tr}(\rho_0[[\dots[A^0_{\mathcal{H}_0}(t_0),A^1_{\mathcal{H}_0}(t_1)],\dots],A^n_{\mathcal{H}_0}(t_n)]),
\end{split}
\end{equation}

and letting $\mathcal{A}(\mathsf{H})$ be the space of operators on the Hilbert space $\mathsf{H}$, we have a function $C:\mathcal{A}(\mathsf{H})\to\mathbb{C}$ with $C(\mathcal{O})=\text{tr}(\mathcal{O})$ for operators $\mathcal{O}\in \mathcal{A}(\mathsf{H})$. Let the space of such functions be $\mathsf{C}_{\mathcal{A}(\mathsf{H})}$ and the space of operators on $\mathsf{C}_{\mathcal{A}(\mathsf{H})}$ be $\mathcal{A}(\mathsf{C}_{\mathcal{A}(\mathsf{H})})$. On this microscopic level, we represent the group $\mathcal{G}_{\mathfrak{T}}$ on the Hilbert space $\mathsf{H}$ as $\pi_{\mathsf{H}}:\mathcal{G}_{\mathfrak{T}}\to\mathcal{A}(\mathsf{H})$, with the action of time-reversal $\mathfrak{T}$ being  $\pi_{\mathsf{H}}(\mathfrak{T})|\psi\rangle=\mathcal{T}_{\mathsf{H}}|\psi\rangle$, where $\mathcal{T}_{\mathsf{H}}$ is an anti-unitary operator, i.e., it is anti-linear and satisfies $\mathcal{T}^{-1}_{\mathsf{H}}=\mathcal{T}_{\mathsf{H}}^{\dagger}$, where the adjoint is defined as $\langle\psi |\mathcal{T}_{\mathsf{H}}|\phi\rangle=\langle\phi| \mathcal{T}_{\mathsf{H}}^{\dagger}|\psi\rangle$ \cite{RammerQT}. Note that for the case of non-integer spin, this representation, denoted $\pi^{\sigma}_{\mathsf{H}}(\mathfrak{T})$, is projective, i.e., $\pi^{\sigma}_{\mathsf{H}}(\mathfrak{T})\pi^{\sigma}_{\mathsf{H}}(\mathfrak{T})=-\pi^{\sigma}_{\mathsf{H}}(\mathfrak{T}^2)=-\pi^{\sigma}_{\mathsf{H}}(\mathfrak{I})$ meaning $(\mathcal{T}^{\sigma}_{\mathsf{H}})^2=-1$ as required. The representation of $\mathcal{G}_{\mathfrak{T}}$ on $\mathsf{H}$ induces a representation $\pi_{\mathcal{A}(\mathsf{H})}$ on the operators $\mathcal{A}(\mathsf{H})$ on $\mathsf{H}$ with the action of time reversal $\mathfrak{T}$ becoming $\pi_{\mathcal{A}(\mathsf{H})}(\mathfrak{T})\mathcal{O}=\mathcal{T}_{\mathsf{H}}\mathcal{O}\mathcal{T}^{\dagger}_{\mathsf{H}}$. Note that this representation is not projective even in the non-integer spin case since $\pi^{\sigma}_{\mathcal{A}(\mathsf{H})}(\mathfrak{T})\pi^{\sigma}_{\mathcal{A}(\mathsf{H})}(\mathfrak{T})\mathcal{O}=\mathcal{T}^{\sigma}_{\mathsf{H}}\mathcal{T}^{\sigma}_{\mathsf{H}}\mathcal{O}(\mathcal{T}^{\sigma}_{\mathsf{H}})^{\dagger}(\mathcal{T}^{\sigma}_{\mathsf{H}})^{\dagger}=\mathcal{O}$ and thereby $\pi^{\sigma}_{\mathcal{A}(\mathsf{H})}(\mathfrak{T})\pi^{\sigma}_{\mathcal{A}(\mathsf{H})}(\mathfrak{T})=\pi^{\sigma}_{\mathcal{A}(\mathsf{H})}(\mathfrak{T}^2)=\pi^{\sigma}_{\mathcal{A}(\mathsf{H})}(\mathfrak{I})$. The representation $\pi_{\mathcal{A}(\mathsf{H})}$ on operators, in turn, induces a representation $\pi_{\mathsf{C}_{\mathcal{A}(\mathsf{H})}}:\mathcal{G}_{\mathfrak{T}}\to \mathcal{A}(\mathsf{C}_{\mathcal{A}(\mathsf{H})})$ on the correlators themselves, with the action of time-reversal $\mathfrak{T}$ being

\begin{equation}
\label{eq:TRrep2}
\begin{split}
(\pi_{\mathsf{C}_{\mathcal{A}(\mathsf{H})}}(\mathfrak{T})C)(\mathcal{O})&=C(\pi^{-1}_{\mathcal{A}(\mathsf{H})}(\mathfrak{T})\mathcal{O})
=C(\mathcal{T}^{\dagger}_{\mathsf{H}}\mathcal{O}\mathcal{T}_{\mathsf{H}}).
\end{split}
\end{equation}

In order to compare this representation with that of \eqref{eq:TRrep1}, we take $\mathcal{O}=\rho_0[[\dots[A^0_{\mathcal{H}_0}(t_0),A^1_{\mathcal{H}_0}(t_1)],\dots],A^n_{\mathcal{H}_0}(t_n)]$ in \eqref{eq:TRrep2} and define the behaviour of the observable operator $A^{i}$ under time-reversal as $\mathcal{T}^{\dagger}_{\mathsf{H}}A^i\mathcal{T}_{\mathsf{H}}=\epsilon_{A^i}^{\mathcal{T}}A^i$ with $\epsilon_{A^i}^{\mathcal{T}}=\pm 1$ depending on what physical quantity $A^i$ describes. Furthermore, since $\mathcal{H}_0(\textbf{m})$ is not dependent on time explicitly, microscopic reversibility requires the action of time-reversal $\mathcal{T}_{\mathsf{H}}$ on it to be $\mathcal{T}^{\dagger}_{\mathsf{H}}\mathcal{H}_0(\textbf{m})\mathcal{T}_{\mathsf{H}}=\mathcal{H}_0(-\textbf{m})$. We thus have

\begin{widetext}
\begin{equation}
\label{eq:corrTimereversal1}
\begin{split}
&C(\mathcal{T}^{\dagger}_{\mathsf{H}}\rho_0[[\dots[A^0_{\mathcal{H}_0}(t_0),A^1_{\mathcal{H}_0}(t_1)],\dots],A^n_{\mathcal{H}_0}(t_n)]\mathcal{T}_{\mathsf{H}})
\\
&=
\epsilon_{A^0}^{\mathcal{T}}\cdots\epsilon_{A^n}^{\mathcal{T}}C\left(\mathcal{T}^{\dagger}_{\mathsf{H}}\rho_0\mathcal{T}_{\mathsf{H}}[[\dots[A^0_{\mathcal{T}_{\mathsf{H}}^{\dagger}\mathcal{H}_0\mathcal{T}_{\mathsf{H}}}(-t_0),A^1_{\mathcal{T}_{\mathsf{H}}^{\dagger}\mathcal{H}_0\mathcal{T}_{\mathsf{H}}}(-t_1)],\dots],A^n_{\mathcal{T}_{\mathsf{H}}^{\dagger}\mathcal{H}_0\mathcal{T}_{\mathsf{H}}}(-t_n)]\right)
\\
&=\epsilon_{A^0}^{\mathcal{T}}\cdots\epsilon_{A^n}^{\mathcal{T}}C_{[A^0\dots A^n]}(-t_0,\dots,-t_n;-\textbf{m}),
\end{split}
\end{equation}
\end{widetext}

where we made repeated use of the fact that $\mathcal{T}_{\mathsf{H}}$ is anti-unitary. The only difference from the representation of $\mathfrak{T}$ in \eqref{eq:TRrep1} is the appearance of the $\epsilon_{A^i}^{\mathcal{T}}$ factors describing the time-reversal behaviour of the observables and we can define a `full' representation $\pi^{\text{full}}_{\mathsf{C}}:\mathcal{G}_{\mathfrak{T}}\to \mathcal{A}(\mathsf{C})$ on the correlator considered as a scalar-valued function of $t_0,\dots,t_n$ and \textbf{m} with action 

\begin{equation}
\label{eq:TfullStripped}
\begin{split}
&(\pi^{\text{full}}_{\mathsf{C}}(\mathfrak{T})C_{[A^0\dots A^n]})(t_0,\dots,t_n;\textbf{m})
\\
&\qquad=\epsilon_{A^0}^{\mathcal{T}}\cdots\epsilon_{A^n}^{\mathcal{T}}C_{[A^0\dots A^n]}(-t_0,\dots,-t_n;-\textbf{m}),
\end{split}
\end{equation}

that probes the temporal properties of the observables; something we found \eqref{eq:TRrep1} to be lacking. We can extend the `full' representation to a representation $\pi^{\text{full}}_{\mathsf{C}}:\mathcal{G}_{\mathfrak{K}}\times \mathcal{G}_{\mathfrak{K}^{\textbf{m}}}\to \mathcal{A}(\mathsf{C})$, of the complete time-reversal group with the action of the remaining non-trivial group elements becoming

\begin{equation}
\begin{split}
&(\pi^{\text{full}}_{\mathsf{C}}(\mathfrak{K})C_{[A^0\dots A^n]})(t_0,\dots,t_n;\textbf{m})
\\
&\qquad\qquad\qquad=\epsilon_{A^0}^{\mathcal{T}}\cdots\epsilon_{A^n}^{\mathcal{T}}C_{[A^0\dots A^n]}(-t_0,\dots,-t_n;\textbf{m}),
\\
&(\pi^{\text{full}}_{\mathsf{C}}(\mathfrak{K}^{\textbf{m}})C_{[A^0\dots A^n]})(t_0,\dots,t_n;\textbf{m})
\\
&\qquad\qquad\qquad=C_{[A^0\dots A^n]}(t_0,\dots,t_n;-\textbf{m}).
\end{split}
\end{equation}

Henceforth, we label $\mathcal{K}^{\text{full}}\equiv\pi^{\text{full}}_{\mathsf{C}}(\mathfrak{K})$ and it is clear that $\mathcal{K}^{\text{full}}=\epsilon_{A^0}^{\mathcal{T}}\cdots\epsilon_{A^n}^{\mathcal{T}}\mathcal{K}$, where $\mathcal{K}$ is the corresponding time-inversion representation in \eqref{eq:TRrep1}. Similarly, we define $\mathcal{T}^{\text{full}}\equiv\pi^{\text{full}}_{\mathsf{C}}(\mathcal{T})$ and have  $\mathcal{T}^{\text{full}}=\epsilon_{A^0}^{\mathcal{T}}\cdots\epsilon_{A^n}^{\mathcal{T}}\mathcal{T}$. On the other hand, both representations of magnetic-inversion are the same and we can write $\mathcal{K}^{\textbf{m}}\equiv\pi^{\text{full}}_{\mathsf{C}}(\mathfrak{K}^{\textbf{m}})=\pi_{\mathsf{C}}(\mathfrak{K}^{\textbf{m}})$. In summary, we have two representations $\{I,\mathcal{K},\mathcal{K}^{\textbf{m}},\mathcal{T}\}$ and $\{I,\mathcal{K}^{\text{full}},\mathcal{K}^{\textbf{m}},\mathcal{T}^{\text{full}}\}$ of the abstract time-reversal group $\mathcal{G}_{\mathfrak{K}}\times \mathcal{G}_{\mathfrak{K}^{\textbf{m}}}$ on the correlator, with the latter representation describing the physical content, fully probing the correlator's temporal properties.\\

The stripped correlator is given by the trace of operators and the trace has the cyclic property that we have not used upto this point. The latter property provides a constraint between the actions of the different elements of $\mathcal{G}_{\mathfrak{K}}\times \mathcal{G}_{\mathfrak{K}^{\textbf{m}}}$. Indeed, by virtue of the fact that $\mathcal{T}_{\mathsf{H}}$ is anti-unitary we can use the cyclicity of the trace to find

\begin{equation}
\label{eq:trT}
C(\mathcal{O})=\text{tr}(\mathcal{O})= \text{tr}(\mathcal{T}^{\dagger}_{\mathsf{H}}\mathcal{O}^{\dagger}\mathcal{T}_{\mathsf{H}})=C(\mathcal{T}^{\dagger}_{\mathsf{H}}\mathcal{O}^{\dagger}\mathcal{T}_{\mathsf{H}}).
\end{equation}

Taking $\mathcal{O}=\rho_0[[\dots[A^0_{\mathcal{H}_0}(t_0),A^1_{\mathcal{H}_0}(t_1)],\dots],A^n_{\mathcal{H}_0}(t_n)]$ and using $[[\dots[A^0,A^1],\dots],A^n]^{\dagger}=(-1)^{n}[[\dots[A^0,A^1],\dots],A^n]$ for Hermitian operators $A^0,A^1,\dots A^n$ together with \eqref{eq:corrTimereversal1} and the corresponding `full' action \eqref{eq:TfullStripped}, we have

\begin{equation}
\label{eq:StrippedCTRId}
\begin{split}
&\mathcal{T}^{\text{full}}C_{[A^0\dots A^n]}(t_0,\dots,t_n;\textbf{m})
\\
&\qquad=(-1)^n C_{[A^0\dots A^n]}(t_0,\dots,t_n;\textbf{m}).
\end{split}
\end{equation}

 The argument leading to such an identity is usually presented in the frequency domain \cite{Rabin1975} without the consideration of group representations. Since $\mathcal{T}^{\text{full}}=\mathcal{K}^{\text{full}}\mathcal{K}^{\textbf{m}}$, we can also rewrite identity \eqref{eq:StrippedCTRId} as
 
\begin{equation}
\label{eq:StrippedCKKmId}
\begin{split} 
&\mathcal{K}^{\text{full}}C_{[A^0\dots A^n]}(t_0,\dots,t_n;\textbf{m})
\\
&\qquad=(-1)^n \mathcal{K}^{\textbf{m}}C_{[A^0\dots A^n]}(t_0,\dots,t_n;\textbf{m}).
\end{split}
\end{equation}

We can notice that the cyclicity of the trace provides an intimate connection between the action of time-\textit{inversion} $\mathcal{K}^{\text{full}}$ and magnetic-inversion $\mathcal{K}^{\textbf{m}}$ on the stripped correlator and, furthermore, as is clear from \eqref{eq:StrippedCTRId}, fixes the stripped correlator as an eigenfunction of time-\textit{reversal}. On the other hand, the physical response is given by the retarded correlator \eqref{eq:Retn} and not the stripped correlator, meaning that we have to lift these identities to the retarded correlator. In order to do  this, consider multiplying the stripped correlator acted on by time-reversal or time-inversion by $\theta_{t_{n}t_{n-1}}\cdots\theta_{t_2t_1}\theta_{t_1t_0}$. Pulling the product of step functions under $\mathcal{K}^{\text{full}}$ or $\mathcal{T}^{\text{full}}$ reverses all their time arguments and consequently the time-ordering enforced by them

\begin{equation}
\begin{split}
&\theta_{t_nt_{n-1}}\cdots\theta_{t_{1}t_0}\mathcal{T}C_{[A^0\dots A^n]}(t_0,\dots,t_n;\textbf{m})
\\
&=\mathcal{T}\theta_{t_0t_{1}}\cdots\theta_{t_{n-1}t_n}C_{[A^0\dots A^n]}(t_0,\dots,t_n;\textbf{m}).
\end{split}
\end{equation} 

This would mean that \eqref{eq:StrippedCKKmId} and \eqref{eq:StrippedCTRId} would connect correlators with opposite time-ordering. Indeed, now consider all the permutations $\pi$ of $\{1,\dots,n\}$ labelling the operators and time arguments in \eqref{eq:StrippedCKKmId} and \eqref{eq:StrippedCTRId} and multiply both sides of the latter two identities corresponding to the given permutation by the appropriate $(-i)^{n}/(\hbar^n n!)\theta_{t_{\pi(n)}t_{\pi(n-1)}}\cdots\theta_{t_{\pi(2)}t_{\pi(1)}}\theta_{t_{\pi(1)}t_0}$, followed by a summation over the permutations. This leads to
 
\begin{align}
\nonumber
&\mathcal{K}^{\text{full}}C^r_{A^0\dots A^n}(t_0,\dots,t_n;\textbf{m})
\\
\label{eq:RetCKKmId}
&\qquad=\mathcal{K}^{\textbf{m}}C^a_{A^0\dots A^n}(t_0,\dots,t_n;\textbf{m}),
\\\nonumber
\\\nonumber
&\mathcal{T}^{\text{full}}C^r_{A^0\dots A^n}(t_0,\dots,t_n;\textbf{m})
\\
\label{eq:RetCTId}
&\qquad=C^a_{A^0\dots A^n}(t_0,\dots,t_n;\textbf{m}).
\end{align}

Using the involutive property $(\mathcal{K}^{\text{full}})^2=\mathcal{K}^2=(\mathcal{K}^{\textbf{m}})^2=1$, we can also obtain the same relations with retarded and advanced correlators swapped and finally arrive at

\begin{align}
\nonumber
&\mathcal{K}^{\text{full}}C^{r(a)}_{A^0\dots A^n}(t_0,\dots,t_n;\textbf{m})
\\
\label{eq:RetAdvCKKmId}
&\qquad=\mathcal{K}^{\textbf{m}}C^{a(r)}_{A^0\dots A^n}(t_0,\dots,t_n;\textbf{m}),
\\\nonumber
\\\nonumber
&\mathcal{T}^{\text{full}}C^{r(a)}_{A^0\dots A^n}(t_0,\dots,t_n;\textbf{m})
\\
\label{eq:RetAdvCTId}
&\qquad=C^{a(r)}_{A^0\dots A^n}(t_0,\dots,t_n;\textbf{m}).
\end{align}

Crucially, in the form of \eqref{eq:RetAdvCKKmId}, we are provided a link between the time-inverted retarded correlator and magnetic-inverted advanced one, and vice-versa. This link is important, because it allows us to analyze the behaviour of the correlator under magnetic-inversion \textit{without} actually knowing the details of the equilibrium Hamiltonian's \textbf{m}-dependence. Indeed, all we need is the explicit time-dependence of the correlator which is known since it is provided by the interaction picture evolution.\\

On the one hand, the full time-inversion $\mathcal{K}^{\text{full}}$ and magnetic-inversion $\mathcal{K}^{\textbf{m}}$ operations, together with their product time-reversal $\mathcal{T}^{\text{full}}$, act as a representation of the group $\mathcal{G}_{\mathfrak{K}}\times\mathcal{G}_{\mathfrak{K}^{\textbf{m}}}$ on scalar-valued functions and were introduced as convenient tools to analyze the \textit{correlator}, a macroscopic quantity, whereas, on the other hand, $\mathcal{T}_{\mathsf{H}}$ represents time-reversal as an element of $\mathcal{G}_{\mathfrak{T}}$ on the Hilbert space and extends to actions on operators including the equilibrium Hamiltonian, $\mathcal{T}^{\dagger}_{\mathsf{H}}\mathcal{H}_0(\textbf{m})\mathcal{T}_{\mathsf{H}}=\mathcal{H}_0(-\textbf{m})$, thereby probing the microscopic realm. An important question we can ask pertains to the physical significance of these operations. The correlator eigenfunctions of magnetic-inversion $\mathcal{K}^{\textbf{m}}$ are linked to the `time-reversal' symmetry properties of the \textit{equilibrium Hamiltonian}: the positive eigenvalued eigenfunction gives the response should $\mathcal{H}_0$ be even in $\textbf{m}$, i.e., time-reversal symmetric if $\textbf{m}=0$, whereas the negative eigenvalued response requires $\mathcal{H}_0$ to be odd in $\textbf{m}$, and consequently the breaking of time-reversal symmetry. By virtue of the connection \eqref{eq:RetAdvCKKmId} between time-inversion and magnetic-inversion elucidated above, we can extract the latter eigenfunctions by means of the former. Note the different uses of `time-reversal'. In the context of the \textit{equilibrium Hamiltonian}, due to the latter's lack of explicit time-dependence, time reversal $\mathcal{T}_{\mathsf{H}}$ probes \textbf{m}-dependence, whereas in the case of the correlator, `time-reversal' $\mathcal{T}^{\text{full}}$ probes \textit{both} time and \textbf{m}-dependence, and it is magnetic-inversion $\mathcal{K}^{\textbf{m}}$ that looks at only \textbf{m}-dependence. Thus, the macroscopic counterpart of `time-reversal' when probing only $\mathcal{H}_0$ is magnetic-inversion and, by virtue of \eqref{eq:RetAdvCKKmId}, time-\textit{inversion}. How about the physical significance of $\mathcal{T}^{\text{full}}$? This operation is the macroscopic counterpart of $\mathcal{T}_{\mathsf{H}}$ when looking at all operators forming the correlator, not only $\mathcal{H}_0$. This means that it fully takes into account all explicit and implicit time-dependences of the full correlator and is thereby closely related to the direction of time. Indeed, the eigenfunctions of $\mathcal{T}^{\text{full}}$ have a definite sign under it and are thereby probes of dissipative properties. A dissipative response is sensitive to the arrow of time whereas a reactive response is not, meaning that the two eigenfunctions of $\mathcal{T}^{\text{full}}$ should correspond to these two responses.\\

In order to extract all of the eigenfunctions discussed in the previous paragraph, we have to take a look at \eqref{eq:RetAdvCKKmId} and \eqref{eq:RetAdvCTId}. The physical response is given by the retarded correlator, however, the latter is not an eigenfunction of either operation in the representation, so we have to decompose it into terms that are.  This is done by simply projecting the retarded correlator onto the irreducible representations of $\mathcal{G}_{\mathfrak{K}}\times\mathcal{G}_{\mathfrak{K}^{\textbf{m}}}$. Indeed, the Klein four-group is commutative, has four 1-dimensional irreducible representations, and, using its characters from Table \ref{tab:Z2Z2}. in the projectors \eqref{eq:ProjIrrep} in Appendix \ref{GroupTools}, we have

\begin{equation}
\begin{split}
&C^r_{A^0\dots A^n}(t_0,\dots,t_n;\textbf{m})
\\
&\qquad=\sum_{s,s_m\in\{+,-\}}C^{r,\mathcal{T}_s^{\text{full}},\mathcal{K}^{\textbf{m}}_{s_m}}_{A^0\dots A^n}(t_0,\dots,t_n;\textbf{m}),
\end{split}
\end{equation}

where

\begin{align}
\nonumber
C^{r,\mathcal{T}^{\text{full}}_{\pm},\mathcal{K}^{\textbf{m}}_{+}}_{A^0\dots A^n}=\frac{1}{4}(&C^r_{A^0\dots A^n}\pm\mathcal{K}^{\text{full}}C^r_{A^0\dots A^n}\\
&+\mathcal{K}^{\textbf{m}}C^r_{A^0\dots A^n}\pm\mathcal{T}^{\text{full}}C^r_{A^0\dots A^n}),
\\\nonumber
C^{r,\mathcal{T}^{\text{full}}_{\mp},\mathcal{K}^{\textbf{m}}_{-}}_{A^0\dots A^n}=\frac{1}{4}(&C^r_{A^0\dots A^n}\pm\mathcal{K}^{\text{full}}C^r_{A^0\dots A^n}
\\
&-\mathcal{K}^{\textbf{m}}C^r_{A^0\dots A^n}\mp\mathcal{T}^{\text{full}}C^r_{A^0\dots A^n}),
\end{align}

and we suppressed the arguments of the correlator for brevity. It is straightforward to see that $C^{r,\mathcal{T}_{\pm},\mathcal{K}^{\textbf{m}}_{\pm}}_{A^0\dots A^n}$ are indeed eigenfunctions of $\mathcal{K}^{\text{full}},\,\mathcal{K}^{\textbf{m}}$ and $\mathcal{T}^{\text{full}}$, with the superscripts indicating the behaviour under the latter. The utility of relations \eqref{eq:RetAdvCKKmId} and \eqref{eq:RetAdvCTId} now becomes apparent. The eigenfunctions $C^{r,\mathcal{T}^{\text{full}}_{\pm},\mathcal{K}^{\textbf{m}}_{\pm}}_{A^0\dots A^n}$ depend explicitly on the correlator with a reversed \textbf{m} meaning that we would need to know the detailed \textbf{m}-dependence of $\mathcal{H}_0$ to proceed with any general analysis, however, by virtue of the aforementioned relations, we can swap magnetic-inversion with a time-inversion, albeit, of the advanced correlator. We are thus left with

\begin{equation}
\label{eq:RDTRSTRB}
\begin{split}
&C^{r,\mathcal{T}^{\text{full}}_{\pm},\mathcal{K}^{\textbf{m}}_{+}}_{A^0\dots A^n}=\frac{1}{4}\left(C^{r\pm a}_{A^0\dots A^n}\pm\mathcal{K}^{\text{full}}C^{r\pm a}_{A^0\dots A^n}\right),
\\
&C^{r,\mathcal{T}^{\text{full}}_{\mp},\mathcal{K}^{\textbf{m}}_{-}}_{A^0\dots A^n}=\frac{1}{4}\left(C^{r\mp a}_{A^0\dots A^n}\pm\mathcal{K}^{\text{full}}C^{r\mp a}_{A^0\dots A^n}\right),
\end{split}
\end{equation}

where we defined

\begin{equation}
C^{r\pm a}_{A^0\dots A^n}\equiv C^{r}_{A^0\dots A^n}\pm C^a_{A^0\dots A^n}.
\end{equation}

Decomposition \eqref{eq:RDTRSTRB} highlights the facts that the dissipative response is always described by the difference $C^{r-a}$ whereas the reactive response by the sum $C^{r+a}$. We can also rewrite \eqref{eq:RDTRSTRB} by keeping manifest the behaviour under magnetic-inversion; a formulation that will be of use to us when moving to the frequency-domain. Indeed, defining

\begin{equation}
\label{eq:RetMagInv}
\begin{split}
C^{r,\mathcal{K}^{\textbf{m}}_{\pm}}_{A^0\dots A^n}&=C^{r,\mathcal{T}^{\text{full}}_{+},\mathcal{K}^{\textbf{m}}_{\pm}}_{A^0\dots A^n}+C^{r,\mathcal{T}^{\text{full}}_{-},\mathcal{K}^{\textbf{m}}_{\pm}}_{A^0\dots A^n}
\\
&=\frac{1}{2}\left(C^{r}_{A^0\dots A^n}\pm\mathcal{K}^{\text{full}}C^{a}_{A^0\dots A^n}\right),
\end{split}
\end{equation}

decomposition \eqref{eq:RDTRSTRB} becomes

\begin{equation}
\label{eq:RDTRSTRBmagInv}
\begin{split}
&C^{r,\mathcal{T}^{\text{full}}_{\pm},\mathcal{K}^{\textbf{m}}_{+}}_{A^0\dots A^n}=\frac{C^{r,\mathcal{K}^{\textbf{m}}_{+}}_{A^0\dots A^n}\pm \mathcal{K}^{\text{full}}C^{r,\mathcal{K}^{\textbf{m}}_{+}}_{A^0\dots A^n}}{2},
\\
&C^{r,\mathcal{T}^{\text{full}}_{\mp},\mathcal{K}^{\textbf{m}}_{-}}_{A^0\dots A^n}=\frac{C^{r,\mathcal{K}^{\textbf{m}}_{-}}_{A^0\dots A^n}\pm \mathcal{K}^{\text{full}}C^{r,\mathcal{K}^{\textbf{m}}_{-}}_{A^0\dots A^n}}{2}.
\end{split}
\end{equation}

The interplay between dissipative/reactive responses and the time-reversal symmetry/breaking of $\mathcal{H}_0$, summarized in Table \ref{tab:RDTRSB},  is thus laid bare and their correspondence can be explicitly formulated via the irreducible representations of $\mathcal{G}_{\mathfrak{K}}\times\mathcal{G}_{\mathfrak{K}^{\textbf{m}}}$  with the specific responses expressed using combinations of the retarded and advanced correlators. 

\begin{table}[h]
\label{tab:RDTRSB}
\begin{center}
\renewcommand{\arraystretch}{1.8}
\begin{tabular}{c|c|c|}
\cline{2-3}
&Reactive&Dissipative\\\hline
\multicolumn{1}{|c|}{TRS}&$C^{r,\mathcal{T}^{\text{full}}_{+},\mathcal{K}^{\textbf{m}}_{+}}_{A^0\dots A^n}$&$C^{r,\mathcal{T}^{\text{full}}_{-},\mathcal{K}^{\textbf{m}}_{+}}_{A^0\dots A^n}$\\\hline
\multicolumn{1}{|c|}{TRB}&$C^{r,\mathcal{T}^{\text{full}}_{+},\mathcal{K}^{\textbf{m}}_{-}}_{A^0\dots A^n}$&$C^{r,\mathcal{T}^{\text{full}}_{-},\mathcal{K}^{\textbf{m}}_{-}}_{A^0\dots A^n}$\\\hline
\end{tabular}
\renewcommand{\arraystretch}{1}
\end{center}
\caption{Decomposition of the retarded correlator. TRS(TRB) refers to whether $\mathcal{H}_0$ is time-reversal symmetric(time-reversal breaking).}
\end{table}

While the time-domain analysis presented above is formally satisfying, practical calculations are most straightforward in the frequency-domain. Making use of the correlators' time-translation invariance (see Appendix \ref{RetPlain}.\ref{TimeTrans}),

\begin{equation}
\nonumber
\begin{split}
C^{r(a)}_{A^0A^1\dots A^n}&(t_0,t_1,\dots,t_n;\textbf{m})
\\
&=C^{r(a)}_{A^0A^1\dots A^n}(0,t_1-t_0,\dots,t_n-t_0;\textbf{m}),
\end{split}
\end{equation}

and relabeling the time arguments, $t_0-t_i\to t_i$, we can define

\begin{equation}
\nonumber
\begin{split}
\tilde{C}^{r(a)}_{A^0A^1\dots A^n}&(t_1,\dots,t_n;\textbf{m})
\\
&\equiv C^{r(a)}_{A^0A^1\dots A^n}(0,-t_1,\dots,-t_n;\textbf{m}),
\end{split}
\end{equation}

and take the Fourier transform

\begin{equation}
\label{eq:Fourier}
\begin{split}
&\tilde{C}^{r(a)}_{A^0\dots A^n}(\omega_1,\dots,\omega_n;\textbf{m})
\\
&=\int dt_1\cdots\int dt_n \tilde{C}^{r(a)}_{A^0\dots A^n}(t_1,\dots,t_n;\textbf{m})e^{i\omega_1 t_1}\cdots e^{i\omega_n t_n}.
\end{split}
\end{equation}

The reality of the time-domain correlator $\tilde{C}^{r(a)}_{A^0\dots A^n}(t_1,\dots,t_n;\textbf{m})$ requires the latter's Fourier transform to satisfy $\tilde{C}^{r(a)^*}_{A^0\dots A^n}(\omega_1,\dots,\omega_n;\textbf{m})=\tilde{C}^{r(a)}_{A^0\dots A^n}(-\omega_1,\dots,-\omega_n;\textbf{m})$. The frequency-domain expression of \eqref{eq:RetMagInv} thus becomes

\begin{equation}
\label{eq:RetMagInvFreq}
\begin{split}
\tilde{C}^{r,\mathcal{K}^{\textbf{m}}_{\pm}}_{A^0\dots A^n}&=\frac{1}{2}\left(\tilde{C}^{r}_{A^0\dots A^n}\pm\epsilon_{A^0}^{\mathcal{T}}\cdots\epsilon_{A^n}^{\mathcal{T}}\tilde{C}^{a^*}_{A^0\dots A^n}\right),
\end{split}
\end{equation}

where we dropped the frequency arguments for brevity, and decomposition \eqref{eq:RDTRSTRBmagInv} can be written as

\begin{equation}
\label{eq:RDTRSTRBmagInvFreq}
\begin{split}
\tilde{C}^{r,\mathcal{T}^{\text{full}}_{\pm},\mathcal{K}^{\textbf{m}}_{+}}_{A^0\dots A^n}&=\frac{\tilde{C}^{r,\mathcal{K}^{\textbf{m}}_{+}}_{A^0\dots A^n}\pm \epsilon_{A^0}^{\mathcal{T}}\cdots\epsilon_{A^n}^{\mathcal{T}}(\tilde{C}^{r,\mathcal{K}^{\textbf{m}}_{+}}_{A^0\dots A^n})^*}{2}
\\
&\equiv\hat{\mathcal{P}}^{(\pm\text{sgn}(\epsilon))}_{\mathcal{K}^*}\tilde{C}^{r,\mathcal{K}^{\textbf{m}}_{+}}_{A^0\dots A^n} ,
\\
\tilde{C}^{r,\mathcal{T}^{\text{full}}_{\mp},\mathcal{K}^{\textbf{m}}_{-}}_{A^0\dots A^n}&=\frac{\tilde{C}^{r,\mathcal{K}^{\textbf{m}}_{-}}_{A^0\dots A^n}\pm \epsilon_{A^0}^{\mathcal{T}}\cdots\epsilon_{A^n}^{\mathcal{T}}(\tilde{C}^{r,\mathcal{K}^{\textbf{m}}_{-}}_{A^0\dots A^n})^*}{2}
\\
&\equiv \hat{\mathcal{P}}^{(\pm\text{sgn}(\epsilon))}_{\mathcal{K}^*}\tilde{C}^{r,\mathcal{K}^{\textbf{m}}_{-}}_{A^0\dots A^n},
\end{split}
\end{equation}

where we defined $\text{sgn}(\epsilon)\equiv\text{sign}(\epsilon_{A^0}^{\mathcal{T}}\cdots\epsilon_{A^n}^{\mathcal{T}})$ and the projector to the real or imaginary parts of a function $f$

\begin{equation}
\label{eq:ReImProj}
\hat{\mathcal{P}}^{(\pm\text{sgn}(\epsilon))}_{\mathcal{K}^*}f=\frac12(f\pm\text{sgn}(\epsilon)f^*).
\end{equation}

It is thus clear that the reactive and dissipative properties of the responses are, \textit{not} respectively, linked to the real and imaginary parts of the \textit{magnetic-inversion eigenfunctions}.\\

\subsubsection{Application to charge conductivity}

We now look at some examples. Consider the first order charge current response to a spatially uniform, time-varying electric field. Then, $A^0=J_{i_0},\,A^1=J_{i_1}$ where $J_{i_0},\,J_{i_1}$ are charge current operators in the directions $i_0,i_1\in\{x,y,z\}$, and $\epsilon_{J_{i_0}}^{\mathcal{T}}=\epsilon_{J_{i_1}}^{\mathcal{T}}=-1\rightarrow \epsilon_{J_{i_0}}^{\mathcal{T}}\epsilon_{J_{i_1}}^{\mathcal{T}}=1$. The first order charge conductivity $\sigma^r_{i_0i_1}(\omega_1)\equiv\sigma_{i_0i_1}(\omega_1)$ is given in terms of the 2-point retarded correlator by the Kubo formula \cite{Kubo1957}

\begin{equation}
\label{eq:Cond1}
-i\omega_1\sigma^r_{i_0i_1}(\omega_1)=\tilde{C}^{r}_{J_{i_0}J_{i_1}}(\omega_1)-\tilde{C}^{r}_{J_{i_0}J_{i_1}}(0),
\end{equation}

and by defining the `advanced conductivity',

\begin{equation}
-i\omega_1\sigma^a_{i_0i_1}(\omega_1)=\tilde{C}^{a}_{J_{i_0}J_{i_1}}(\omega_1)-\tilde{C}^{a}_{J_{i_0}J_{i_1}}(0),
\end{equation}

we can express the magnetic-inversion eigenfunctions corresponding to \eqref{eq:RetMagInvFreq} as

\begin{equation}
\label{eq:MagInvCond1}
\sigma^{\mathcal{K}^{\textbf{m}}_{\pm}}_{i_0i_1}(\omega_1)=\frac{\sigma^r_{i_0i_1}(\omega_1)\mp\sigma^{a^*}_{i_0i_1}(\omega_1)}{2}.
\end{equation}

By \eqref{eq:RDTRSTRBmagInvFreq}, the dissipative and reactive parts of the first order conductivity then reduce to

\begin{equation}
\label{eq:RDCond1}
\begin{split}
&\sigma^{\mathcal{T}^{\text{full}}_{\pm},\mathcal{K}^{\textbf{m}}_{+}}_{i_0i_1}(\omega_1)=\frac{\sigma^{\mathcal{K}^{\textbf{m}}_{+}}_{i_0i_1}(\omega_1)\mp \left(\sigma^{\mathcal{K}^{\textbf{m}}_{+}}_{i_0i_1}(\omega_1)\right)^*}{2},
\\
&\sigma^{\mathcal{T}^{\text{full}}_{\mp},\mathcal{K}^{\textbf{m}}_{-}}_{i_0i_1}(\omega_1)=\frac{\sigma^{\mathcal{K}^{\textbf{m}}_{-}}_{i_0i_1}(\omega_1)\mp \left(\sigma^{\mathcal{K}^{\textbf{m}}_{-}}_{i_0i_1}(\omega_1)\right)^*}{2}.
\end{split}
\end{equation}

Thus, to obtain a reactive response from a time-reversal symmetric equilibrium Hamiltonian---that manifests on the correlator level as invariance under magnetic-inversion---, we need $\sigma^{\mathcal{T}^{\text{full}}_{+},\mathcal{K}^{\textbf{m}}_{+}}_{i_0i_1}(\omega_1)$ to be non-vanishing. However, this is only fulfilled if the imaginary part of $\sigma^{\mathcal{K}^{\textbf{m}}_{+}}_{i_0i_1}(\omega_1)$ is non-zero, which, since $(\sigma^{r(a)}(\omega_1))^*=\sigma^{r(a)}(-\omega_1)$, requires $\omega_1\neq 0$. This means that, in the static $\omega_1\to 0$ limit, the charge current response of a time-reversal invariant system to an electric field is necessarily dissipative. On the other hand, the reactive response $\sigma^{\mathcal{T}^{\text{full}}_{+},\mathcal{K}^{\textbf{m}}_{-}}_{i_0i_1}(\omega_1)$ is given by the real part of $\sigma^{\mathcal{K}^{\textbf{m}}_{-}}_{i_0i_1}(\omega_1)$, and can thus remain finite even in the static limit, albeit the time-reversal symmetry of $\mathcal{H}_0$ must be broken. These are well-known results for the charge conductivity (see ref. \cite{NagaosaAHErev} for example), but are usually presented via energy arguments and not group representations.\\

Now let us look at the second order charge conductivity $\sigma^r_{i_0i_1i_2}(\omega_1,\omega_2)\equiv\sigma_{i_0i_1i_2}(\omega_1,\omega_2)$. In \cite{Bonbien2021b}, we show that this can be expressed using frequency-domain retarded 2 and 3-point correlators as

\begin{widetext}
\begin{equation}
\label{eq:Cond2}
\begin{split}
\omega_1\omega_2\sigma^r_{i_0i_1i_2}(\omega_1,\omega_2)=&
\tilde{C}^{r}_{J_{i_0}J_{i_1i_2}}(\omega_1+\omega_2)-\tilde{C}^{r}_{J_{i_0}J_{i_1i_2}}(\omega_1)-\tilde{C}^{r}_{J_{i_0}J_{i_1i_2}}(\omega_2)+\tilde{C}^{r}_{J_{i_0}J_{i_1i_2}}(0)
\\
&-(\tilde{C}^{r}_{J_{i_0}J_{i_1}J_{i_2}}(\omega_1,\omega_2)-\tilde{C}^{r}_{J_{i_0}J_{i_1}J_{i_2}}(\omega_1,0)-\tilde{C}^{r}_{J_{i_0}J_{i_1}J_{i_2}}(0,\omega_2)+\tilde{C}^{r}_{J_{i_0}J_{i_1}J_{i_2}}(0,0)),
\end{split}
\end{equation}
\end{widetext}

where $J_{i_1i_2}=\frac{ie}{4\hbar}([r_{i_1},J_{i_2}]+[r_{i_2},J_{i_1}])$ with $r_k$ being the $k$-th coordinate component of the position operator. We clearly have $\epsilon_{J_{i_1i_2}}^{\mathcal{T}}=1$, meaning that  $\epsilon_{J_{i_0}}^{\mathcal{T}}\epsilon_{J_{i_1i_2}}^{\mathcal{T}}=\epsilon_{J_{i_0}}^{\mathcal{T}}\epsilon_{J_{i_1}}^{\mathcal{T}}\epsilon_{J_{i_2}}^{\mathcal{T}}=-1$. Defining the `advanced second order charge conductivity' $\sigma^{a}_{i_0i_1i_2}(\omega_1,\omega_2)$ by exchanging all the retarded correlators to advanced ones in \eqref{eq:Cond2}, and using \eqref{eq:RetMagInvFreq}, we obtain

\begin{equation}
\label{eq:TRCond2}
\sigma^{\mathcal{K}^{\textbf{m}}_{\pm}}_{i_0i_1i_2}(\omega_1,\omega_2)=\frac{\sigma^r_{i_0i_1i_2}(\omega_1,\omega_2)\mp\sigma^{a^*}_{i_0i_1i_2}(\omega_1,\omega_2)}{2},
\end{equation}

with the reactive and dissipative conductivities becoming

\begin{equation}
\label{eq:RDCond2}
\begin{split}
&\sigma^{\mathcal{T}^{\text{full}}_{\pm},\mathcal{K}^{\textbf{m}}_{+}}_{i_0i_1i_2}(\omega_1,\omega_2)=\frac{\sigma^{\mathcal{K}^{\textbf{m}}_{+}}_{i_0i_1i_2}(\omega_1,\omega_2)\mp \left(\sigma^{\mathcal{K}^{\textbf{m}}_{+}}_{i_0i_1i_2}(\omega_1,\omega_2)\right)^*}{2},
\\
&\sigma^{\mathcal{T}^{\text{full}}_{\mp},\mathcal{K}^{\textbf{m}}_{-}}_{i_0i_1i_2}(\omega_1,\omega_2)=\frac{\sigma^{\mathcal{K}^{\textbf{m}}_{-}}_{i_0i_1i_2}(\omega_1,\omega_2)\mp \left(\sigma^{\mathcal{K}^{\textbf{m}}_{-}}_{i_0i_1i_2}(\omega_1,\omega_2)\right)^*}{2},
\end{split}
\end{equation}

where we used \eqref{eq:RDTRSTRBmagInvFreq}. We can note the similarity to the corresponding expression \eqref{eq:RDCond1} for the first order charge conductivity. In the special case of a time-reversal symmetric $\mathcal{H}_0$, we have $\textbf{m}=0$ and $\sigma^{\mathcal{K}^{\textbf{m}}_{-}}_{i_0i_1i_2}(\omega_1,\omega_2)=0 \leftrightarrow \sigma^{\mathcal{K}^{\textbf{m}}_{+}}_{i_0i_1i_2}(\omega_1,\omega_2)=\sigma^{r}_{i_0i_1i_2}(\omega_1,\omega_2)$ meaning that the reactive (dissipative) conductivities correspond directly to the imaginary (real) part of the total conductivity $\sigma^{r}_{i_0i_1i_2}(\omega_1,\omega_2)$. In the context of nonlinear optics, the second order electrical susceptibility $\chi_{i_0i_1i_2}(\omega_1,\omega_2)$ providing the polarization response \cite{BoydNLO} is the preferred quantity of interest and is related to the conductivity as $\sigma^r_{i_0i_1i_2}(\omega_1,\omega_2)\propto -i(\omega_1+\omega_2)\chi_{i_0i_1i_2}(\omega_1,\omega_2)$. For  $\textbf{m}=0$, this simply means that the reactive (dissipative) contribution will be from the real (imaginary) part of $\chi_{i_0i_1i_2}(\omega_1,\omega_2)$ in complete consistency with the energy arguments present in the literature \cite{BoydNLO,SipeShkrebtii2000}. Note, however, that the relation of the total response function's reality to reaction and dissipation is not particularly straightforward in the time-reversal breaking case; in fact, the real and imaginary parts of the magnetic-inversion eigenfunctions are the ones that we should consider.\\

We can thus draw the conclusion that, when performing practical calculations in the frequency-domain, we should first find the eigenfunctions of magnetic-inversion according to \eqref{eq:RetMagInvFreq} and the real/imaginary parts of the result will directly yield the responses that are reactive or dissipative.\\

As a final note, we look at the static limit. In this case, the correlator $\tilde{C}_{A^0\dots A^n}^r(\textbf{m})$ is independent of time explicitly and the representation $\mathcal{K}$ of time-\textit{inversion} $\mathfrak{K}$ acts on it as the identity, i.e., $\mathcal{K}\tilde{C}_{A^0\dots A^n}^r(\textbf{m})=\tilde{C}_{A^0\dots A^n}^r(\textbf{m})$. This means that the $\mathcal{T}$ representation of time-\textit{reversal} $\mathfrak{T}$ on the correlator becomes equivalent to the $\mathcal{K}^{\textbf{m}}$ representation of magnetic-inversion $\mathfrak{K}^{\textbf{m}}$, i.e., $\mathcal{T}\tilde{C}_{A^0\dots A^n}^r(\textbf{m})=\mathcal{K}^{\textbf{m}}\tilde{C}_{A^0\dots A^n}^r(\textbf{m})=\tilde{C}_{A^0\dots A^n}^r(-\textbf{m})$. This can lead to significant confusion since both of the latter operations can be considered as a `time-reversal'. Added to this, we also have the `full' representation of time-reversal $\mathcal{T}^{\text{full}}$ acting as $\mathcal{T}^{\text{full}}\tilde{C}_{A^0\dots A^n}^r(\textbf{m})=\epsilon^{\mathcal{T}}_{A^0}\dots\epsilon^{\mathcal{T}}_{A^n}\tilde{C}_{A^0\dots A^n}^r(-\textbf{m})$ that takes into account the properties of the physical observables and can likewise be dubbed `time-reversal'. In fact, as discussed at length throughout this section, it is the latter $\mathcal{T}^{\text{full}}$ representation that probes the `complete' time-dependence, and thereby directly relates to irreversibility and dissipation. The $\mathcal{T}$ representation is then, in this static case, best thought of as magnetic-inversion. For example, should the \textbf{m}-dependence of our correlators label a magnetic structure, the latter's symmetry is, in a large number of cases, conveniently described by one of the magnetic groups \cite{DresselhausSymm,BurnsGlazer}. The point operation of the magnetic groups that reverses the magnetic moments i.e., takes $\textbf{m}$ to $-\textbf{m}$,---which, in many instances, is also referred to as `time-reversal'---then corresponds to precisely the action of magnetic-inversion $\mathcal{K}^{\textbf{m}}$ on the correlator (see also footnote \cite{Note1}) . Since we are in the static limit, this is equivalent to $\mathcal{T}$ and provides a realization of the $\mathcal{T}$ representation of time-reversal, albeit \textit{only} in this limit.\\

This `abundance' of time-reversal operations led to significant confusion and controversy for decades \cite{Grimmer1993} and clarity for general correlators is yet to be widely reflected within the literature. The primary source of confusion is related to the constitutive relation---$J_i=\sigma_{ij}E_j$ in the case of first order charge current. Naïvely, we might then consider the charge current $J_i$ to be odd under full time-reversal, the electric field to be even, hence the conductivity to be odd and the response to be necessarily be dissipative. However, this constitutive relation stems from a calculation of the charge current operator's expectation value to first order in the electric field (see \eqref{eq:Obsn} with $n=1$, for the general case) and is therefore, by definition, dependent on $\textbf{m}$ through the Hamiltonian. Thus, maintaining consistency, we should be writing $J_i(\textbf{m})=\sigma_{ij}(\textbf{m})E_j$, and the sign under full time-reversal is no longer odd, i.e., the left hand side becomes $J_i(\textbf{m})\to -J_i(-\textbf{m})$ and consequently $\sigma_{ij}(\textbf{m})\to-\sigma_{ij}(-\textbf{m})$ giving rise to both reactive and dissipative responses, whereas under magnetic-inversion $J_i(\textbf{m})\to J_i(-\textbf{m})$ leading to $\sigma_{ij}(\textbf{m})\to\sigma_{ij}(-\textbf{m})$, providing a probe of only the \textbf{m}-dependence. The natural need to separately look at magnetic-inversion and full time-reversal becomes clearly apparent (see also \cite{BonbienRev2021} for an elementary account of this). Conversely, for example, the author of \cite{Grimmer1993} asserts that differentiating between time-reversal and magnetic-inversion, the latter referred to as `magnetic-reversal', leads to ``unnecessary complications" and maintains the widespread view that Onsager's reciprocity principal has to be invoked for a consistent discussion. It is, however, abundantly clear that, in general, all of these operations \textit{must} be considered in their own right, \textit{cannot} be conflated, and, moreover, notwithstanding the fact that Onsager's reciprocity is only valid for 2-point correlators (see Section \ref{OnsagerSection} below), as demonstrated throughout this subsection, the latter does not even have to be mentioned for a general discussion of temporal effects. The group theoretical perspective of this section was aimed at offering clarity to the meaning and proper use of each operation.\\

\subsection{Kramers-Kronig relations}

Both the retarded and advanced correlators, as respectively defined in \eqref{eq:Retn} and \eqref{eq:Advn}, are time-ordered combinations of stripped correlators and only differ in their time-orderings, the latter being ordered in reverse with respect to the former. The reactive and dissipative responses \eqref{eq:RDTRSTRB} are respectively given in terms of the sum and difference of the retarded and advanced correlators. Owing to the fact that both the sum and difference of the retarded and advanced correlators contain both forward and reverse time-orderings of the same stripped correlators, we might expect a relation between the sum and the difference thereby providing a close link between the reactive and dissipative responses. This link is the Kramers-Kronig relation \cite{Peiponen2009}, which we present in the time-domain and show that it is trivial.\par

From \eqref{eq:Retn} and \eqref{eq:Advn} we have for the sum and difference of the retarded and advanced correlators

\begin{equation}
\label{eq:CRpmA}
\begin{split}
&C^{r\pm a}_{A^0\dots A^n}=\frac{(-i)^{n}}{\hbar^nn!}\sum_{\pi\in S_n}\bigg(\theta_{t_0 t_{\pi(1)}}\cdots\theta_{t_{\pi(n-1)}t_{\pi(n)}}
\\
&\qquad\pm(-1)^{n}\theta_{t_{\pi(n)}t_{\pi(n-1)}}\cdots\theta_{t_{\pi(1)}t_0}\bigg)C_{[A^0A^{\pi(1)}\dots A^{\pi(n)}]},
\end{split}
\end{equation}

where, keeping brevity in sight, we suppressed the time arguments of the correlators but, specifically for the stripped correlators, these are also being permuted, i.e., $C_{[A^0A^{\pi(1)}\dots A^{\pi(n)}]}\equiv C_{[A^0A^{\pi(1)}\dots A^{\pi(n)}]}(t_0,t_{\pi(1)},\dots,t_{\pi(n)})$. Now suppose we multiply both sides of \eqref{eq:CRpmA} with $\theta_{t_0t_i}-\theta_{t_it_0}$ for some $i\in\{1,\dots,n\}$. Then, we have for the products of step functions

\begin{equation}
\nonumber
\begin{split}
&(\theta_{t_0t_i}-\theta_{t_it_0})\bigg(\theta_{t_0 t_{\pi(1)}}\cdots\theta_{t_{\pi(n-1)}t_{\pi(n)}}
\\
&\qquad\qquad\qquad\qquad\pm(-1)^{n}\theta_{t_{\pi(n)}t_{\pi(n-1)}}\cdots\theta_{t_{\pi(1)}t_0}\bigg)
\\
&=\theta_{t_0 t_{\pi(1)}}\cdots\theta_{t_{\pi(n-1)}t_{\pi(n)}}\mp(-1)^{n}\theta_{t_{\pi(n)}t_{\pi(n-1)}}\cdots\theta_{t_{\pi(1)}t_0},
\end{split}
\end{equation}

where we used the fact that $\theta_{t_0t_i}\theta_{t_it_0}=0$--- more precisely, in the forward time-order, $t_0>t_i$ for all $i\in\{1,\dots,n\}$, whereas in the reverse time-order, $t_i>t_0$, so, when multiplying with $\theta_{t_0t_i}-\theta_{t_it_0}$, the `mixed' time-orders vanish and we are left with the initial forward and reverse time-orders, albeit with a sign change. Similarly, multiplying both sides of \eqref{eq:CRpmA} by $(\theta_{t_0t_{i_1}}-\theta_{t_{i_1}t_0})(\theta_{t_0t_{i_2}}-\theta_{t_{i_2}t_0})\cdots (\theta_{t_0t_{i_{2k+1}}}-\theta_{t_{i_{2k+1}}t_0})$, where $k=0,1,\dots,\left\lfloor{\frac{n-1}{2}}\right\rfloor$, we get the same result. Thus

\begin{equation}
\label{eq:KKnRA}
\begin{split}
&C^{r\pm a}_{A^0\dots A^n}(t_0,\dots,t_n)=(\theta_{t_0t_{i_1}}-\theta_{t_{i_1}t_0})(\theta_{t_0t_{i_2}}-\theta_{t_{i_2}t_0})\times\cdots
\\
&\qquad\times\cdots(\theta_{t_0t_{i_{2k+1}}}-\theta_{t_{i_{2k+1}}t_0})C^{r\mp a}_{A^0\dots A^n}(t_0,\dots,t_n).
\end{split}
\end{equation}

These are all general Kramers-Kronig relations connecting $C^{r+a}$ with $C^{r-a}$ and applying them to \eqref{eq:RDTRSTRB} provides links between the reactive and dissipative responses. Moving to the frequency-domain yields their usual form and, most importantly, in general, by virtue of \eqref{eq:RDTRSTRBmagInvFreq}, they only connect the real and imaginary parts of the magnetic-inversion eigenfunctions, not of the total response function.\\

It is apparent that, contrary to the usual treatment \cite{Peiponen2009}, the Kramers-Kronig relations are merely corollaries of the definition of retarded and advanced correlators and can be expressed perfectly well in the time-domain without reference to complex analysis and the frequency-domain. In fact, it is the former perspective that arms us with the insights necessary for a natural interpretation of their general physical meaning.\\
 
\section{Fluctuation-dissipation, fluctuation-reaction, and sum rules}
\label{FlucDissReact}

The retarded and advanced correlators \eqref{eq:Retn} and \eqref{eq:Advn} are expressed as time-ordered combinations of stripped correlators

\begin{equation}
\nonumber
\begin{split}
C_{A^0A^1A^2\dots}^{r,a}\propto
&\theta(\dots)\cdots\theta(\dots) C_{[A^0A^1A^2\dots]}
\\
&+\theta(\dots)\cdots\theta(\dots)C_{[A^0A^2A^1\dots]}
\\
&+\cdots,
\end{split}
\end{equation}

in other words, $C^r\sim f_r(C_{[\dots]}),C^a\sim f_a(C_{[\dots]})$, where $f_r,f_a$ indicate that $C_r,C_a$ are functions of the stripped correlators $C_{[\dots]}$.  Thus, a natural question we can ask is whether it is possible to express the stripped correlators as a function of retarded and advanced ones $C_{[\dots]}\propto^{?} f(C^{r\pm a})$? This is precisely what leads to a fluctuation-dissipation or fluctuation-reaction theorem, since the stripped correlator without time-ordering can be considered as a `fluctuation' and the sum and difference of retarded and advanced correlators contribute to, respectively, the reactive and dissipative responses. That such an expression, $C_{[A^0A^1]}\propto f_2(C^{r-a})$ exists for 2-point correlators has been widely known throughout the last century and is, in fact, the statement of the standard `linear' fluctuation-dissipation theorem \cite{Kubo1957,Kubo1966, RadiJishi}. The fact that a fluctuation-reaction relation $C_{[A^0A^1A^2]}\propto f_3(C^{r+a})$ for 3-point correlators, analogous to the fluctuation-dissipation relation $C_{[A^0A^1]}\propto f_2(C^{r-a})$ for 2-point correlators, exists \cite{Kalman1987} has, on the other hand, not received comparable attention. In order to remedy this rather tatterdemalion state of affairs, following an exceptionally simple derivation of this relation, we relate it to a specific important contribution to the second order conductivity expression \eqref{eq:Cond2}. An important further perspective that arises when moving from 2 to 3-point correlators is the possibility of `partial' time ordering. Indeed, a 3-point correlator, such as the stripped correlator $C_{[A^0A^1A^2]}(t_0,t_1,t_2)$, has three time arguments and ordering only two of them, say $t_0>t_1$, to give $\theta_{t_0t_1}C_{[A^0A^1A^2]}(t_0,t_1,t_2)$ leaves the third one, $t_2$, unconstrained. We can then obtain a `weak' fluctuation-\textit{dissipation} relation by enforcing the partial time-ordering on the fluctuation-reaction relation $C_{[A^0A^1A^2]}\propto f_3(C^{r+a})$ to be presented below. Bringing to forefront our guinea pig, the second order charge conductivity, we find that this `weak' fluctuation-dissipation relation is far less restrictive than the fluctuation-reaction relation and underlies a large number of dissipative second order conductivity responses.\\

An important consequence of the fluctuation-dissipation and fluctuation-reaction theorems is the existence of certain frequency sum rules. The fundamental reason the latter arise is due to the fact that the time arguments of the stripped correlator are not ordered and thereby have free rein on what values to take. Indeed, heuristically, on a technical level, the time arguments arise through the interaction picture time evolution factors $e^{-\frac{i}{\hbar}\mathcal{H}_0t}Ae^{\frac{i}{\hbar}\mathcal{H}_0t}$ for some observable operator $A$, and without time-ordering each of them can simply be written as

\begin{equation}
e^{-\frac{i}{\hbar}\mathcal{H}_0t}A e^{\frac{i}{\hbar}\mathcal{H}_0t}=\int d\varepsilon'\int d\varepsilon \delta(\varepsilon'-\mathcal{H}_0)A\delta(\varepsilon-\mathcal{H}_0)e^{-\frac{i}{\hbar}(\varepsilon'-\varepsilon)t},
\end{equation}

which, upon moving to the frequency domain, becomes

\begin{equation}
\int dt \left( e^{-\frac{i}{\hbar}\mathcal{H}_0t}A e^{\frac{i}{\hbar}\mathcal{H}_0t}\right) e^{i\omega t} \propto \int d\varepsilon \delta(\varepsilon+\hbar\omega-\mathcal{H}_0)A\delta(\varepsilon-\mathcal{H}_0).
\end{equation}

The stripped correlator is the trace of such expressions and upon expanding the trace in a basis of $\mathcal{H}_0$ eigenstates, we can perform the $\varepsilon$ integration leaving a delta function dependent on frequency. The delta function restricts the frequency thereby allowing it to be integrated out. On the other hand, should the time arguments be ordered, we would have, instead, to deal with Green's functions, e.g. $\theta(t)e^{-\frac{i}{\hbar}\mathcal{H}_0t}Ae^{\frac{i}{\hbar}\mathcal{H}_0t}=i\hbar G^r(t)Ae^{\frac{i}{\hbar}\mathcal{H}_0t}$ (see Appendix \ref{SpectRep} for details) with

\begin{equation}
\begin{split}
&\theta(t)e^{-\frac{i}{\hbar}\mathcal{H}_0t}Ae^{\frac{i}{\hbar}\mathcal{H}_0t}=i\hbar G^r(t)Ae^{\frac{i}{\hbar}\mathcal{H}_0t}
\\
&=i\hbar\int\frac{d\varepsilon'}{2\pi}\int d\varepsilon\frac{1}{\varepsilon'-\mathcal{H}_0+i\eta}A\delta(\varepsilon-\mathcal{H}_0)e^{-\frac{i}{\hbar}(\varepsilon'-\varepsilon) t}
\\
&=i\hbar\int\frac{d\varepsilon'}{2\pi}\int d\varepsilon \,G^r(\varepsilon')A\delta(\varepsilon-\mathcal{H}_0)e^{-\frac{i}{\hbar}(\varepsilon'-\varepsilon) t}.
\end{split}
\end{equation}

Moving to the frequency domain,

\begin{equation}
\begin{split}
\int dt&\left(\theta(t)e^{-\frac{i}{\hbar}\mathcal{H}_0t}Ae^{\frac{i}{\hbar}\mathcal{H}_0t}\right) e^{i\omega t}
\\
& = i\hbar^2\int d\varepsilon \,G^r(\varepsilon+\hbar\omega)A\delta(\varepsilon-\mathcal{H}_0),
\end{split}
\end{equation}

expanding the trace in a basis of $\mathcal{H}_0$ eigenstates, and performing the $\varepsilon$ integration, the frequency will, in general, no longer be an argument of a delta function and we cannot integrate it out. In this way, the frequency integral of the dissipation or reaction `side' of the theorems is equivalent to a frequency-independent expression on the fluctuation `side'. In the partially ordered or `weak' case, since one of the time arguments in the stripped correlator is still unconstrained, moving to the frequency-domain will continue to leave us with a delta function with a frequency argument that will succumb to frequency integration, and thus provide us with a family of novel sum rules.

\subsection{Fluctuation-dissipation and the 2-point correlator}

In order to prepare our approach to the 3-point correlator, we first review the standard 2-point expression from our point of view. We have for the step function $\theta(t)+\theta(-t)=1$, meaning that \textit{any} function can be expressed as $f(t)=(\theta(t)+\theta(-t))f(t)=\theta(t)f(t)+\theta(-t)f(t)$. This means that

\begin{equation}
\label{eq:2FlucDiss}
\begin{split}
&C_{[A^0A^1]}(t_0,t_1)=(\theta_{t_0t_1}+\theta_{t_1t_0})C_{[A^0A^1]}(t_0,t_1)
\\
&=i\hbar\bigg(\frac{-i}{\hbar}\theta_{t_0t_1}C_{[A^0A^1]}(t_0,t_1)-\frac{i}{\hbar}\theta_{t_1t_0}C_{[A^0A^1]}(t_0,t_1)\bigg)
\\
&=i\hbar\bigg(C^r_{A^0A^1}(t_0,t_1)-C^a_{A^0A^1}(t_0,t_1)\bigg)
\equiv i\hbar C^{r-a}_{A^0A^1}(t_0,t_1).
\end{split}
\end{equation}

We have expressed the stripped correlator as the difference of the retarded and advanced one, which, as we can see from \eqref{eq:RDTRSTRB}, contributes to the dissipative response. What we see is that a specific combination $C^{r-a}$ of time-ordered expressions, $C^r$ and $C^a$, providing a dissipative response, is, overall, not time-ordered and thereby equivalent to a fluctuation---this is the fundamental principle underlying the existence of the fluctuation-dissipation relation. Using time-translation invariance and taking the Fourier transform of both sides of \eqref{eq:2FlucDiss} according to \eqref{eq:Fourier}, we get

\begin{equation}
\label{eq:2pointFDFreq0}
\tilde{C}^{r-a}_{A^0A^1}(\omega_1)=-\frac{i}{\hbar}\tilde{C}_{[A^0A^1]}(\omega_1),
\end{equation}

which means that the frequency domain version of the dissipative response in \eqref{eq:RDTRSTRB} for $n=1$ can be expressed using the stripped correlator as

\begin{equation}
\label{eq:2pointFDFreq}
\begin{split}
\tilde{C}^{r,\mathcal{T}^{\text{full}}_{-},\mathcal{K}^{\textbf{m}}_{\pm}}_{A^0A^1}(\omega_1)&=-\frac{i}{4\hbar}\left(\tilde{C}_{[A^0A^1]}(\omega_1)\pm\epsilon^{\mathcal{T}}_{A^0}\epsilon^{\mathcal{T}}_{A^1}\tilde{C}^*_{[A^0A^1]}(\omega_1)\right)
\\
&=-\frac{i}{2\hbar}\hat{\mathcal{P}}^{(\pm\text{sgn}(\epsilon))}_{\mathcal{K}^*}\tilde{C}_{[A^0A^1]}(\omega_1),
\end{split}
\end{equation}

where the projector $\hat{\mathcal{P}}^{(\pm\text{sgn}(\epsilon))}_{\mathcal{K}^*}$ to the real and imaginary parts was defined in \eqref{eq:ReImProj}.

We can arrive at a concrete result via the spectral representation of the stripped correlator (see Appendix \ref{SpectRep}.\ref{stripped2specApp} for the derivation)

\begin{equation}
\label{eq:2strippedSpec}
\begin{split}
\tilde{C}_{[A^0A^1]}(\omega_1)=&-2\pi\hbar\int d\varepsilon(\rho_0(\varepsilon+\hbar\omega_1)-\rho_0(\varepsilon))
\\
&\times\text{tr}(A^0\delta(\varepsilon+\hbar\omega_1-\mathcal{H}_0)A^1\delta(\varepsilon-\mathcal{H}_0)),
\end{split}
\end{equation}

meaning that \eqref{eq:2pointFDFreq} can be written as

\begin{equation}
\label{eq:2pointFDFreqSpec}
\begin{split}
\tilde{C}^{r,\mathcal{T}^{\text{full}}_{-},\mathcal{K}^{\textbf{m}}_{\pm}}_{A^0A^1}(\omega_1)=&i\pi\hat{\mathcal{P}}^{(\Gamma_1^{\pm\text{sgn}(\epsilon)})}_{A^0A^1}\int d\varepsilon(\rho_0(\varepsilon+\hbar\omega_1)-\rho_0(\varepsilon))
\\
&\times\text{tr}(A^0\delta(\varepsilon+\hbar\omega_1-\mathcal{H}_0)A^1\delta(\varepsilon-\mathcal{H}_0)),
\end{split}
\end{equation}

where we introduced the notation

\begin{equation}
\hat{\mathcal{P}}^{(\Gamma_1^{\pm})}_{A^0A^1}f_{A^0A^1}=\frac{f_{A^0A^1}\pm f_{A^1A^0}}{2},
\end{equation}

to describe symmetrization and antisymmetrization with respect to $A^0$ and $A^1$ for some function $f$ dependent on both. The notation is related to the irreducible representations of permutation groups which we present in detail in Section \ref{pstruct} below. It is clear that we could rewrite all of the spectral representations using Green's operators via identity \eqref{eq:GrGaEps}, i.e., $G^r(\varepsilon)-G^a(\varepsilon)=-i2\pi\delta(\varepsilon-\mathcal{H}_0)$, and go on to perform explicit calculations, but we kept the delta functions for illustrative purposes.\\

As a concrete example, we look at the first order conductivity given in terms of correlators by \eqref{eq:Cond1}. We have, $A^0=J_{i_0}$ and $A^1=J_{i_1}$ with $\text{sgn}(\epsilon)=\text{sign}\left(\epsilon^{\mathcal{T}}_{J_{i_0}}\epsilon^{\mathcal{T}}_{J_{i_1}}\right)=+1$, moreover, from \eqref{eq:2pointFDFreqSpec}, $\tilde{C}^{r,\mathcal{T}^{\text{full}}_{-},\mathcal{K}^{\textbf{m}}_{\pm}}_{A^0A^1}(\omega_1=0)=0$, hence

\begin{equation}
\label{eq:Cond1FlucDiss}
\begin{split}
\sigma^{\mathcal{T}^{\text{full}}_{-},\mathcal{K}^{\textbf{m}}_{\pm}}_{i_0i_1}(\omega_1)=&-\pi\hbar\hat{\mathcal{P}}^{(\Gamma_1^{\pm})}_{i_0i_1}\int d\varepsilon\frac{\rho_0(\varepsilon+\hbar\omega_1)-\rho_0(\varepsilon)}{\hbar\omega_1}
\\
&\times\text{tr}(J_{i_0}\delta(\varepsilon+\hbar\omega_1-\mathcal{H}_0)J_{i_1}\delta(\varepsilon-\mathcal{H}_0))
\\
=&\frac{\hbar}{4\pi}\hat{\mathcal{P}}^{(\Gamma_1^{\pm})}_{i_0i_1}\int d\varepsilon\frac{\rho_0(\varepsilon+\hbar\omega_1)-\rho_0(\varepsilon)}{\hbar\omega_1}
\\
&\times\text{tr}(J_{i_0}G^{r-a}_{\omega_1}J_{i_1}G^{r-a}),
\end{split}
\end{equation}

where we used the short-hands \eqref{eq:GreenShorthand} introduced for Green's operators in Appendix \ref{SpectRep}.\ref{Green}.\\

Referring to our discussion of sum rules in the introduction to this section, upon expanding the trace on the right hand side using a complete set of $\mathcal{H}_0$ eigenstates, we can perform the $\varepsilon$ integration leaving us with a delta function containing the frequency. Integration with respect to frequency then provides us with a general version of the sum rules originally devised by Kubo \cite{Kubo1957}. Together with the sum rules for the 3-point case, we present the details in paper III \cite{Bonbien2021c} of our series.

\subsection{Fluctuation-reaction and the 3-point correlator}

Following the strategy \eqref{eq:2FlucDiss} for the 2-point correlator above, we first find a combination of step functions that covers all possible orderings for three time variables and amend it to the 3-point stripped correlator. We have

\begin{equation}
\begin{split}
&C_{[A^0A^1A^2]}(t_0,t_1,t_2)
=(\theta_{t_0t_1}\theta_{t_1t_2}+\theta_{t_0t_2}\theta_{t_2t_1}+\theta_{t_1t_0}\theta_{t_0t_2}
\\
&\quad+\theta_{t_1t_2}\theta_{t_2t_0}+\theta_{t_2t_0}\theta_{t_0t_1}+\theta_{t_2t_1}\theta_{t_1t_0})C_{[A^0A^1A^2]}(t_0,t_1,t_2).
\end{split}
\end{equation}

At this point, for the 2-point stripped correlator above, we could recognize the retarded and advanced correlators straight away, however, this is no longer amenable and we need to use the 3-point stripped correlator's explicit nested commutator structure and the Jacobi identity to transform the right hand side into a combination of 3-point retarded and advanced correlators. We show the explicit steps in Appendix \ref{FluctDiss} and arrive at

\begin{equation}
\label{eq:3FlucDiss}
\begin{split}
&C_{[A^0A^1A^2]}(t_0,t_1,t_2)
\\
&\qquad=2\hbar^2(C^{r+a}_{A^1A^0A^2}(t_1,t_0,t_2)-C^{r+a}_{A^0A^1A^2}(t_0,t_1,t_2)).
\end{split}
\end{equation}

We have expressed the non time-ordered 3-point stripped correlator using a combination of time-ordered 3-point retarded and advanced correlators, however, unlike the 2-point case, we have the sum $C^{r+a}$ that contributes to the reactive response and have thereby obtained a `fluctuation-\textit{reaction}' theorem. Taking a closer look, there is a further twist when compared with the 2-point case: the two contributions on the right hand side have permuted observables. More precisely, different kind of response phenomena can be connected to provide a 3-point `fluctuation'. For a heuristic example, looking at the thermoelectric responses, let $A^0=J_x^T$ be the heat current operator in the $x$ direction and $A^1=J_x,\,A^2=J_y$ be charge current operators in the $x,\,y$ directions. Then, $C_{[J_x^T J_xJ_y]}$ is related to $C^{r+a}_{J_xJ_x^T J_y}$ and $C^{r+a}_{J_x^TJ_xJ_y}$. The former can contribute to a second order charge current response in the $x$ direction resulting from an applied electric field in the $y$ direction and a thermal gradient in the $x$ direction, whereas the latter can contribute to a second order heat current response in the $x$ direction resulting from an applied electric field that is circularly polarized in the $x-y$ plane.\\

Applying time-translation invariance and moving to the frequency-domain, \eqref{eq:3FlucDiss} becomes

\begin{equation}
\label{eq:3FlucDissFreq}
\begin{split}
&\tilde{C}^{r+a}_{A^1A^0A^2}(-\omega_1-\omega_2,\omega_2)-\tilde{C}^{r+a}_{A^0A^1A^2}(\omega_1,\omega_2)
\\
&\qquad\qquad=\frac{1}{2\hbar^2}\tilde{C}_{[A^0A^1A^2]}(\omega_1,\omega_2),
\end{split}
\end{equation}

hence, the frequency domain version of the reactive response in \eqref{eq:RDTRSTRB} for $n=2$ can be expressed using the stripped correlator as

\begin{equation}
\label{eq:3FlucDissFreq1}
\begin{split}
&\tilde{C}^{r,\mathcal{T}^{\text{full}}_{+},\mathcal{K}^{\textbf{m}}_{\pm}}_{A^1A^0A^2}(-\omega_1-\omega_2,\omega_2)-\tilde{C}^{r,\mathcal{T}^{\text{full}}_{+},\mathcal{K}^{\textbf{m}}_{\pm}}_{A^0A^1A^2}(\omega_1,\omega_2)
\\
&=\frac{1}{8\hbar^2}(\tilde{C}_{[A^0A^1A^2]}(\omega_1,\omega_2)\pm\epsilon^{\mathcal{T}}_{A^0}\epsilon^{\mathcal{T}}_{A^1}\epsilon^{\mathcal{T}}_{A^2}\tilde{C}^*_{[A^0A^1A^2]}(\omega_1,\omega_2))
\\
&=\frac{1}{4\hbar^2}\hat{\mathcal{P}}^{(\pm\text{sgn}(\epsilon))}_{\mathcal{K}^*}\tilde{C}_{[A^0A^1A^2]}(\omega_1,\omega_2),
\end{split}
\end{equation}

with the latter's spectral representation being (see Appendix \ref{SpectRep}.\ref{stripped3specApp} for the proof)

\begin{widetext}
\begin{equation}
\label{eq:3strippedSpec}
\begin{split}
\tilde{C}_{[A^0A^1A^2]}(\omega_1,\omega_2)=&4\pi^2\hbar^2\int d\varepsilon (\rho_0(\varepsilon)-\rho_0(\varepsilon+\hbar\omega_2))\text{tr}(A^0\delta(\varepsilon+\hbar(\omega_1+\omega_2)-\mathcal{H}_0)A^1\delta(\varepsilon+\hbar\omega_2-\mathcal{H}_0)A^2\delta(\varepsilon-\mathcal{H}_0))
\\
&+4\pi^2\hbar^2\int d\varepsilon (\rho_0(\varepsilon)-\rho_0(\varepsilon-\hbar\omega_2))\text{tr}(A^1\delta(\varepsilon-\hbar(\omega_1+\omega_2)-\mathcal{H}_0)A^0\delta(\varepsilon-\mathcal{H}_0)A^2\delta(\varepsilon-\hbar\omega_2-\mathcal{H}_0)).
\end{split}
\end{equation}
\end{widetext}

The frequency domain version of the commutator identity $C_{[A^0A^1A^2]}(t_0,t_1,t_2)=-C_{[A^1A^0A^2]}(t_1,t_0,t_2)$ follows from \eqref{eq:3strippedSpec} in a straightforward way 

\begin{equation}
\label{eq:3commFreq}
\tilde{C}_{[A^0A^1A^2]}(\omega_1,\omega_2)=-\tilde{C}_{[A^1A^0A^2]}(-\omega_1-\omega_2,\omega_2).
\end{equation}

Note that in the case of the observables being dipole operators $d_i$, the 3-point retarded correlator is just the second order electric susceptibility, and, assuming only non-resonant frequencies, we can read-off from \eqref{eq:3strippedSpec} that  $\tilde{C}_{[d_{i_0}d_{i_1}d_{i_2}]}(\omega_1,\omega_2)\to 0$, meaning $\tilde{C}^{r,\mathcal{T}^{\text{full}}_{+},\mathcal{K}^{\textbf{m}}_{\pm}}_{d_{i_1}d_{i_0}d_{i_2}}(-\omega_1-\omega_2,\omega_2)=\tilde{C}^{r,\mathcal{T}^{\text{full}}_{+},\mathcal{K}^{\textbf{m}}_{\pm}}_{d_{i_0}d_{i_1}d_{i_2}}(\omega_1,\omega_2)$, and the reactive susceptibility obeys `full' permutation symmetry \cite{BoydNLO}.\\

We now consider special cases of these relations and show that their combination yields a contribution to the second order charge conductivity $\sigma_{i_0i_1i_2}(\omega_1,\omega_2)$ given in terms of correlators by \eqref{eq:Cond2}. We fix $A^0=J_{i_0},\,A^1=J_{i_1},\,A^2=J_{i_2}$, meaning $\epsilon^{\mathcal{T}}_{J_{i_0}}\epsilon^{\mathcal{T}}_{J_{i_1}}\epsilon^{\mathcal{T}}_{J_{i_2}}=-1$, so $\text{sgn}(\epsilon)=-1$. In order to proceed, we take a closer look at the second order charge conductivity expression \eqref{eq:Cond2}, and make two observations: first, that it consists of 2 and 3-point correlators, but our fluctuation-reaction relation \eqref{eq:3FlucDissFreq} is only valid for 3-point correlators; we can resolve this conundrum by realizing that all of the 2-point correlator terms in \eqref{eq:Cond2} are symmetric in $i_1$ and $i_2$, hence antisymmetrizing with respect to these two indices will yield an expression containing only 3-point correlators; second, that it contains 3-point correlators of the form $\tilde{C}^r_{J_{i_0}J_{i_1}J_{i_2}}(\omega_1,\omega_2)$, $\tilde{C}^r_{J_{i_0}J_{i_1}J_{i_2}}(\omega_1,0)$, etc., meaning that the fluctuation-reaction relation \eqref{eq:3FlucDissFreq} should relate correlators of these sort. We can obtain relations from \eqref{eq:3FlucDissFreq} along the lines of the second observation by first taking $\omega_1=\omega$ and $\omega_2=-\omega$ to get,

\begin{equation}
\label{eq:3FlucDissFreqCurr1}
\begin{split}
&\tilde{C}^{r+a}_{J_{i_1}J_{i_0}J_{i_2}}(0,-\omega)-\tilde{C}^{r+a}_{J_{i_0}J_{i_1}J_{i_2}}(\omega,-\omega)
\\
&\qquad\qquad=\frac{1}{2\hbar^2}\tilde{C}_{[J_{i_0}J_{i_1}J_{i_2}]}(\omega,-\omega),
\end{split}
\end{equation}

and then, by taking $\omega_1=\omega$ and $\omega_2=0$ to get

\begin{equation}
\label{eq:3FlucDissFreqCurr2}
\begin{split}
\tilde{C}^{r+a}_{J_{i_1}J_{i_0}J_{i_2}}(-\omega,0)-\tilde{C}^{r+a}_{J_{i_0}J_{i_1}J_{i_2}}(\omega,0)=0,
\end{split}
\end{equation}

where we used the fact that, in general, $\tilde{C}_{[A^0A^1A^2]}(\omega,0)=0$ (see \eqref{eq:3strippedSpec}). Taking into account the first observation, we can thus anticipate a contribution to the antisymmetric in $i_1,\,i_2$ combination $\sigma_{i_0i_1i_2}(\omega,-\omega)-\sigma_{i_0i_2i_1}(\omega,-\omega)$. By definition, all retarded and advanced correlators satisfy the so-called intrinsic permutation symmetry, which is $\tilde{C}^{r(a)}_{A^0A^1A^2}(\omega_1,\omega_2)=\tilde{C}^{r(a)}_{A^0A^2A^1}(\omega_2,\omega_1)$ for the 3-point case, and $\tilde{C}^{r(a)}_{A^0A^1A^2}(\omega_1,\omega_2)=\tilde{C}^{r(a)^*}_{A^0A^1A^2}(-\omega_1,-\omega_2)$ due to being Fourier transforms of real quantities, meaning that we also have for the conductivity $\sigma^*_{i_0i_1i_2}(\omega,-\omega)=\sigma_{i_0i_1i_2}(-\omega,\omega)=\sigma_{i_0i_2i_1}(\omega,-\omega)$. In other words, the antisymmetric combination corresponds to the imaginary part $\sigma_{i_0i_1i_2}(\omega,-\omega)-\sigma_{i_0i_2i_1}(\omega,-\omega)=\sigma_{i_0i_1i_2}(\omega,-\omega)-\sigma^*_{i_0i_1i_2}(\omega,-\omega)$. By virtue of \eqref{eq:RDCond2}, the reactive imaginary response corresponds to the $\mathcal{K}^{\textbf{m}}_+$ eigenfunction of magnetic-inversion, $\sigma^{\mathcal{T}^{\text{full}}_+,\mathcal{K}^{\textbf{m}}_+}_{i_0i_1i_2}(\omega,-\omega)$, which, using \eqref{eq:Cond2}, can be written as

\begin{equation}
\label{eq:Cond2FlucDiss1}
\begin{split}
&2\omega^2\sigma^{\mathcal{T}^{\text{full}}_+,\mathcal{K}^{\textbf{m}}_+}_{i_0i_1i_2}(\omega,-\omega)=
\hat{\mathcal{P}}^{(\Gamma_1^{-})}_{i_1i_2}(\tilde{C}^{r+a}_{J_{i_0}J_{i_1}J_{i_2}}(\omega,-\omega)
\\
&-\tilde{C}^{r+a}_{J_{i_0}J_{i_1}J_{i_2}}(\omega,0)-\tilde{C}^{r+a}_{J_{i_0}J_{i_1}J_{i_2}}(0,-\omega)).
\end{split}
\end{equation}

However, relations \eqref{eq:3FlucDissFreqCurr1} and \eqref{eq:3FlucDissFreqCurr2} connect correlators with different first indices. We thus need to project $\sigma^{\mathcal{T}^{\text{full}}_+,\mathcal{K}^{\textbf{m}}_+}_{i_0i_1i_2}(\omega,-\omega)$ onto a permutation class of the indices $i_0,i_1,i_2$, where, under permutation class, we mean an irreducible representation of the corresponding permutation group of three objects $P(3)$---in our case acting via permutations of the three indices (see Section \ref{pstruct} for a general analysis of correlators' permutation structure and more on permutation groups). $P(3)$ has three irreducible representations \cite{DresselhausSymm}: the 1-dimensional totally symmetric $\Gamma_1^+$ and totally antisymmetric $\Gamma_1^-$ representations, and the 2-dimensional `mixed symmetric' $\Gamma_2$ representation. Since $i_1,\,i_2$ are already antisymmetrized, we have only two options: projection onto $\Gamma_1^-$ or $\Gamma_2$. It turns out that upon applying the latter projection to \eqref{eq:Cond2FlucDiss1}, the resulting combination of retarded and advanced correlators allows us to use the fluctuation-reaction relations \eqref{eq:3FlucDissFreqCurr1}, \eqref{eq:3FlucDissFreqCurr2} and reduce the result into a combination of stripped correlators, thereby providing a second order analogue of the first order result \eqref{eq:Cond1FlucDiss}. Concretely, using the characters from Table \ref{tab:P3} in the projector \eqref{eq:ProjIrrep}, we find that the projection $\hat{\mathcal{P}}^{(\Gamma_2)}_{i_0i_1i_2}$ onto $\Gamma_2$ acts on a function $f_{i_0i_1i_2}$ as

\begin{equation}
\label{eq:S3Gamma2Proj}
\hat{\mathcal{P}}^{(\Gamma_2)}_{i_0i_1i_2}f_{i_0i_1i_2}=\frac{1}{3}(2f_{i_0i_1i_2}-f_{i_2i_0i_1}-f_{i_1i_2i_0}).
\end{equation}

Defining 

\begin{equation}
\sigma^{\mathcal{T}^{\text{full}}_+,\mathcal{K}^{\textbf{m}}_+,\Gamma_2^-}_{i_0i_1i_2}(\omega,-\omega)\equiv\hat{\mathcal{P}}^{(\Gamma_2)}_{i_0i_1i_2}\sigma^{\mathcal{T}^{\text{full}}_+,\mathcal{K}^{\textbf{m}}_+}_{i_0i_1i_2}(\omega,-\omega),
\end{equation}

applying $\hat{\mathcal{P}}^{(\Gamma_2)}_{i_0i_1i_2}$ to \eqref{eq:Cond2FlucDiss1}, and utilizing \eqref{eq:3FlucDissFreqCurr1} together with \eqref{eq:3FlucDissFreqCurr2}, we find (see Appendix \ref{FluctDiss}.\ref{FluctDissCondProof} for the proof)

\begin{equation}
\label{eq:Cond2FlucDiss2}
\begin{split}
&2\omega^2\sigma^{\mathcal{T}^{\text{full}}_+,\mathcal{K}^{\textbf{m}}_+,\Gamma_2^-}_{i_0i_1i_2}(\omega,-\omega)=
-\frac{1}{2\hbar^2}\hat{\mathcal{P}}^{(\Gamma_2^{-})}_{i_0i_1i_2}\tilde{C}_{[J_{i_0}J_{i_1}J_{i_2}]}(\omega,-\omega),
\end{split}
\end{equation}

where $\hat{\mathcal{P}}^{(\Gamma_2^{-})}_{i_0i_1i_2}=\hat{\mathcal{P}}^{(\Gamma_2)}_{i_0i_1i_2}\hat{\mathcal{P}}^{(\Gamma_1^{-})}_{i_1i_2}$. We can arrive at a concrete expression via the spectral representation \eqref{eq:3strippedSpec}

\begin{equation}
\label{eq:Cond2FlucDissFinal}
\begin{split}
&\sigma^{\mathcal{T}^{\text{full}}_+,\mathcal{K}^{\textbf{m}}_+,\Gamma_2^-}_{i_0i_1i_2}(\omega,-\omega)
\\
&=-2\pi^2\hbar^2\hat{\mathcal{P}}^{(\Gamma_2^{-})}_{i_0i_1i_2}\hat{\mathcal{P}}^{(\Gamma_1^{-})}_{(\omega,-\omega)}\int d\varepsilon\frac{\rho_0(\varepsilon+\hbar\omega)-\rho_0(\varepsilon)}{\hbar^2\omega^2}
\\
&\quad\times\text{tr}(J_{i_0}\delta(\varepsilon-\mathcal{H}_0)J_{i_1}\delta(\varepsilon+\hbar\omega-\mathcal{H}_0)J_{i_2}\delta(\varepsilon-\mathcal{H}_0))
\\
&=\frac{i\hbar^2}{4\pi}\hat{\mathcal{P}}^{(\Gamma_2^{-})}_{i_0i_1i_2}\hat{\mathcal{P}}^{(\Gamma_1^{-})}_{(\omega,-\omega)}\int d\varepsilon\frac{\rho_0(\varepsilon+\hbar\omega)-\rho_0(\varepsilon)}{\hbar^2\omega^2}
\\
&\quad\times\text{tr}(J_{i_0}G^{r-a}J_{i_1}G^{r-a}_{\omega}J_{i_2}G^{r-a}),
\end{split}
\end{equation}

where $\hat{\mathcal{P}}^{(\Gamma_1^{-})}_{(\omega,-\omega)}f(\omega)=(f(\omega)-f(-\omega))/2$ for some function $f$.\\
An important question is what response does this particular conductivity expression represent? First of all, it is a direct current response and it does not require the breaking of time-reversal symmetry; then, it is antisymmetric in the driving field directions, $i_1$ and $i_2$, meaning that it can be induced by a circularly polarized electric field; and, finally, since $i_0$ can take on the value of $i_1$ or $i_2$ without rendering the expression null (see the projection \eqref{eq:S3Gamma2Proj} with antisymmetrized $i_1,i_2$) it \textit{can be} an in-plane, i.e., `longitudinal', response. Note that the `transverse' response would require $i_0$ to be `transverse' to $i_1$ and $i_2$, meaning that the corresponding response would be the one that is totally antisymmetric in all three indices, and would not yield an in-plane component, i.e., it would correspond to a projection onto the totally antisymmetric $\Gamma_1^-$ representation. (This is a special case of the more general permutation decomposition to be introduced in Section \ref{pstruct}.) In semiconductors, such a response corresponds to the non-transverse---in the sense discussed above---component of the so-called `injection current' or `circular photocurrent' \cite{SipeShkrebtii2000}. In paper III \cite{Bonbien2021c} of our series, we show that, for single electrons in the clean limit, \eqref{eq:Cond2FlucDissFinal} reduces to the standard expression for the injection current present in the literature.\\

Another important consequence of \eqref{eq:Cond2FlucDissFinal} is the existence of new sum rules. Indeed, reflecting on our earlier discussion of sum rules, upon expanding the trace with respect to a complete set of $\mathcal{H}_0$ eigenstates, the frequency will be the argument of a delta function on the right hand side which means that we can integrate it and are left with sum rules. We present the explicit sum rules in paper III \cite{Bonbien2021c}.

\subsection{Weak fluctuation-dissipation and the 3-point correlator}

The fluctuation-reaction relation \eqref{eq:3FlucDiss} for the 3-point correlator establishes an expression between the non-time-ordered 3-point stripped correlator and combinations of time-ordered 3-point retarded and advanced correlators. We can also obtain a slightly weaker relation by taking a `partial' time-ordering of the stripped correlator. Indeed, the latter 3-point stripped correlator, $C_{[A^0A^1A^2]}(t_0,t_1,t_2)$, has 3 time arguments, $t_0,\,t_1$, and $t_2$, and ordering only the first two, say, $t_0>t_1$ or $t_0<t_1$ yielding $\theta_{t_0t_1}C_{[A^0A^1A^2]}(t_0,t_1,t_2)$ or $\theta_{t_1t_0}C_{[A^0A^1A^2]}(t_0,t_1,t_2)$, still leaves $t_2$ unconstrained with respect to $t_0$ and $t_1$, thereby providing a partial time-ordering. These partially time-ordered 3-point stripped correlators can be related to a combination of 3-point retarded and advanced correlators, leading to a `weak' fluctuation-dissipation relation. In order to obtain this relation, let us first impose $t_0>t_1$ on the fluctuation-reaction relation \eqref{eq:3FlucDiss}

\begin{equation}
\label{eq:3FlucDissW1}
\begin{split}
&\theta_{t_0t_1}C_{[A^0A^1A^2]}(t_0,t_1,t_2)
\\
&\qquad=2\hbar^2(C^{a}_{A^1A^0A^2}(t_1,t_0,t_2)-C^{r}_{A^0A^1A^2}(t_0,t_1,t_2)),
\end{split}
\end{equation}

where we used the facts that $\theta_{t_0t_1}C^{r}_{A^1A^0A^2}(t_1,t_0,t_2)=\theta_{t_0t_1}C^{a}_{A^0A^1A^2}(t_0,t_1,t_2)=0$, since both correlators satisfy $t_1>t_0$.\\
Next, we impose the opposite ordering $t_0<t_1$ on \eqref{eq:3FlucDiss} and get

\begin{equation}
\label{eq:3FlucDissW2}
\begin{split}
&\theta_{t_1t_0}C_{[A^0A^1A^2]}(t_0,t_1,t_2)
\\
&\qquad=2\hbar^2(C^{r}_{A^1A^0A^2}(t_1,t_0,t_2)-C^{a}_{A^0A^1A^2}(t_0,t_1,t_2)).
\end{split}
\end{equation}

Substracting \eqref{eq:3FlucDissW1} from \eqref{eq:3FlucDissW2} results in

\begin{equation}
\label{eq:3FlucDissW}
\begin{split}
&(\theta_{t_1t_0}-\theta_{t_0t_1})C_{[A^0A^1A^2]}(t_0,t_1,t_2)
\\
&\qquad=2\hbar^2(C^{r-a}_{A^1A^0A^2}(t_1,t_0,t_2)+C^{r-a}_{A^0A^1A^2}(t_0,t_1,t_2)).
\end{split}
\end{equation}

This relation establishes a connection between partially time-ordered 3-point stripped correlators---which can still be considered as `fluctuations' since not all of their time arguments are constrained---and a combination of $C^{r-a}$ factors, with the latter contributing to \textit{dissipative} responses. We refer to it as a `weak' fluctuation-dissipation relation. Note that we could have also obtained this relation directly from \eqref{eq:3FlucDiss} by using the Kramers-Kronig relations \eqref{eq:KKnRA} with $n=2$, however, we believe that our demonstration emphasizes the underlying time-ordering issue better.\\

Utilizing time-translation invariance; the integral representation,
\begin{equation}
\theta(t)=\lim_{\eta\to 0}\int\frac{d\omega}{2\pi}\frac{i\hbar}{\hbar\omega+i\eta}e^{-i\omega t},
\end{equation}
of Heaviside's step function; and taking the Fourier transform, \eqref{eq:3FlucDissW} can be written as

\begin{equation}
\label{eq:3FlucDissWFreq}
\begin{split}
&\tilde{C}^{r-a}_{A^1A^0A^2}(-\omega_1-\omega_2,\omega_2)+\tilde{C}^{r-a}_{A^0A^1A^2}(\omega_1,\omega_2)
\\
&=\frac{i\hbar}{2\hbar^2\pi}\lim_{\eta\to 0}\int d\omega'\frac{\hbar\omega'}{\hbar^2\omega'^2+\eta^2}\tilde{C}_{[A^0A^1A^2]}(\omega'+\omega_1,\omega_2)
\\
&\equiv \frac{i}{2\hbar^2\pi}\mathtt{pv}\int\frac{d\omega'}{\omega'}\tilde{C}_{[A^0A^1A^2]}(\omega'+\omega_1,\omega_2),
\end{split}
\end{equation}

where $\mathtt{pv}(\dots)$ refers to the Cauchy principal value. We can thus express the frequency domain version of the dissipative response \eqref{eq:RDTRSTRB} for $n=2$ as

\begin{equation}
\label{eq:3FlucDissWFreq1}
\begin{split}
&\tilde{C}^{r,\mathcal{T}^{\text{full}}_{-},\mathcal{K}^{\textbf{m}}_{\pm}}_{A^1A^0A^2}(-\omega_1-\omega_2,\omega_2)+\tilde{C}^{r,\mathcal{T}^{\text{full}}_{-},\mathcal{K}^{\textbf{m}}_{\pm}}_{A^0A^1A^2}(\omega_1,\omega_2)
\\
&=\frac{i}{4\hbar^2\pi}\mathtt{pv}\int\frac{d\omega'}{\omega'}\hat{\mathcal{P}}^{(\pm\text{sgn}(\epsilon))}_{\mathcal{K}^*}\tilde{C}_{[A^0A^1A^2]}(\omega'+\omega_1,\omega_2).
\end{split}
\end{equation}

We now show that these relations are less restrictive than the fluctuation-reaction relations \eqref{eq:3FlucDissFreq1} and allow us to express a broader range of responses via partially ordered stripped correlators.\\

Let us exchange $A^0$ and $A^2$ in \eqref{eq:3FlucDissWFreq} and use intrinsic permutation symmetry to get

\begin{equation}
\label{eq:3FlucDissWFreq02}
\begin{split}
&\tilde{C}^{r-a}_{A^1A^0A^2}(\omega_2,-\omega_1-\omega_2)+\tilde{C}^{r-a}_{A^2A^0A^1}(\omega_2,\omega_1)
\\
&= \frac{i}{2\hbar^2\pi}\mathtt{pv}\int\frac{d\omega'}{\omega'}\tilde{C}_{[A^2A^1A^0]}(\omega'+\omega_1,\omega_2).
\end{split}
\end{equation}

Now, again in \eqref{eq:3FlucDissWFreq}, we exchange $A^1$ and $A^2$ together with $\omega_1$ and $\omega_2$, use intrinsic permutation symmetry, and add the result to, yet again, \eqref{eq:3FlucDissWFreq}. We arrive at

\begin{widetext}
\begin{equation}
\label{eq:3FlucDissWFreq12sum}
\begin{split}
&\tilde{C}^{r-a}_{A^1A^0A^2}(-\omega_1-\omega_2,\omega_2)+\tilde{C}^{r-a}_{A^2A^0A^1}(-\omega_1-\omega_2,\omega_1)+2\tilde{C}^{r-a}_{A^0A^1A^2}(\omega_1,\omega_2)
\\
&=\frac{i}{2\hbar^2\pi}\mathtt{pv}\int\frac{d\omega'}{\omega'}(\tilde{C}_{[A^0A^1A^2]}(\omega'+\omega_1,\omega_2)+\tilde{C}_{[A^0A^2A^1]}(\omega'+\omega_2,\omega_1)).
\end{split}
\end{equation}
\end{widetext}

We can note that the left hand sides of both \eqref{eq:3FlucDissWFreq02} and \eqref{eq:3FlucDissWFreq12sum} contain overlapping observable permutations, albeit with different frequency arguments. This means that if we could align said frequency arguments, we could eliminate these terms by subtracting the two equations from each other. There are two ways of accomplishing this. In the first case, we take $\omega_1=\omega$ and $\omega_2=0$ in \eqref{eq:3FlucDissWFreq02}, and $\omega_1=\omega$ together with $\omega_2=-\omega$ in \eqref{eq:3FlucDissWFreq12sum}. The right hand side of \eqref{eq:3FlucDissWFreq02} vanishes since $\tilde{C}_{[A^2A^1A^0]}(\omega'+\omega,0)=0$ (see \eqref{eq:3strippedSpec}), meaning that the overlapping terms satisfy  $\tilde{C}^{r-a}_{A^1A^0A^2}(0,-\omega)+\tilde{C}^{r-a}_{A^2A^0A^1}(0,\omega)=0$, and we can express the remaining term in \eqref{eq:3FlucDissWFreq12sum} as

\begin{equation}
\label{eq:3FlucDissWFreqDC}
\begin{split}
\tilde{C}^{r-a}_{A^0A^1A^2}(\omega,-\omega)=&\frac{i}{4\hbar^2\pi}\mathtt{pv}\int\frac{d\omega'}{\omega'}(\tilde{C}_{[A^0A^1A^2]}(\omega'+\omega,-\omega)
\\
&+\tilde{C}_{[A^0A^2A^1]}(\omega'-\omega,\omega)).
\end{split}
\end{equation}

Injecting this into \eqref{eq:3FlucDissWFreq} with $\omega_1=\omega$ and $\omega_2=-\omega$, and exchanging $A^0$ and $A^1$ results in

\begin{equation}
\label{eq:3FlucDissWFreq01}
\begin{split}
\tilde{C}^{r-a}_{A^0A^1A^2}(0,-\omega)=&\frac{i}{4\hbar^2\pi}\mathtt{pv}\int\frac{d\omega'}{\omega'}(\tilde{C}_{[A^1A^0A^2]}(\omega'+\omega,-\omega)
\\
&-\tilde{C}_{[A^1A^2A^0]}(\omega'-\omega,\omega)).
\end{split}
\end{equation}

Finally, reversing the sign of $\omega$ and exchanging $A^1$ and $A^2$ in \eqref{eq:3FlucDissWFreq01} yields

\begin{equation}
\label{eq:3FlucDissWFreq10}
\begin{split}
\tilde{C}^{r-a}_{A^0A^1A^2}(\omega,0)=&\frac{i}{4\hbar^2\pi}\mathtt{pv}\int\frac{d\omega'}{\omega'}(\tilde{C}_{[A^2A^0A^1]}(\omega'-\omega,\omega)
\\
&-\tilde{C}_{[A^2A^1A^0]}(\omega'+\omega,-\omega)).
\end{split}
\end{equation}

Remarkably, we could express all the dissipative correlators necessary to describe the relevant, $\sigma^{r-a}_{i_0i_1i_2}(\omega,-\omega)$,  second order charge conductivity (cf. formula \eqref{eq:Cond2}). This was not possible with the fluctuation-reaction relation and we could only express a specific combination of reactive correlators using the stripped correlator. This implied severe restrictions on the form of the conductivity to which the relation was applicable. Before going forward with the conductivity analysis, however, we look at the second case in which the `overlap' between \eqref{eq:3FlucDissWFreq02} and \eqref{eq:3FlucDissWFreq12sum} can be eliminated. We take $\omega_1=-2\omega$ and $\omega_2=\omega$ in both \eqref{eq:3FlucDissWFreq02} and \eqref{eq:3FlucDissWFreq12sum}. Then, \eqref{eq:3FlucDissWFreq02} becomes

\begin{equation}
\label{eq:3FlucDissWFreq1121}
\begin{split}
&\tilde{C}^{r-a}_{A^1A^0A^2}(\omega,\omega)+\tilde{C}^{r-a}_{A^2A^0A^1}(\omega,-2\omega)
\\
&= \frac{i}{2\hbar^2\pi}\mathtt{pv}\int\frac{d\omega'}{\omega'}\tilde{C}_{[A^2A^1A^0]}(\omega'-2\omega,\omega),
\end{split}
\end{equation}

and we can use it to eliminate the two overlapping terms in \eqref{eq:3FlucDissWFreq12sum}. We find

\begin{equation}
\label{eq:3FlucDissWFreq21}
\begin{split}
&\tilde{C}^{r-a}_{A^0A^1A^2}(-2\omega,\omega)=\frac{i}{4\hbar^2\pi}\mathtt{pv}\int\frac{d\omega'}{\omega'}(\tilde{C}_{[A^0A^1A^2]}(\omega'-2\omega,\omega)
\\
&\qquad+\tilde{C}_{[A^0A^2A^1]}(\omega'+\omega,-2\omega)-\tilde{C}_{[A^2A^1A^0]}(\omega'-2\omega,\omega)).
\end{split}
\end{equation}

Exchanging $A^0$ and $A^2$, utilizing intrinsic permutation symmetry, putting the result back into \eqref{eq:3FlucDissWFreq1121}, and following this by a further exchange of $A^0$ and $A^1$, we arrive at

\begin{equation}
\label{eq:3FlucDissWFreqSHG}
\begin{split}
&\tilde{C}^{r-a}_{A^0A^1A^2}(\omega,\omega)=\frac{i}{4\hbar^2\pi}\mathtt{pv}\int\frac{d\omega'}{\omega'}(\tilde{C}_{[A^2A^0A^1]}(\omega'-2\omega,\omega)
\\
&\qquad+\tilde{C}_{[A^1A^0A^2]}(\omega'-2\omega,\omega)-\tilde{C}_{[A^1A^2A^0]}(\omega'+\omega,-2\omega)),
\end{split}
\end{equation}

where we also used \eqref{eq:3commFreq} to rewrite the last term.\\

Taking a look at the second order charge conductivity expression \eqref{eq:Cond2}, we see that the dissipative contributions; $\sigma^{r-a}_{i_0i_1i_2}(\omega,-\omega)$, $\sigma^{r-a}_{i_0i_1i_2}(\omega,\omega)$, and $\sigma^{r-a}_{i_0i_1i_2}(\omega,-2\omega)$ to, respectively, the direct current response, the second-harmonic current response, and a difference-frequency current response; can all be expressed using partially time-ordered stripped correlators. We next present the explicit expressions for the first two cases. 

\begin{widetext}
\subsubsection{Dissipative, second order, direct current charge conductivity}

We take $\omega_1=\omega$ and $\omega_2=-\omega$ in \eqref{eq:Cond2} to get

\begin{equation}
\label{eq:Cond2FDWDC}
\begin{split}
\omega^2\sigma^{r-a}_{i_0i_1i_2}(\omega,-\omega)=&\tilde{C}^{r-a}_{J_{i_0}J_{i_1i_2}}(\omega)+\tilde{C}^{r-a}_{J_{i_0}J_{i_1i_2}}(-\omega)
\\
&+\tilde{C}^{r-a}_{J_{i_0}J_{i_1}J_{i_2}}(\omega,-\omega)-\tilde{C}^{r-a}_{J_{i_0}J_{i_1}J_{i_2}}(\omega,0)-\tilde{C}^{r-a}_{J_{i_0}J_{i_1}J_{i_2}}(0,-\omega),
\end{split}
\end{equation}

and we used $\tilde{C}^{r-a}_{J_{i_0}J_{i_1i_2}}(0)=0$ together with $\tilde{C}^{r-a}_{J_{i_0}J_{i_1}J_{i_2}}(0,0)=0$, both of which follow from, respectively, \eqref{eq:2pointFDFreq0} combined with \eqref{eq:2strippedSpec} and, say, \eqref{eq:3FlucDissWFreq10} combined with \eqref{eq:3strippedSpec}.\\

From \eqref{eq:TRCond2} and \eqref{eq:RDCond2}, it is clear that the real and imaginary parts of \eqref{eq:Cond2FDWDC} contribute to, respectively, the dissipative magnetic-inversion even $\sigma^{\mathcal{T}^{\text{full}}_{-},\mathcal{K}^{\textbf{m}}_{+}}_{i_0i_1i_2}(\omega,-\omega)$ and dissipative magnetic-inversion odd $\sigma^{\mathcal{T}^{\text{full}}_{+},\mathcal{K}^{\textbf{m}}_{-}}_{i_0i_1i_2}(\omega,-\omega)$ conductivities. Furthermore, by intrinsic permutation symmetry, we have $(\sigma^{r-a}_{i_0i_1i_2}(\omega,-\omega))^*=\sigma^{r-a}_{i_0i_2i_1}(\omega,-\omega)$, in other words, the real and imaginary parts correspond to, respectively, the symmetric in $i_1,i_2$ and antisymmetric in $i_1,i_2$ parts. Thus, the 2-point correlators on the right hand side of \eqref{eq:Cond2FDWDC} only affect $\sigma^{\mathcal{T}^{\text{full}}_{-},\mathcal{K}^{\textbf{m}}_{+}}_{i_0i_1i_2}(\omega,-\omega)$. We can then write for the magnetic-inversion odd contribution

\begin{equation}
\label{eq:Cond2FDWDC1}
\begin{split}
&2\omega^2\sigma^{\mathcal{T}^{\text{full}}_-,\mathcal{K}^{\textbf{m}}_-}_{i_0i_1i_2}(\omega,-\omega)=
\hat{\mathcal{P}}^{(\Gamma_1^{-})}_{i_1i_2}(\tilde{C}^{r-a}_{J_{i_0}J_{i_1}J_{i_2}}(\omega,-\omega)-\tilde{C}^{r-a}_{J_{i_0}J_{i_1}J_{i_2}}(\omega,0)-\tilde{C}^{r-a}_{J_{i_0}J_{i_1}J_{i_2}}(0,-\omega)).
\end{split}
\end{equation}

Utilizing the partially time-ordered 3-point stripped correlator expressions \eqref{eq:3FlucDissWFreqDC}, \eqref{eq:3FlucDissWFreq01}, and \eqref{eq:3FlucDissWFreq10} we get

\begin{equation}
\begin{split}
\omega^2\sigma^{\mathcal{T}^{\text{full}}_-,\mathcal{K}^{\textbf{m}}_-}_{i_0i_1i_2}(\omega,-\omega)=\frac{i}{4\hbar^2\pi}\hat{\mathcal{P}}^{(\Gamma_1^{-})}_{i_1i_2}\hat{\mathcal{P}}^{(\Gamma_1^{-})}_{(\omega,-\omega)}\mathtt{pv}\int\frac{d\omega'}{\omega'}(&\tilde{C}_{[J_{i_0}J_{i_1}J_{i_2}]}(\omega'+\omega,-\omega)
+\tilde{C}_{[J_{i_2}J_{i_0}J_{i_1}]}(\omega'+\omega,-\omega)
\\
&-\tilde{C}_{[J_{i_1}J_{i_2}J_{i_0}]}(\omega'+\omega,-\omega)).
\end{split}
\end{equation}

We can now use the spectral representation \eqref{eq:3strippedSpec} and a Green's operator expression

\begin{equation}
\label{eq:GrpGa}
\begin{split}
\mathtt{pv}\int\frac{d\omega'}{\omega'}\delta(\varepsilon+\hbar\omega'-\mathcal{H}_0)\to&\int d(\hbar\omega')\frac{\hbar\omega'}{\hbar^2\omega'^2+\eta^2}\delta(\varepsilon+\hbar\omega'-\mathcal{H}_0)=-\frac{\varepsilon-\mathcal{H}_0}{(\varepsilon-\mathcal{H}_0)^2+\eta^2}
\\
&=-\frac{1}{2}\left(\frac{1}{\varepsilon-\mathcal{H}_0+i\eta}+\frac{1}{\varepsilon-\mathcal{H}_0-i\eta}\right)=-\frac{1}{2}(G^{r}(\varepsilon)+G^a(\varepsilon))\equiv-\frac{1}{2}G^{r+a},
\end{split}
\end{equation}

to get

\begin{equation}
\begin{split}
\sigma^{\mathcal{T}^{\text{full}}_-,\mathcal{K}^{\textbf{m}}_-}_{i_0i_1i_2}(\omega,-\omega)=&-\frac{i\pi\hbar^2}{2}\hat{\mathcal{P}}^{(\Gamma_1^{-})}_{i_1i_2}\hat{\mathcal{P}}^{(\Gamma_1^{-})}_{(\omega,-\omega)}\int d\varepsilon\frac{\rho_0(\varepsilon+\hbar\omega)-\rho_0(\varepsilon)}{\hbar^2\omega^2}
\\
&\quad\times(\text{tr}(J_{i_0}G^{r+a}J_{i_1}\delta(\varepsilon+\hbar\omega-\mathcal{H}_0)J_{i_2}\delta(\varepsilon-\mathcal{H}_0)-J_{i_0}\delta(\varepsilon-\mathcal{H}_0)J_{i_1}\delta(\varepsilon+\hbar\omega-\mathcal{H}_0)J_{i_2}G^{r+a})
\\
&\qquad\,+\text{tr}(J_{i_2}G^{r+a}J_{i_0}\delta(\varepsilon+\hbar\omega-\mathcal{H}_0)J_{i_1}\delta(\varepsilon-\mathcal{H}_0)+J_{i_2}\delta(\varepsilon-\mathcal{H}_0)J_{i_0}\delta(\varepsilon+\hbar\omega-\mathcal{H}_0)J_{i_1}G^{r+a})
\\
&\qquad\,-\text{tr}(J_{i_1}G^{r+a}J_{i_2}\delta(\varepsilon+\hbar\omega-\mathcal{H}_0)J_{i_0}\delta(\varepsilon-\mathcal{H}_0)+J_{i_1}\delta(\varepsilon-\mathcal{H}_0)J_{i_2}\delta(\varepsilon+\hbar\omega-\mathcal{H}_0)J_{i_0}G^{r+a}))
\\
=&\frac{i\hbar^2}{8\pi}\hat{\mathcal{P}}^{(\Gamma_1^{-})}_{i_1i_2}\hat{\mathcal{P}}^{(\Gamma_1^{-})}_{(\omega,-\omega)}\int d\varepsilon\frac{\rho_0(\varepsilon+\hbar\omega)-\rho_0(\varepsilon)}{\hbar^2\omega^2}
\\
&\times(\text{tr}(J_{i_0}G^{r+a}J_{i_1}G^{r-a}_{\omega}J_{i_2}G^{r-a}-J_{i_0}G^{r-a}J_{i_1}G^{r-a}_{\omega}J_{i_2}G^{r+a})
\\
&\quad\,+\text{tr}(J_{i_2}G^{r+a}J_{i_0}G^{r-a}_{\omega}J_{i_1}G^{r-a}+J_{i_2}G^{r-a}J_{i_0}G^{r-a}_{\omega}J_{i_1}G^{r+a})
\\
&\quad\,-\text{tr}(J_{i_1}G^{r+a}J_{i_2}G^{r-a}_{\omega}J_{i_0}G^{r-a}+J_{i_1}G^{r-a}J_{i_2}G^{r-a}_{\omega}J_{i_0}G^{r+a})).
\end{split}
\end{equation}

This is a relatively `simple' expression, giving the relevant conductivity purely in terms of Green's operators and also gives rise to sum rules due to the frequency being an argument of a delta function. We elucidate this in paper III \cite{Bonbien2021c} and also show that, for single electrons in a clean semiconductor, this formula reduces to the standard expression for the `circular' shift current \cite{Ahn2020,Watanabe2021}, a contribution to the bulk photovoltaic effect requiring the breaking of time-reversal symmetry.\\

Next, we look at the magnetic-inversion even contribution. According to \eqref{eq:TRCond2} and \eqref{eq:RDCond2}, in this case, \eqref{eq:Cond2FDWDC} becomes

\begin{equation}
\label{eq:Cond2FDWDC2}
\begin{split}
2\omega^2\sigma^{\mathcal{T}^{\text{full}}_-,\mathcal{K}^{\textbf{m}}_+}_{i_0i_1i_2}(\omega,-\omega)=&\tilde{C}^{r-a}_{J_{i_0}J_{i_1i_2}}(\omega)+\tilde{C}^{r-a}_{J_{i_0}J_{i_1i_2}}(-\omega)
\\
&+\hat{\mathcal{P}}^{(\Gamma_1^{+})}_{i_1i_2}(\tilde{C}^{r-a}_{J_{i_0}J_{i_1}J_{i_2}}(\omega,-\omega)-\tilde{C}^{r-a}_{J_{i_0}J_{i_1}J_{i_2}}(\omega,0)-\tilde{C}^{r-a}_{J_{i_0}J_{i_1}J_{i_2}}(0,-\omega)).
\end{split}
\end{equation}

The 2-point fluctuation-dissipation relation \eqref{eq:2pointFDFreq} and the 3-point weak fluctuation-dissipation relations \eqref{eq:3FlucDissWFreqDC}, \eqref{eq:3FlucDissWFreq01}, and \eqref{eq:3FlucDissWFreq10} let us write \eqref{eq:Cond2FDWDC2} as

\begin{equation}
\begin{split}
\omega^2\sigma^{\mathcal{T}^{\text{full}}_-,\mathcal{K}^{\textbf{m}}_+}_{i_0i_1i_2}(\omega,-\omega)=&-\frac{i}{\hbar}\hat{\mathcal{P}}^{(\Gamma_1^{+})}_{(\omega,-\omega)}\tilde{C}_{[J_{i_0}J_{i_1i_2}]}(\omega)
\\
&+\frac{i}{4\hbar^2\pi}\hat{\mathcal{P}}^{(\Gamma_1^{+})}_{i_1i_2}\hat{\mathcal{P}}^{(\Gamma_1^{+})}_{(\omega,-\omega)}\mathtt{pv}\int\frac{d\omega'}{\omega'}(\tilde{C}_{[J_{i_0}J_{i_1}J_{i_2}]}(\omega'+\omega,-\omega)
-\tilde{C}_{[J_{i_2}J_{i_0}J_{i_1}]}(\omega'+\omega,-\omega)
\\
&\qquad\qquad\qquad\qquad\qquad\qquad\qquad+\tilde{C}_{[J_{i_1}J_{i_2}J_{i_0}]}(\omega'+\omega,-\omega)).
\end{split}
\end{equation}

Making use of the spectral representations \eqref{eq:2strippedSpec} and \eqref{eq:3strippedSpec} and also \eqref{eq:GrpGa} we finally get

\begin{equation}
\begin{split}
\sigma^{\mathcal{T}^{\text{full}}_-,\mathcal{K}^{\textbf{m}}_+}_{i_0i_1i_2}(\omega,-\omega)
=&-\frac{i\hbar^2}{2\pi}\hat{\mathcal{P}}^{(\Gamma_1^{+})}_{(\omega,-\omega)}\int d\varepsilon\frac{\rho_0(\varepsilon+\hbar\omega)-\rho_0(\varepsilon)}{\hbar^2\omega^2}
\text{tr}(J_{i_0}G^{r-a}_{\omega}J_{i_1i_2}G^{r-a})
\\
&-\frac{i\hbar^2}{8\pi}\hat{\mathcal{P}}^{(\Gamma_1^{+})}_{i_1i_2}\hat{\mathcal{P}}^{(\Gamma_1^{+})}_{(\omega,-\omega)}\int d\varepsilon\frac{\rho_0(\varepsilon+\hbar\omega)-\rho_0(\varepsilon)}{\hbar^2\omega^2}
\\
&\quad\times(\text{tr}(J_{i_0}G^{r+a}J_{i_1}G^{r-a}_{\omega}J_{i_2}G^{r-a}-J_{i_0}G^{r-a}J_{i_1}G^{r-a}_{\omega}J_{i_2}G^{r+a})
\\
&\qquad\,-\text{tr}(J_{i_2}G^{r+a}J_{i_0}G^{r-a}_{\omega}J_{i_1}G^{r-a}+J_{i_2}G^{r-a}J_{i_0}G^{r-a}_{\omega}J_{i_1}G^{r+a})
\\
&\qquad\,+\text{tr}(J_{i_1}G^{r+a}J_{i_2}G^{r-a}_{\omega}J_{i_0}G^{r-a}+J_{i_1}G^{r-a}J_{i_2}G^{r-a}_{\omega}J_{i_0}G^{r+a})).
\end{split}
\end{equation}

This is a formula for the dissipative contribution to the direct current conductivity that remains non-vanishing even in the presence of time-reversal symmetry---it is even under magnetic-inversion. In the literature, this is referred to as the shift current \cite{SipeShkrebtii2000,Ahn2020} and we show in paper III \cite{Bonbien2021c} that our formula reduces to the standard expressions in the clean limit. Reflecting on our discussion of sum rules, it is clear that we can integrate out the frequency and, yet again, are provided with a sum rule. In fact, special versions of this particular sum rule have appeared recently \cite{Patankar2018,QRSR}, however, it was not recognized that, fundamentally, they stem from our weak fluctuation-dissipation theorem. We present the details in paper III \cite{Bonbien2021c}.\\

\subsubsection{Dissipative, second-harmonic charge conductivity}

Now we take $\omega_1=\omega_2=\omega$ in \eqref{eq:Cond2} to get

\begin{equation}
\label{eq:Cond2FDWSHG}
\begin{split}
\omega^2\sigma^{r-a}_{i_0i_1i_2}(\omega,\omega)=&\tilde{C}^{r-a}_{J_{i_0}J_{i_1i_2}}(2\omega)-2\tilde{C}^{r-a}_{J_{i_0}J_{i_1i_2}}(\omega)
\\
&-\tilde{C}^{r-a}_{J_{i_0}J_{i_1}J_{i_2}}(\omega,\omega)+\tilde{C}^{r-a}_{J_{i_0}J_{i_1}J_{i_2}}(\omega,0)+\tilde{C}^{r-a}_{J_{i_0}J_{i_1}J_{i_2}}(0,\omega),
\end{split}
\end{equation}

and, yet again, we used $\tilde{C}^{r-a}_{J_{i_0}J_{i_1i_2}}(0)=0$ together with $\tilde{C}^{r-a}_{J_{i_0}J_{i_1}J_{i_2}}(0,0)=0$. Note that intrinsic permutation symmetry requires $\sigma^{r-a}_{i_0i_1i_2}(\omega,\omega)$ to be symmetric in $i_1,i_2$ and, contrary to the direct current responses discussed above, there is no connection between the real (imaginary) part and symmetry (antisymmetry) in $i_1,i_2$. By \eqref{eq:TRCond2} and \eqref{eq:RDCond2} the magnetic inversion eigenfunctions are then

\begin{equation}
\label{eq:Cond2FDWSHG2}
\begin{split}
2\omega^2\sigma^{\mathcal{T}^{\text{full}}_-,\mathcal{K}^{\textbf{m}}_{\pm}}_{i_0i_1i_2}(\omega,\omega)=&\hat{\mathcal{P}}^{(\Gamma_1^{\pm})}_{(\omega,-\omega)}(\tilde{C}^{r-a}_{J_{i_0}J_{i_1i_2}}(2\omega)-2\tilde{C}^{r-a}_{J_{i_0}J_{i_1i_2}}(\omega))
\\
&-\hat{\mathcal{P}}^{(\Gamma_1^{+})}_{i_1i_2}\hat{\mathcal{P}}^{(\Gamma_1^{\pm})}_{(\omega,-\omega)}(\tilde{C}^{r-a}_{J_{i_0}J_{i_1}J_{i_2}}(\omega,\omega)-2\tilde{C}^{r-a}_{J_{i_0}J_{i_1}J_{i_2}}(\omega,0)).
\end{split}
\end{equation}

We invoke the 2-point fluctuation-dissipation relation \eqref{eq:2pointFDFreq} and the 3-point weak fluctuation-dissipation relations \eqref{eq:3FlucDissWFreqSHG} and \eqref{eq:3FlucDissWFreq10} to express the result in terms of stripped correlators as

\begin{equation}
\begin{split}
\omega^2\sigma^{\mathcal{T}^{\text{full}}_-,\mathcal{K}^{\textbf{m}}_{\pm}}_{i_0i_1i_2}(\omega,\omega)=&-\frac{i}{2\hbar}\hat{\mathcal{P}}^{(\Gamma_1^{\pm})}_{(\omega,-\omega)}(\tilde{C}_{[J_{i_0}J_{i_1i_2}]}(2\omega)-2\tilde{C}_{[J_{i_0}J_{i_1i_2}]}(\omega))
\\
&-\frac{i}{8\hbar^2\pi}\hat{\mathcal{P}}^{(\Gamma_1^{+})}_{i_1i_2}\hat{\mathcal{P}}^{(\Gamma_1^{\pm})}_{(\omega,-\omega)}\mathtt{pv}\int\frac{d\omega'}{\omega'}(2\tilde{C}_{[J_{i_2}J_{i_0}J_{i_1}]}(\omega'-2\omega,\omega)-\tilde{C}_{[J_{i_1}J_{i_2}J_{i_0}]}(\omega'+\omega,-2\omega)
\\
&\qquad\qquad\qquad\qquad\qquad\qquad\qquad+2\tilde{C}_{[J_{i_1}J_{i_2}J_{i_0}]}(\omega'+\omega,-\omega)-2\tilde{C}_{[J_{i_2}J_{i_0}J_{i_1}]}(\omega'-\omega,\omega)).
\end{split}
\end{equation}

Applying the spectral representations \eqref{eq:2strippedSpec} and \eqref{eq:3strippedSpec} together with \eqref{eq:GrpGa} we obtain

\begin{equation}
\begin{split}
&\sigma^{\mathcal{T}^{\text{full}}_-,\mathcal{K}^{\textbf{m}}_{\pm}}_{i_0i_1i_2}(\omega,\omega)
\\
&=-\frac{i\hbar^2}{2\pi}\hat{\mathcal{P}}^{(\Gamma_1^{\pm})}_{(\omega,-\omega)}\int d\varepsilon\bigg(2\frac{\rho_0(\varepsilon+2\hbar\omega)-\rho_0(\varepsilon)}{4\hbar^2\omega^2}
\text{tr}(J_{i_0}G^{r-a}_{2\omega}J_{i_1i_2}G^{r-a})-\frac{\rho_0(\varepsilon+\hbar\omega)-\rho_0(\varepsilon)}{\hbar^2\omega^2}
\text{tr}(J_{i_0}G^{r-a}_{\omega}J_{i_1i_2}G^{r-a})\bigg)
\\
&\quad-\frac{i\hbar^2}{8\pi}\hat{\mathcal{P}}^{(\Gamma_1^{+})}_{i_1i_2}\hat{\mathcal{P}}^{(\Gamma_1^{\pm})}_{(\omega,-\omega)}\int d\varepsilon\bigg\{2\frac{\rho_0(\varepsilon+2\hbar\omega)-\rho_0(\varepsilon)}{4\hbar^2\omega^2}\text{tr}(J_{i_1}G^{r+a}_{-\omega}J_{i_2}G^{r-a}_{-2\omega}J_{i_0}G^{r-a}\mp J_{i_2}G^{r-a}J_{i_0}G^{r-a}_{-2\omega}J_{i_1}G^{r+a}_{-\omega})
\\
&\qquad\qquad\qquad\qquad\qquad\qquad\,-\frac{\rho_0(\varepsilon+\hbar\omega)-\rho_0(\varepsilon)}{\hbar^2\omega^2}\bigg(
\text{tr}(J_{i_2}G^{r+a}_{-\omega}J_{i_0}G^{r-a}_{\omega}J_{i_1}G^{r-a}\mp J_{i_1}G^{r-a}J_{i_2}G^{r-a}_{\omega}J_{i_0}G^{r+a}_{-\omega})
\\
&\qquad\qquad\qquad\qquad\qquad\qquad\quad\qquad\qquad\qquad\qquad\quad\,
-\text{tr}(J_{i_2}G^{r+a}J_{i_0}G^{r-a}_{\omega}J_{i_1}G^{r-a}+J_{i_2}G^{r-a}J_{i_0}G^{r-a}_{\omega}J_{i_1}G^{r+a})
\\
&\qquad\qquad\qquad\qquad\qquad\qquad\quad\qquad\qquad\qquad\qquad\quad\,\pm\text{tr}(J_{i_1}G^{r+a}J_{i_2}G^{r-a}_{\omega}J_{i_0}G^{r-a}+J_{i_1}G^{r-a}J_{i_2}G^{r-a}_{\omega}J_{i_0}G^{r+a})\bigg)\bigg\}.
\end{split}
\end{equation}

While this expression looks particularly daunting, conceptually, it is, in fact, relatively simple and provides us with an \textit{exact} formula for all dissipative second-harmonic charge current responses, moreover, it also allows us to derive new sum rules since the frequency can be integrated out.
\end{widetext}

\section{Permutation structure of the retarded correlator}
\label{pstruct}
The correlator depends on multiple time-arguments and observable operators and thus carries the action of several realizations of permutation groups. If we consider a correlator $C_{A^0A^1\dots A^n}(t_0,t_1,\dots,t_n)$, we can permute the time arguments and observables separately and can also permute them simultaneously. These three realizations of the permutation group, that we denote as $P^{t_0\dots t_n}(n+1),\,P^{A^1\dots A^n}(n+1)$ and $P^{A^0_{t_0}\dots A^n_{t_n}}(n+1)$ respectively, and their representations on the correlator will be the focus of this section. The behaviour of the retarded correlator under permutation is closely linked to the physical phenomena it can describe and consequently decomposing it into terms transforming in the irreducible representations of the above-mentioned permutation groups leads to a natural way of attributing the given physical phenomena to specific terms in the decomposition. Owing to the fact that the permutation group $P(n)$ of $n$ objects, dubbed the symmetric group $S_n$ in the mathematics literature, is non-commutative for $n>2$ and has $n!$ elements, we will only perform the decomposition for the retarded 2, 3 and 4-point correlators acted on by $P(2),P(3)$ and $P(4)$, contributing to first, second and third order responses.\\

The notation for the permutations in this section will be as follows. The group elements represented on the correlator are labelled as $\Pi^{t_0\dots t_n},\,\Pi^{A^0\dots A^n}$ and $\Pi^{A^0_{t_0}\dots A^n_{t_n}}$ for the three groups respectively. For example, $P^{t_0t_1}(2)\to\{\Pi^{t_0t_1},\Pi^{t_1t_0}\}$ and $P^{t_0t_1t_2}(3)\to \{\Pi^{t_0t_1t_2},\Pi^{t_1t_0t_2},\Pi^{t_2t_1t_0},\Pi^{t_0t_2t_1},\Pi^{t_1t_2t_0},\Pi^{t_2t_0t_1}\}$ with the actions being

\begin{equation}
\nonumber
\Pi^{t_1t_0t_2}C_{A^0A^1A^2}(t_0,t_1,t_2)=C_{A^0A^1A^2}(t_1,t_0,t_2),
\end{equation}

and similarly for all group elements. The action of the other two groups is denoted analogously with example actions being

\begin{equation}
\nonumber
\begin{split}
&\Pi^{A_2A_0A_1}C_{A^0A^1A^2}(t_0,t_1,t_2)=C_{A^2A^0A^1}(t_0,t_1,t_2),
\\
&\Pi^{A^1_{t_1}A^2_{t_2}A^0_{t_0}}C_{A^0A^1A^2}(t_0,t_1,t_2)=C_{A^1A^2A^0}(t_1,t_2,t_0),
\end{split}
\end{equation}

and so on. We can also consider the action of $P(n)$ on an $n+1$-point correlator by fixing one observable or time-argument and permuting the rest; an example for the 3-point correlator is

\begin{equation}
\nonumber
\Pi^{A^2A^1}C_{A^0A^1A^2}(t_0,t_1,t_2)=C_{A^0A^2A^1}(t_1,t_0,t_2).
\end{equation}

We present the $4!=24$ elements of $ P(4)$ and summarize the character tables and irreducible representations of $P(2),\,P(3)$ and $P(4)$ in Appendix \ref{GroupTools}.\\

We first look at the interplay between the permutation of the time-arguments, observables and the time-inversion, \textbf{m}-inversion. Following this, we present the formal decomposition into irreducible representations of the different permutation groups.

\subsection{Permutation, time-inversion and magnetic-inversion}
\label{OnsagerSection}

We introduced the groups $P^{t_0\dots t_n}(n+1)$ permuting time-labels and $P^{A^0\dots A^n}(n+1)$ permuting the observables, however the actions of these on the correlators of interest to us are closely related. Consider the exchange of $A^0$ and $A^1$ in the stripped correlator. We have

\begin{equation}
\label{eq:PermA=t}
\begin{split}
&\Pi^{A^1A^0A^2\dots A^n}C_{[A^0A^1\dots A^n]}(t_0,t_1,\dots,t_n)
\\
&\,=C_{[A^1A^0\dots A^n]}(t_0,t_1,\dots,t_n)
\\
&\,=\text{tr}(\rho_0[[\dots[A^1_{\mathcal{H}_0}(t_0),A^0_{\mathcal{H}_0}(t_1)],\dots],A^n_{\mathcal{H}_0}(t_n)])
\\
&\,=-\text{tr}(\rho_0[[\dots[A^0_{\mathcal{H}_0}(t_1),A^1_{\mathcal{H}_0}(t_0)],\dots],A^n_{\mathcal{H}_0}(t_n)])
\\
&\,=-\Pi^{t_1t_0t_2\dots t_n}C_{[A^0A^1\dots A^n]}(t_0,t_1,t_2,\dots,t_n).
\end{split}
\end{equation} 

The commutator structure ensures that we can permute the observables and time-labels simultaneously, so permuting only observables can be compensated by permuting time-labels. We next elaborate on the implications of this.\par

Using the time-translation invariance of the stripped correlator stemming from the cyclic property of the trace,  we have

\begin{equation}
\begin{split}
&C_{[A^0A^1\dots A^n]}(t_0,t_1,t_2,\dots,t_n;\textbf{m})
\\
&\,=C_{[A^0A^1\dots A^n]}(0,t_1-t_0,t_2-t_0,\dots,t_n-t_0;\textbf{m}).
\end{split}
\end{equation}

Now suppose we exchange $t_0$ and $t_1$ to get

\begin{equation}
\nonumber
\begin{split}
&\Pi^{t_1t_0t_2\dots t_n}C_{[A^0A^1\dots A^n]}(t_0,t_1,t_2,\dots,t_n;\textbf{m})
\\
&=C_{[A^0A^1\dots A^n]}(0,t_0-t_1,t_2-t_1,\dots,t_n-t_1;\textbf{m}).
\end{split}
\end{equation}

It is apparent that if $n=1$, i.e., we are looking at a 2-point correlator with two time arguments, this exchange is equivalent to a time-inversion, but for $n>1$ it is not. In fact, no matter how we permute the time labels for $n>1$ we cannot, in general, make the permutation equivalent to a time-inversion. Indeed, through permutation we can obtain time arguments as $t_{i_1}-t_i,t_{i_2}-t_i,\dots, t_{i_n}-t_i$ where we are subtracting the same $t_i$ from the rest, whereas a time-inversion would yield $t_j-t_0,t_j-t_1,\dots,t_j-t_n$, and we subtract \textit{from} the same $t_j$.\\

For the stripped 2-point correlator we thus have

\begin{equation}
\Pi^{t_1t_0}C_{[A^0A^1]}(t_0,t_1;\textbf{m})=\mathcal{K}C_{[A^0A^1]}(t_0,t_1;\textbf{m}).
\end{equation}

However, we showed in \eqref{eq:PermA=t} above, that the exchange of time labels is related to the exchange of observables, meaning that 

\begin{equation}
\label{eq:PermA=K}
\mathcal{K}C_{[A^0A^1]}(t_0,t_1;\textbf{m})=-\Pi^{A^1A^0}C_{[A^0A^1]}(t_0,t_1;\textbf{m}).
\end{equation}

Thus, for the special case of the 2-point stripped correlator, time-inversion is, upto a sign, equivalent to the exchange of the observables. This identity was derived by Kubo in his seminal paper \cite{Kubo1957}, however not from the general, group representation point of view presented here. As recognized by Kubo, the identity has crucial significance for first-order responses and is the underlying reason for the existence of Onsager's reciprocity relations. Suppose we decompose $C_{[A^0A^1]}(t_0,t_1)$ into terms $C^{\mathcal{K}_{\pm}}_{[A^0A^1]}(t_0,t_1)$  transforming in the irreducible representations of the time-inversion group $\mathcal{G}_{\mathfrak{K}}$, or, in other words, terms that are even and odd under time-inversion and consequently eigenfunctions of $\mathcal{K}$. Then, owing to this identity, these terms will also be eigenfunctions of the permutation $\Pi^{A^1A^0}$ and transform in the irreducible representations $\Gamma_1^{+}\,(\Gamma_1^{-})$ of the group $P^{A^0A^1}(2)$ corresponding to terms that are symmetric (antisymmetric) under the exchange of $A^0$ and $A^1$. Accordingly, we can add a further label to the decomposed stripped correlator $C^{\mathcal{K}_{\pm},\Gamma_1^{\mp}}_{[A^0A^1]}(t_0,t_1)$, displaying the fact that the negative sign in the identity results in the time-inversion even (odd) eigenfunction being \textit{anti}symmetric (symmmetric). \par

The physical response is given by the retarded correlator so we have to translate identity \eqref{eq:PermA=K} to it. Multiplying both sides of \eqref{eq:PermA=K} by $i\theta_{t_0t_1}$ and seperately by $-i\theta_{t_1t_0}$ yields

\begin{equation}
\label{eq:PermA=KRA}
\begin{split}
&\mathcal{K}C^{r(a)}_{A^0A^1}(t_0,t_1;\textbf{m})=\Pi^{A^1A^0}C^{a(r)}_{A^0A^1}(t_0,t_1;\textbf{m}).
\end{split}
\end{equation} 

We see that, in stark contrast to the stripped correlator, if we decompose the retarded and advanced correlators into eigenfunctions of time-inversion, the latter eigenfunctions will not simultaneously be eigenfunctions of permutation. On the other hand, time-inversion is closely related to magnetic-inversion through \eqref{eq:RetAdvCKKmId}, which we can then combine with identities \eqref{eq:PermA=KRA} to find

\begin{equation}
\label{eq:PermA=KmRR}
\begin{split}
&\Pi^{A^1A^0}C^{r(a)}_{A^0A^1}(t_0,t_1;\textbf{m})=\epsilon^{\mathcal{T}}_{A^0}\epsilon^{\mathcal{T}}_{A^1}\mathcal{K}^{\textbf{m}}C^{r(a)}_{A^0A^1}(t_0,t_1;\textbf{m}),
\end{split}
\end{equation} 

and we can recognize Onsager's reciprocity relations. The consequence of these relations is that we can decompose the 2-point retarded (and advanced) correlators into simultaneous eigenfunctions of magnetic-inversion and permutation as 

\begin{equation}
\label{eq:PermMagInv}
\begin{split}
C^{r,\mathcal{K}^{\textbf{m}}_{\pm},\Gamma^{\pm\text{sgn}(\epsilon)}_1}_{A^0A^1}=&
\frac{C^r_{A^0A^1}\pm\mathcal{K}^{\textbf{m}}C^{r}_{A^0A^1}}{2}
\\
=&\frac{C^r_{A^0A^1}\pm \epsilon^{\mathcal{T}}_{A^0}\epsilon^{\mathcal{T}}_{A^1}\Pi^{A^1A^0}C^{r}_{A^0A^1}}{2},
\end{split}
\end{equation}

where we defined $\text{sgn}(\epsilon)\equiv\text{sign}(\epsilon^{\mathcal{T}}_{A^0}\epsilon^{\mathcal{T}}_{A^1})$ and dropped the time and $\textbf{m}$ arguments for brevity. This also means that, when projecting the retarded correlator onto the irreducible representations of $\mathcal{G}_{\mathfrak{K}}\times\mathcal{G}_{\mathfrak{K}^{\textbf{m}}}$, we can add a further label corresponding to the irreducible representations of $P^{A^0A^1}(2)$. Indeed, injecting identity \eqref{eq:PermMagInv} into the $n=1$ case of \eqref{eq:RDTRSTRBmagInv}, the reactive and dissipative responses can be written as

\begin{equation}
\label{eq:RDTRSTRBmagInvPerm}
\begin{split}
&C^{r,\mathcal{T}^{\text{full}}_{\pm},\mathcal{K}^{\textbf{m}}_{+},\Gamma^{\text{sgn}(\epsilon)}_1}_{A^0A^1}=\frac{C^{r,\mathcal{K}^{\textbf{m}}_{+},\Gamma^{\text{sgn}(\epsilon)}_1}_{A^0A^1}\pm \mathcal{K}^{\text{full}}C^{r,\mathcal{K}^{\textbf{m}}_{+},\Gamma^{\text{sgn}(\epsilon)}_1}_{A^0A^1}}{2},
\\
&C^{r,\mathcal{T}^{\text{full}}_{\mp},\mathcal{K}^{\textbf{m}}_{-},\Gamma^{-\text{sgn}(\epsilon)}_1}_{A^0A^1}=\frac{C^{r,\mathcal{K}^{\textbf{m}}_{-},\Gamma^{-\text{sgn}(\epsilon)}_1}_{A^0A^1}\pm \mathcal{K}^{\text{full}}C^{r,\mathcal{K}^{\textbf{m}}_{-},\Gamma^{-\text{sgn}(\epsilon)}_1}_{A^0A^1}}{2}.
\end{split}
\end{equation}

Thus, the behaviour under permutation is directly related to whether $\mathcal{H}_0$ is time-reversal symmetric or time-reversal breaking with the reactive and dissipative responses behaving accordingly. We can further obtain the frequency domain version of \eqref{eq:RDTRSTRBmagInvPerm}, by adding a permutation label to the $n=1$ case of \eqref{eq:RDTRSTRBmagInvFreq} and realize that the reactive and dissipative responses correspond to, \textit{not} respectively, the real or imaginary parts of the symmetric or antisymmetric retarded correlators. In the special case of the first order charge conductivity $\sigma^r_{i_0i_1}(\omega_1)$, we have $\text{sgn}(\epsilon)=\text{sign}(\epsilon^{\mathcal{T}}_{J_{i_0}}\epsilon^{\mathcal{T}}_{J_{i_1}})=+1$, meaning $\sigma^{\mathcal{K}^{\textbf{m}}_{\pm}}_{i_0i_1}(\omega_1)\to \sigma^{\mathcal{K}^{\textbf{m}}_{\pm},\Gamma^{\pm}_1}_{i_0i_1}$ and the reactive/dissipative responses found in \eqref{eq:RDCond1} become

\begin{equation}
\label{eq:RDCond1Perm}
\begin{split}
&\sigma^{\mathcal{T}^{\text{full}}_{\pm},\mathcal{K}^{\textbf{m}}_{+},\Gamma^{+}_1}_{i_0i_1}(\omega_1)=\frac{\sigma^{r,\mathcal{K}^{\textbf{m}}_{+},\Gamma^{+}_1}_{i_0i_1}(\omega_1)\mp \left(\sigma^{r,\mathcal{K}^{\textbf{m}}_{+},\Gamma^{+}_1}_{i_0i_1}(\omega_1)\right)^*}{2},
\\
&\sigma^{\mathcal{T}^{\text{full}}_{\mp},\mathcal{K}^{\textbf{m}}_{-},\Gamma^{-}_1}_{i_0i_1}(\omega_1)=\frac{\sigma^{r,\mathcal{K}^{\textbf{m}}_{-},\Gamma^{-}_1}_{i_0i_1}(\omega_1)\mp \left(\sigma^{r,\mathcal{K}^{\textbf{m}}_{-},\Gamma^{-}_1}_{i_0i_1}(\omega_1)\right)^*}{2}.
\end{split}
\end{equation}

It is clear that the real symmetric and imaginary antisymmetric parts are dissipative, whereas the real antisymmetric and imaginary symmetric parts are reactive, in complete consistency with the energy arguments in the literature \cite{LandauLifshitzEoCM,NagaosaAHErev}.\\

While the results presented in this subsection are widely known, we want to emphasize that, from the correlator's point of view, the reciprocity relations are more of a `coincidental' result stemming from the intimate connections between the representations of different groups on the correlator, with the crucial connection being that, for the 2-point stripped correlator, the permutation of time-labels is equivalent to time-inversion, the latter being closely related to magnetic-inversion. It is also apparent why the reciprocity relations cannot be extended to higher order correlators: for the stripped correlator, in general, permutations of the time labels cannot be made equivalent to a time-inversion!\\

\subsection{Permutation decomposition}
\label{PermDecomp}

Having decomposed the 2-point retarded correlator into the irreducible representations of the permutation group $P^{A^0A^1}(2)$, we can think of extending this decomposition to retarded $n+1$-point correlators acted on by $P^{A^0\dots A^n}(n+1)$. All permutation groups have two 1-dimensional irreducible representations---the totally symmetric $\Gamma_1^+$ and antisymmetric $\Gamma_1^-$ representations \cite{DresselhausSymm}. A correlator transforming in these representations has the property that the exchange of any two of its relevant arguments leaves it either unchanged in the symmetric case or flips its sign in the antisymmetric case. Thus, $\Gamma_1^+$ corresponds to the trivial or scalar representation and leaves the correlator invariant.  The rest of the irreducible representations, such as $\Gamma_2$ of $P^{A^0A^1A^2}(3)$, correspond to mixed behaviour, meaning that the correlator transforming in these does not have a definite sign under exchanges of arguments. This line of thought will prove to be fruitful since it will provide a further classification of physical responses. For example, in the 2-point case, the transverse response, that could be the anomalous Hall or spin Hall response, is given by the $\Gamma_1^-$ representation of $P^{A^0A^1}(2)$, whereas in the 3-point case, as we shall expound below, the transverse response to a single driving field in a fixed direction is given by the $\Gamma_2$ representation of $P^{A^0A^1A^2}(3)$. On the other hand, contrary to the 2-point case, as alluded to above, the terms in the decomposition of higher-order retarded correlators are not intimately linked to time-inversion and magnetic-inversion and, consequently, there is no stringent, $\mathcal{H}_0$ time-reversal symmetry dependent restriction on which irreducible representation the correlator can transform in.\\

For $n+1$-point correlators with $n>1$ we can also think of leaving their first observable and time arguments unchanged and permuting only the remaining $n$. This corresponds to actions of $P^{A^1\dots A^n}(n)$, $P^{t_1\dots t_n}(n)$ or $P^{A^1_{t_1}\dots A^{n}_{t_n}}(n)$. For example, these can be represented on the 3-point retarded correlator $C^r_{A^0A^1A^2}(t_0,t_1,t_2)$ with elements $\Pi^{A^1A^2},\Pi^{t_1t_2}$ or $\Pi^{A^1_{t_1}A^2_{t_2}}$ and the identity. The retarded correlator can then be further decomposed into the irreducible representations of these groups. The physical importance of this can be emphasized by, for example, thinking of the second-harmonic response. This is described by the 3-point correlator and corresponds to a term that is symmetric in the exchange of $t_1,t_2$.\par
We can now recognize that the behaviour of the retarded correlator under $P^{A^1_{t_1}\dots A^{n}_{t_n}}(n)$ is already fixed. Indeed, looking at the definition \eqref{eq:Retn} of $C^r_{A^0A^1\dots A^n}(t_0,t_1,\dots,t_n)$ we see that the latter is totally symmetric under the simultaneous exchange of the observables $A^1\dots A^n$ and time arguments $t_1\dots t_n$, and thus transforms trivially, i.e., in the totally symmetric irreducible representation $\Gamma_1^+$ of $P^{A^1_{t_1}\dots A^{n}_{t_n}}(n)$, and is known as the 'intrinsic permutation symmetry' in the literature \cite{Rabin1975,BoydNLO}. This provides restrictions on how the retarded correlator can transform under $P^{A^1\dots A^n}(n)$ and $P^{t_1\dots t_n}(n)$. For example in the 3-point case, if the retarded correlator is antisymmetric under the exchange of $t_1,t_2$ it also has to be antisymmetric under the exchange of $A^1,A^2$, otherwise it will not be symmetric under the simultaneous exchange. Formally, $P^{A^1_{t_1}\dots A^{n}_{t_n}}(n)$ is a subgroup of the product group $P^{A^1\dots A^n}(n)\times P^{t_1\dots t_n}(n)$ and they are both represented on the correlator. An irreducible representation of $P^{A^1\dots A^n}(n)\times P^{t_1\dots t_n}(n)$ is not necessarily an irreducible representation of its subgroup $P^{A^1_{t_1}\dots A^{n}_{t_n}}(n)$ but is definitely a \textit{reducible} representation of the latter and we can thus decompose the irreducible representations of $P^{A^1\dots A^n}(n)\times P^{t_1\dots t_n}(n)$ into the irreducible representations of $P^{A^1_{t_1}\dots A^{n}_{t_n}}(n)$. We want to find those irreducible representations of $P^{A^1\dots A^n}(n)\times P^{t_1\dots t_n}(n)$ whose latter decompositions contain the trivial representation $\Gamma_1^+$ of $P^{A^1_{t_1}\dots A^{n}_{t_n}}(n)$. The irreducible representations of $P^{A^1\dots A^n}(n)\times P^{t_1\dots t_n}(n)$ can be written as $\Gamma_i\otimes\Gamma_j$ where $\Gamma_i$ on the left(right) of the product  is the $i$-th irreducible representation of $P^{A^1\dots A^n}(n)(P^{t_1\dots t_n}(n))$. The only way the product $\Gamma_i\otimes\Gamma_j$ will contain $\Gamma_1^+$ of $P^{A^1_{t_1}\dots A^{n}_{t_n}}(n)$, is if $\Gamma_i=\Gamma_j$. The reason for this is, roughly speaking, that if purely observable and purely time label permutations are represented differently they cannot compensate each other to yield a quantity transforming trivially under their simultaneous action. For example, the correlator cannot be totally antisymmetric under $P^{A^1\dots A^n}(n)$ but totally symmetric under $P^{t_1\dots t_n}(n)$ while being totally symmetric under $P^{A^1_{t_1}\dots A^{n}_{t_n}}(n)$. Thus, apart from the decomposition into the irreducible representations of $P^{A^0\dots A^n}(n+1)$, we can also decompose the retarded correlator $C^r_{A^0A^1\dots A^n}(t_0,t_1,\dots,t_n)$ into terms transforming in $\Gamma_i\otimes\Gamma_i$ that are certain irreducible representations of $P^{A^1\dots A^n}(n)\times P^{t_1\dots t_n}(n)$.\par

These decompositions are made explicit by utilizing projections onto the irreducible representations which we denote as $\hat{\mathcal{P}}^{(\Gamma_i)}_{A^{j_1}\dots A^{j_n}},\hat{\mathcal{P}}^{(\Gamma_i)}_{t_{j_1}\dots t_{j_n}}$ and $\hat{\mathcal{P}}^{(\Gamma_i)}_{A^{j_1}_{t_{j_1}}\dots A^{j_n}_{t_{j_n}}}$ for the $\Gamma_i$ irreducible representation of $P^{A^{j_1}\dots A^{j_n}}(n), P^{t_{j_1}\dots t_{j_n}}(n)$ or $P^{A^{j_1}_{t_{j_1}}\dots A^{j_n}_{t_{j_n}}}(n)$. Based on the above discussion, we have
\begin{equation}
\label{eq:NPermDecomp}
\begin{split}
&C^r_{A^0A^1\dots A^n}(t_0,t_1,\dots,t_n)
\\
&=\sum_{j}\hat{\mathcal{P}}^{(\Gamma_j)}_{A^{0}\dots A^{n}}\hat{\mathcal{P}}^{(\Gamma_1^+)}_{A^{1}_{t_{1}}\dots A^{n}_{t_{n}}}C^r_{A^0A^1\dots A^n}(t_0,t_1,\dots,t_n)
\\
&=\sum_{j}\hat{\mathcal{P}}^{(\Gamma_j)}_{A^{0}\dots A^{n}}\sum_i \hat{\mathcal{P}}^{(\Gamma_i)}_{A^{1}\dots A^{n}}\hat{\mathcal{P}}^{(\Gamma_i)}_{t_{1}\dots t_{n}}C^r_{A^0A^1\dots A^n}(t_0,t_1,\dots,t_n),
\end{split}
\end{equation} 

where, in the first equality, we used the fact that

\begin{equation}
\label{eq:RetIntrinsicPerm}
\begin{split}
C^r_{A^0A^1\dots A^n}&(t_0,t_1,\dots,t_n)
\\
&=\hat{\mathcal{P}}^{(\Gamma_1^+)}_{A^{1}_{t_{1}}\dots A^{n}_{t_{n}}}C^r_{A^0A^1\dots A^n}(t_0,t_1,\dots,t_n),
\end{split}
\end{equation}

which is the statement of the so-called intrinsic permutation symmetry and follows from the definition \eqref{eq:Retn} of the retarded correlator that can be written using the projector as

\begin{equation}
\nonumber
\begin{split}
&C^r_{A^0A^1\dots A^n}(t_0,t_1,\dots,t_n)
\\
&=(-i)^n\hat{\mathcal{P}}^{(\Gamma_1^+)}_{A^{1}_{t_{1}}\dots A^{n}_{t_{n}}}\theta_{t_0t_1}\cdots\theta_{t_{n-1}t_n}C_{[A^0A^1\dots A^n]}(t_0,t_1,\dots,t_n),
\end{split}
\end{equation}

whereas in the second equality, we utilized that only the $\Gamma_i\otimes\Gamma_i$ representations of $P^{A^1\dots A^n}(n)\times P^{t_1\dots t_n}(n)$ contain the $\Gamma_1^+$ representation of $P^{A^1_{t_1}\dots A^{n}_{t_n}}(n)$.\par

All the results of this subsection can, of course, be transferred to the frequency domain. We can utilize time-translation invariance, consider the Fourier transforms, and replace the redefined time-labels by frequency labels, i.e., the equivalent of \eqref{eq:RetIntrinsicPerm} becomes

\begin{equation}
\label{eq:RetIntrinsicPermFreq}
\begin{split}
\tilde{C}^r_{A^0A^1\dots A^n}&(\omega_1,\dots,\omega_n)
\\
&=\hat{\mathcal{P}}^{(\Gamma_1^+)}_{A^{1}_{\omega_{1}}\dots A^{n}_{\omega_{n}}}\tilde{C}^r_{A^0A^1\dots A^n}(\omega_1,\dots,\omega_n),
\end{split}
\end{equation}

where $\hat{\mathcal{P}}^{(\Gamma_1^+)}_{A^{1}_{\omega_{1}}\dots A^{n}_{\omega_{n}}}$ is the the projector to the totally symmetric irreducible representation of the permutation group $P^{A^1_{\omega_1}\dots A^n_{\omega_n}}(n)$ that acts by simultaneously permuting observables and frequency labels.\\

This is as far as we go on general grounds and now specialize to the cases of $n=2,3,4$ corresponding to retarded 2, 3 and 4-point correlators. The character tables for the permutation groups $P(2),\,P(3)$, and $P(4)$ listing their irreducible representations and facilitating the explicit construction of the projectors to the latter can all be found in Appendix \ref{GroupTools}.\par

\subsubsection{The 2-point retarded correlator}
The decomposition of the 2-point retarded correlator $C^{r}_{A^0A^1}(t_0,t_1)$ is the simplest. We have $P^{A^1}(1)\times P^{t_1}(1)$ which is the trivial group with a single element, the unit, and $P^{A^0A^1}(2)$ which is the permutation group with two elements. The latter has only two irreducible representations $\Gamma_1^+$ and $\Gamma_1^-$ meaning that \eqref{eq:NPermDecomp} becomes

\begin{equation}
C^{r}_{A^0A^1}(t_0,t_1)=\hat{\mathcal{P}}^{(\Gamma_1^+)}_{A^0A^1}C^{r}_{A^0A^1}(t_0,t_1)+\hat{\mathcal{P}}^{(\Gamma_1^-)}_{A^0A^1}C^{r}_{A^0A^1}(t_0,t_1),
\end{equation}

which is just the decomposition into symmetric and antisymmetric parts. Using time-translation invariance, and moving to the frequency domain, we can write the corresponding decomposition as

\begin{equation}
\tilde{C}^{r}_{A^0A^1}(\omega_1)=\hat{\mathcal{P}}^{(\Gamma_1^+)}_{A^0A^1}\tilde{C}^{r}_{A^0A^1}(\omega_1)+\hat{\mathcal{P}}^{(\Gamma_1^-)}_{A^0A^1}\tilde{C}^{r}_{A^0A^1}(\omega_1).
\end{equation}

Physically, the symmetric term corresponds to a longitudinal response, whereas the antisymmetric term describes a transverse response.\\

\subsubsection{The 3-point retarded correlator}

The 3-point retarded correlator $C^r_{A^0A^1A^2}(t_0,t_1,t_2)$ is more involved. We have $P^{A^1A^2}(2)\times P^{t_1t_2}(2)$ with the relevant irreducible representations $\Gamma_1^+\otimes\Gamma_1^+,\,\Gamma_1^-\otimes\Gamma_1^-$, and also $P^{A^0A^1A^2}(3)$ with 3 irreducible representations $\Gamma_1^+,\Gamma_1^-,\Gamma_2$. Thus, \eqref{eq:NPermDecomp} becomes

\begin{widetext}
\begin{equation}
\label{eq:3PermDecomp}
\begin{split}
C^r_{A^0A^1A^2}(t_0,t_1,t_2)=&\left(\hat{\mathcal{P}}^{(\Gamma_1^+)}_{A^0A^1A^2}+
\hat{\mathcal{P}}^{(\Gamma_1^-)}_{A^0A^1A^2}+
\hat{\mathcal{P}}^{(\Gamma_2)}_{A^0A^1A^2}\right)
\left(\hat{\mathcal{P}}^{(\Gamma_1^+)}_{A^1A^2}\hat{\mathcal{P}}^{(\Gamma_1^+)}_{t_1t_2}+\hat{\mathcal{P}}^{(\Gamma_1^-)}_{A^1A^2}\hat{\mathcal{P}}^{(\Gamma_1^-)}_{t_1t_2}\right)C^r_{A^0A^1A^2}(t_0,t_1,t_2)
\\
=&\bigg(\hat{\mathcal{P}}^{(\Gamma_1^+)}_{A^0A^1A^2}\hat{\mathcal{P}}^{(\Gamma_1^+)}_{A^1A^2}\hat{\mathcal{P}}^{(\Gamma_1^+)}_{t_1t_2}
+
\hat{\mathcal{P}}^{(\Gamma_2)}_{A^0A^1A^2}\hat{\mathcal{P}}^{(\Gamma_1^+)}_{A^1A^2}\hat{\mathcal{P}}^{(\Gamma_1^+)}_{t_1t_2}
\\
&\quad +\hat{\mathcal{P}}^{(\Gamma_2)}_{A^0A^1A^2}\hat{\mathcal{P}}^{(\Gamma_1^-)}_{A^1A^2}\hat{\mathcal{P}}^{(\Gamma_1^-)}_{t_1t_2}
+\hat{\mathcal{P}}^{(\Gamma_1^-)}_{A^0A^1A^2}\hat{\mathcal{P}}^{(\Gamma_1^-)}_{A^1A^2}\hat{\mathcal{P}}^{(\Gamma_1^-)}_{t_1t_2}\bigg)C^r_{A^0A^1A^2}(t_0,t_1,t_2)
,
\end{split}
\end{equation}

where we used $\hat{\mathcal{P}}^{(\Gamma_1^{\pm})}_{A^0A^1A^2}\hat{\mathcal{P}}^{(\Gamma_1^{\mp})}_{A^1A^2}=0$ in the second line. This can be shown using the explicit projections, or we can simply think of it as a consequence of the fact that if we symmetrize (antisymmetrize) $A^1A^2$ in $A^0A^1A^2$, the result cannot be totally antisymmetric (symmetric). The two projections $\hat{\mathcal{P}}^{(\Gamma_2)}_{A^0A^1A^2}\hat{\mathcal{P}}^{(\Gamma_1^+)}_{A^0A^1},\hat{\mathcal{P}}^{(\Gamma_2)}_{A^0A^1A^2}\hat{\mathcal{P}}^{(\Gamma_1^-)}_{A^0A^1}$ can be thought of as two partners for the 2-dimensional representation $\Gamma_2$ and thus yield a natural, physically meaningful basis for the latter.\\

We can utilize time-translation invariance and arrive at the same decomposition for the frequency-domain retarded correlator $\tilde{C}^r_{A^0A^1A^2}(\omega_1,\omega_2)$ by replacing the temporal permutations with frequency permutations and corresponding projections $\hat{\mathcal{P}}^{(\Gamma_1^{\pm})}_{\omega_1\omega_2}$. An important question is the physical interpretation of the terms in the decomposition. In order to proceed with this, let us write out the projections explicitly. Using the characters of $P(3)$ from Table \ref{tab:P3}, the projectors \eqref{eq:ProjIrrep} can be written for a function $f_{A^0A^1A^2}$ dependent on the observables as

\begin{align}
&\hat{\mathcal{P}}^{(\Gamma_1^{\pm})}_{A^0A^1A^2}\hat{\mathcal{P}}^{(\Gamma_1^{\pm})}_{A^1A^2}f_{A^0A^1A^2}=\frac{1}{6}(f_{A^0A^1A^2}+f_{A^2A^0A^1}+f_{A^1A^2A^0}\pm f_{A^0A^2A^1}\pm f_{A^1A^0A^2}\pm f_{A^2A^1A^0}),
\\
\label{eq:Gamma2Proj}
&\hat{\mathcal{P}}^{(\Gamma_2)}_{A^0A^1A^2}\hat{\mathcal{P}}^{(\Gamma_1^{\pm})}_{A^1A^2}f_{A^0A^1A^2}=\frac{1}{6}(2f_{A^0A^1A^2}-f_{A^2A^0A^1}-f_{A^1A^2A^0}\pm 2f_{A^0A^2A^1}\mp f_{A^1A^0A^2}\mp f_{A^2A^1A^0}).
\end{align}
\end{widetext}

Recall that $A^1$ and $A^2$ couple to the external driving field and thereby reflect the latter's configuration, whereas $A^0$ refers to the observable whose expectation value we are interested in measuring. Based on the explicit form of the projectors, we can then interpret the permutation decomposition as providing a hierarchy of `longitudinal' and `transverse' responses. Indeed, suppose $A^1=A^2$; then $\hat{\mathcal{P}}^{(\Gamma_1^{+})}_{A^0A^1A^1}$ and $\hat{\mathcal{P}}^{(\Gamma_2)}_{A^0A^1A^1}\hat{\mathcal{P}}^{(\Gamma_1^{+})}_{A^1A^1}\equiv\hat{\mathcal{P}}^{(\Gamma_2)}_{A^0A^1A^1}$ are the only projections to survive. Now, if we further let $A^0=A^1$, then only $\hat{\mathcal{P}}^{(\Gamma_1^{+})}_{A^1A^1A^1}$ will survive. In this way, we can consider $\hat{\mathcal{P}}^{(\Gamma_1^{+})}_{A^0A^1A^2}$ and $\hat{\mathcal{P}}^{(\Gamma_2)}_{A^0A^1A^2}\hat{\mathcal{P}}^{(\Gamma_1^{+})}_{A^1A^2}$ as providing, respectively, a longitudinal and transverse response pair, in direct analogy with the 2-point case. More geometrically, in the former case, interpreting $A^1=A^2$ as a line, $A^0$ \textit{can be} `in-\textit{line}' with it, i.e., we \text{can} have $A^0=A^1=A^2$, whereas in the latter case it cannot, meaning that it is `transverse'. Now suppose $A^1\neq A^2$; in this case, both $\hat{\mathcal{P}}^{(\Gamma_2)}_{A^0A^1A^2}\hat{\mathcal{P}}^{(\Gamma_1^{-})}_{A^1A^2}$ and $\hat{\mathcal{P}}^{(\Gamma_1^{-})}_{A^0A^1A^2}$ survive.  It is clear that in $\hat{\mathcal{P}}^{(\Gamma_1^{-})}_{A^0A^1A^2}$, the observable $A^0$ cannot coincide with neither $A^1$ nor $A^2$, whereas in $\hat{\mathcal{P}}^{(\Gamma_2)}_{A^0A^1A^2}\hat{\mathcal{P}}^{(\Gamma_1^{-})}_{A^1A^2}$ it can. Thus, we can once again define the latter as describing a `longitudinal' and the former a `transverse' response, although in a more generalized sense: now we interpret $\hat{\mathcal{P}}^{(\Gamma_1^{-})}_{A^1A^2}$ as a plane, and say that $A^0$ \textit{can be} `in-\textit{plane}' with $A^1,\,A^2$ in the longitudinal case, and \textit{cannot be} in the transverse case. In this way, we have a hierarchy of physically distinguishable longitudinal and transverse response pairs directly related to the irreducible representations of the permutation group $P(3)$.\\

While this discussion was rather abstract, there is an exemplary scenario in which we can relate it to a physical response and field configuration directly. Consider the direct current second order charge conductivity $\sigma_{i_0i_1i_2}(\omega,-\omega)$. In this case, we can replace the observables by the directional indices of the relevant charge current operators (see Eq. \eqref{eq:Cond2}) and the permutations act on $i_0, i_1, i_2\in\{x,y,z\}$. Let the driving field be a monochromatic light source with frequency $\omega$; then, the relevant second order driving field configuration is $E_{i_1}(\omega)E_{i_2}(-\omega)=|E_{i_1}(\omega)||E_{i_2}(\omega)|e^{i(\varphi_{i_1}-\varphi_{i_2})}$. From this, we have

\begin{equation}
\begin{split}
&\hat{\mathcal{P}}^{(\Gamma_1^{\pm})}_{i_1i_2}E_{i_1}(\omega)E_{i_2}(-\omega)
\\
&=\frac{1}{2}|E_{i_1}(\omega)||E_{i_2}(\omega)|(e^{i(\varphi_{i_1}-\varphi_{i_2})}\pm e^{-i(\varphi_{i_1}-\varphi_{i_2})}).
\end{split}
\end{equation}

Suppose $\varphi_{i_1}=\varphi_{i_2}$, i.e., the incident light is linearly polarized. We can then see that only $\hat{\mathcal{P}}^{(\Gamma_1^{+})}_{i_1i_2}$ survives and we can rotate our coordinate system in such a way that the polarization direction aligns with one of the axes, say $y$, i.e., we take $i_1=i_2=y$. Then, the `longitudinal' response given by the projection $\hat{\mathcal{P}}^{(\Gamma_1^{+})}_{i_0 yy}\sigma_{i_0yy}(\omega,-\omega)$ \textit{can be} `in-line' with the direction of polarization, whereas the transverse response $\hat{\mathcal{P}}^{(\Gamma_{2})}_{i_0yy}\sigma_{i_0yy}(\omega,-\omega)$ cannot. In other words, $\hat{\mathcal{P}}^{(\Gamma_1^{+})}_{yyy}\sigma_{yyy}(\omega,-\omega)=\sigma_{yyy}(\omega,-\omega)$, whereas $\hat{\mathcal{P}}^{(\Gamma_{2})}_{yyy}\sigma_{yyy}(\omega,-\omega)=0$. Now suppose $\varphi_{i_1}-\varphi_{i_2}=\frac{\pi}{2}$, i.e. the incident light is circularly polarized. In this case, only $\hat{\mathcal{P}}^{(\Gamma_1^{-})}_{i_1i_2}$ survives, and we can align our coordinate system such that the plane of polarization corresponds to, say, the $yz$ plane. Then, the longitudinal response given by $\hat{\mathcal{P}}^{(\Gamma_{2})}_{i_0 yz}\hat{\mathcal{P}}^{(\Gamma_1^{-})}_{yz}\sigma_{i_0yz}(\omega,-\omega)$ \textit{can be} in-plane, i.e., $\hat{\mathcal{P}}^{(\Gamma_{2})}_{yyz}\hat{\mathcal{P}}^{(\Gamma_1^{-})}_{yz}\sigma_{yyz}(\omega,-\omega)\neq 0$ and $\hat{\mathcal{P}}^{(\Gamma_{2})}_{zyz}\hat{\mathcal{P}}^{(\Gamma_1^{-})}_{yz}\sigma_{zyz}(\omega,-\omega)\neq 0$, whereas the transverse response $\hat{\mathcal{P}}^{(\Gamma_1^{-})}_{i_0 yz}\sigma_{i_0yz}(\omega,-\omega)$ can clearly not. This direct link between the polarization state of the incident electric field and the symmetry in $i_1$ and $i_2$ does not carry-over to other responses---except in the static $\omega=0$ case, which we can consider as the degenerate case of the linearly polarized response in which the electric field points in a fixed direction; the transverse response then corresponds to the second order static Hall response. The general decomposition discussed earlier can be looked at as an abstraction of this rather intuitive special case. As an example, we can also look at the second harmonic response $\sigma_{i_0i_1i_2}(\omega,\omega)$, in which case $E_{i_1}(\omega)E_{i_2}(\omega)=|E_{i_1}(\omega)||E_{i_2}(\omega)|e^{i(\varphi_{i_1}+\varphi_{i_2})}$. Clearly, only the symmetric part $\hat{\mathcal{P}}^{(\Gamma_1^{+})}_{i_1i_2}$ survives and we can define the longitudinal $\hat{\mathcal{P}}^{(\Gamma_1^{+})}_{i_0i_1i_2}\sigma_{i_0i_1i_2}(\omega,\omega)$ and transverse $\hat{\mathcal{P}}^{(\Gamma_{2})}_{i_0i_1i_2}\hat{\mathcal{P}}^{(\Gamma_1^{+})}_{i_1i_2}\sigma_{i_0i_1i_2}(\omega,\omega)$ responses, but these are not fundamentally restricted by the polarization state of the incident light.\\

In order to connect our fundamental group theoretical perspective to the more standard tensorial view, consider the rankn3 tensor $C^{\pm}_{i_0i_1i_2}=\varepsilon_{i_0i_1k}C_{ki_2}\pm\varepsilon_{i_0i_2k}C_{ki_1}$, where $C_{ik}$ is a rank 2 tensor, $\varepsilon_{ijk}$ is the totally antisymmetric Levi-Civita symbol, and summation over $k$ is implied. We can see that $C^{+}_{i_0i_1i_2}$ transforms in $\Gamma_2$ under permutations, since, as can be checked in a straightforward manner using the projections, $\hat{\mathcal{P}}^{(\Gamma_{2})}_{i_0i_1i_2}C^+_{i_0i_1i_2}=C^+_{i_0i_1i_2}$, and thereby provides an example of a transverse response of the first kind discussed above. On the other hand, $C^{-}_{i_0i_1i_2}$ does not, in general, transform in $\Gamma_2$, but also has a $\Gamma_1^-$ part and has to be separated accordingly. Our group theoretical analysis allows a direct and unique identification of different terms via projections without having to ponder over tensorial symmetry properties. Furthermore, it is clearly generalizable to tensors of arbitrary order since the representation theory of permutation groups is fully developed (see for example the standard textbook \cite{FultonHarris}).\\

\subsubsection{The 4-point retarded correlator}
\label{PermDecomp4Point}

The true utility of our decomposition starts revealing itself with the retarded 4-point correlator $C^r_{A^0A^1A^2A^3}(t_0,t_1,t_2,t_3)$. We have $P^{A^1A^2A^3}(3)\times P^{t_1t_2t_3}(3)$ with the relevant representations $\Gamma_1^+\otimes\Gamma_1^+,\,\Gamma_1^-\otimes\Gamma_1^-,\,\Gamma_2\otimes\Gamma_2$ and $P^{A^0A^1A^2A^3}(4)$ with 5 irreducible representations $\Gamma_1^+,\Gamma_1^-,\Gamma_2,\Gamma_3,\Gamma_{3'}$. Consequently, \eqref{eq:NPermDecomp} becomes

\begin{widetext}

\begin{equation}
\label{eq:4PermDecomp}
\begin{split}
C^r_{A^0A^1A^2A^3}(t_0,t_1,t_2,t_3)=&\left(\hat{\mathcal{P}}^{(\Gamma_1^+)}_{A^0A^1A^2A^3}+
\hat{\mathcal{P}}^{(\Gamma_1^-)}_{A^0A^1A^2A^3}+
\hat{\mathcal{P}}^{(\Gamma_2)}_{A^0A^1A^2A^3}+\hat{\mathcal{P}}^{(\Gamma_3)}_{A^0A^1A^2A^3}+\hat{\mathcal{P}}^{(\Gamma_{3'})}_{A^0A^1A^2A^3}\right)\times
\\
&\times\left(\hat{\mathcal{P}}^{(\Gamma_1^+)}_{A^1A^2A^3}\hat{\mathcal{P}}^{(\Gamma_1^+)}_{t_1t_2t_3}+\hat{\mathcal{P}}^{(\Gamma_1^-)}_{A^1A^2A^3}\hat{\mathcal{P}}^{(\Gamma_1^-)}_{t_1t_2t_3}+\hat{\mathcal{P}}^{(\Gamma_2)}_{A^1A^2A^3}
\hat{\mathcal{P}}^{(\Gamma_2)}_{t_1t_2t_3}\right)C^r_{A^0A^1A^2A^3}(t_0,t_1,t_2,t_3)
\\
=&\bigg(\hat{\mathcal{P}}^{(\Gamma_1^+)}_{A^0A^1A^2A^3}\hat{\mathcal{P}}^{(\Gamma_1^+)}_{A^1A^2A^3}\hat{\mathcal{P}}^{(\Gamma_1^+)}_{t_1t_2t_3}
+\hat{\mathcal{P}}^{(\Gamma_3)}_{A^0A^1A^2A^3}\hat{\mathcal{P}}^{(\Gamma_1^+)}_{A^1A^2A^3}\hat{\mathcal{P}}^{(\Gamma_1^+)}_{t_1t_2t_3}
\\
&\quad +
\hat{\mathcal{P}}^{(\Gamma_{3})}_{A^0A^1A^2A^3}\hat{\mathcal{P}}^{(\Gamma_2)}_{A^1A^2A^3}\hat{\mathcal{P}}^{(\Gamma_2)}_{t_1t_2t_3}
+\hat{\mathcal{P}}^{(\Gamma_2)}_{A^0A^1A^2A^3}\hat{\mathcal{P}}^{(\Gamma_2)}_{A^1A^2A^3}\hat{\mathcal{P}}^{(\Gamma_2)}_{t_1t_2t_3}
+\hat{\mathcal{P}}^{(\Gamma_{3'})}_{A^0A^1A^2A^3}\hat{\mathcal{P}}^{(\Gamma_2)}_{A^1A^2A^3}\hat{\mathcal{P}}^{(\Gamma_2)}_{t_1t_2t_3}
\\
&\quad
+\hat{\mathcal{P}}^{(\Gamma_{3'})}_{A^0A^1A^2A^3}\hat{\mathcal{P}}^{(\Gamma_1^-)}_{A^1A^2A^3}\hat{\mathcal{P}}^{(\Gamma_1^-)}_{t_1t_2t_3}
+\hat{\mathcal{P}}^{(\Gamma_1^-)}_{A^0A^1A^2A^3}\hat{\mathcal{P}}^{(\Gamma_1^-)}_{A^1A^2A^3}\hat{\mathcal{P}}^{(\Gamma_1^-)}_{t_1t_2t_3}
\bigg)C^r_{A^0A^1A^2A^3}(t_0,t_1,t_2,t_3),
\end{split}
\end{equation}

where, in the second equality, we used $\hat{\mathcal{P}}^{(\Gamma_1^{\pm})}_{A^0A^1A^2A^3}\hat{\mathcal{P}}^{(\Gamma_1^{\mp})}_{A^1A^2A^3}=\hat{\mathcal{P}}^{(\Gamma_1^{\pm})}_{A^0A^1A^2A^3}\hat{\mathcal{P}}^{(\Gamma_2)}_{A^1A^2A^3}=\hat{\mathcal{P}}^{(\Gamma_2)}_{A^0A^1A^2A^3}\hat{\mathcal{P}}^{(\Gamma_1^{\pm})}_{A^1A^2A^3}=0$ and
\end{widetext}
 $\hat{\mathcal{P}}^{(\Gamma_3)}_{A^0A^1A^2A^3}\hat{\mathcal{P}}^{(\Gamma_1^{-})}_{A^1A^2A^3}=\hat{\mathcal{P}}^{(\Gamma_{3'})}_{A^0A^1A^2A^3}\hat{\mathcal{P}}^{(\Gamma_1^{+})}_{A^1A^2A^3}=0$, that can all be proved in a straightforward way by direct application of the projections.\\
 
Once again, we can make use of time-translation invariance and arrive at the same decomposition for the frequency-domain retarded correlator $\tilde{C}^r_{A^0A^1A^2A^3}(\omega_1,\omega_2,\omega_3)$ by replacing the temporal permutations with frequency permutations and looking at the corresponding $\hat{\mathcal{P}}^{(\dots)}_{\omega_1\omega_2\omega_3}$ projections.\\

The interpretation of the terms is more subtle than it was for the 3-point correlator. As before, we can define a hierarchy of `longitudinal' and `transverse' responses, however not for all the terms. Indeed, suppose we take $A^1=A^2=A^3$; then, the only surviving terms are $\hat{\mathcal{P}}^{(\Gamma_1^+)}_{A^0A^1A^1A^1}$ and $\hat{\mathcal{P}}^{(\Gamma_3)}_{A^0A^1A^1A^1}$. If we also take $A^0=A^1$, then only $\hat{\mathcal{P}}^{(\Gamma_1^+)}_{A^1A^1A^1A^1}$ continues to not vanish, and we can look at $\hat{\mathcal{P}}^{(\Gamma_1^+)}_{A^0A^1A^2A^3}$ and $\hat{\mathcal{P}}^{(\Gamma_3)}_{A^0A^1A^2A^3}\hat{\mathcal{P}}^{(\Gamma_1^+)}_{A^1A^2A^3}$ as forming a `longitudinal-transverse' response pair. Note that this pair of responses can survive even in the static limit and we can interpret the transverse part as giving the third order Hall response. We can define another pair by supposing that $A^1\neq A^2\neq A^3$, in which case both $\hat{\mathcal{P}}^{(\Gamma_{3'})}_{A^0A^1A^2A^3}\hat{\mathcal{P}}^{(\Gamma_1^-)}_{A^1A^2A^3}$ and $\hat{\mathcal{P}}^{(\Gamma_1^-)}_{A^0A^1A^2A^3}$ \textit{can} survive. If we allow $A^0$ to be either one of $A^1,\,A^2$ or $A^3$, then only the former projection survives and we can think of it as a `longitudinal' response with respect to the latter. Geometrically, interpreting the totally antisymmetric combination $\hat{\mathcal{P}}^{(\Gamma_1^-)}_{A^1A^2A^3}$ as a `volume', the longitudinal response $\hat{\mathcal{P}}^{(\Gamma_{3'})}_{A^0A^1A^2A^3}\hat{\mathcal{P}}^{(\Gamma_1^-)}_{A^1A^2A^3}$ corresponds to the one that \textit{can be} `in-volume'. What about the three other terms? For these, we cannot define such pairs in the same unambiguous manner. Indeed, the projection $\hat{\mathcal{P}}^{(\Gamma_2)}_{A^1A^2A^3}$ requires at least one of $A^1,\,A^2,\,A^3$ to differ from the other two---for example, $A^1,A^1,A^2$. With this particular example, the projections then survive as: $A^0=A^1$ with $\hat{\mathcal{P}}^{(\Gamma_3)}_{A^1A^1A^1A^2}\hat{\mathcal{P}}^{(\Gamma_2)}_{A^1A^1A^2}$---we \textit{can} have three of a kind; $A^0=A^2$ with $\hat{\mathcal{P}}^{(\Gamma_2)}_{A^2A^1A^1A^2}\hat{\mathcal{P}}^{(\Gamma_2)}_{A^1A^1A^2}$---we \text{can} have two each; and $A^0\neq A^1\neq A^2$ with $\hat{\mathcal{P}}^{(\Gamma_{3'})}_{A^0A^1A^1A^2}\hat{\mathcal{P}}^{(\Gamma_2)}_{A^1A^1A^2}$---we \textit{can} have exactly two identical. Thus, in contrast to the other terms, we cannot provide a geometric interpretation since $\Gamma_2$ is a 2-dimensional representation and can be looked at as a mix of lines and planes. The physics does not provide us with a natural basis for it, and thereby a meaningful sense of `longitudinal' and `transverse'. Note that it is the intrinsic permutation symmetry that gives us the physically meaningful partners (bases) for the representations and for each of the 3-dimensional $\Gamma_3$ and $\Gamma_{3'}$ representations only one of the partners---$\hat{\mathcal{P}}^{(\Gamma_3)}_{A^0A^1A^2A^3}\hat{\mathcal{P}}^{(\Gamma_1^+)}_{A^1A^2A^3}$ and $\hat{\mathcal{P}}^{(\Gamma_{3'})}_{A^0A^1A^2A^3}\hat{\mathcal{P}}^{(\Gamma_1^-)}_{A^1A^2A^3}$---is provided explicitly, with the other two remaining redundancies of the description.\\

In order to somewhat ground the notions discussed, we give an explicit example and show that, in this particular case, it is possible to distinguish between some of the different terms using specific field configurations. Consider the self-focusing third order charge conductivity response $\sigma_{i_0i_1i_2i_3}(\omega,-\omega,\omega)$. We can then replace the observables by the directional indices of the relevant charge current operators and the permutations act on $i_0, i_1, i_2, i_3\in\{x,y,z\}$. Let the driving field be a monochromatic electric field with frequency $\omega$; then, the relevant third order driving field configuration is $E_{i_1}(\omega)E_{i_2}(-\omega)E_{i_3}(\omega)=|E_{i_1}(\omega)||E_{i_2}(\omega)||E_{i_3}(\omega)|e^{i(\varphi_{i_1}-\varphi_{i_2}+\varphi_{i_3})}$. It is clear that the totally antisymmetric combination $\hat{\mathcal{P}}^{(\Gamma_1^-)}_{i_1i_2i_3}$ does not contribute since the driving field configuration is symmetric in $i_1$ and $i_3$. Consequently, we are left with

\begin{widetext}
\begin{align}
\hat{\mathcal{P}}^{(\Gamma_1^{+})}_{i_1i_2i_3}&E_{i_1}(\omega)E_{i_2}(-\omega)E_{i_3}(\omega)
\nonumber
\\
&=\frac{1}{3}|E_{i_1}(\omega)||E_{i_2}(\omega)||E_{i_3}(\omega)|(e^{i(\varphi_{i_1}-\varphi_{i_2}+\varphi_{i_3})}+e^{i(-\varphi_{i_1}+\varphi_{i_2}+\varphi_{i_3})}+e^{i(\varphi_{i_1}+\varphi_{i_2}-\varphi_{i_3})}),
\\
\hat{\mathcal{P}}^{(\Gamma_2)}_{i_1i_2i_3}&E_{i_1}(\omega)E_{i_2}(-\omega)E_{i_3}(\omega)
\nonumber
\\
&=\frac{1}{3}|E_{i_1}(\omega)||E_{i_2}(\omega)||E_{i_3}(\omega)|(2e^{i(\varphi_{i_1}-\varphi_{i_2}+\varphi_{i_3})}-e^{i(-\varphi_{i_1}+\varphi_{i_2}+\varphi_{i_3})}-e^{i(\varphi_{i_1}+\varphi_{i_2}-\varphi_{i_3})}).
\end{align}
\end{widetext}

Suppose that all components of the driving field are in-phase; $\varphi_{i_1}=\varphi_{i_2}=\varphi_{i_3}$---this could be, e.g., a single beam of linearly polarized light. Then, we can see that only $\hat{\mathcal{P}}^{(\Gamma_1^{+})}_{i_1i_2i_3}$ survives and we are able to set our coordinate system such that one of the axes, say $y$, is parallel to the effective direction of the driving field. We thus take $i_1=i_2=i_3=y$ meaning that the corresponding `longitudinal' response given by the projection $\hat{\mathcal{P}}^{(\Gamma_1^{+})}_{i_0 yyy}\sigma_{i_0yyy}(\omega,-\omega,\omega)$ \textit{can be} `in-line' with the direction of polarization, whereas the transverse response $\hat{\mathcal{P}}^{(\Gamma_{3})}_{i_0yyy}\sigma_{i_0yyy}(\omega,-\omega,\omega)$ cannot. In other words, $\hat{\mathcal{P}}^{(\Gamma_1^{+})}_{yyyy}\sigma_{yyyy}(\omega,-\omega,\omega)=\sigma_{yyyy}(\omega,-\omega,\omega)$, whereas $\hat{\mathcal{P}}^{(\Gamma_{3})}_{yyyy}\sigma_{yyyy}(\omega,-\omega,\omega)=0$. Can we find an electric field configuration that eliminates $\hat{\mathcal{P}}^{(\Gamma_1^{+})}_{i_1i_2i_3}$ but keeps $\hat{\mathcal{P}}^{(\Gamma_2)}_{i_1i_2i_3}$? Indeed we can, by choosing the relative phases as, for example, $\varphi_{i_2}-\varphi_{i_1}=\frac{\pi}{3},\,\varphi_{i_1}-\varphi_{i_3}=\frac{\pi}{3}$ and $\varphi_{i_2}-\varphi_{i_3}=\frac{2\pi}{3}$, meaning that, upto a phase, the driving field can be written as $\textbf{E}(\omega)\propto (|E_x(\omega)|e^{i\frac{\pi}{3}},|E_y(\omega)|e^{i\frac{2\pi}{3}},|E_z(\omega)|)^T$. In other words, we need at least two light beams; one of them elliptically polarized, and the other, linearly polarized and orthogonal to the polarization plane of the former with an appropriately shifted phase. We could, of course, also take three mutually orthogonal linearly polarized light beams with the given phase shifts. By construction, this particular field configuration restricts the responses to $\hat{\mathcal{P}}^{(\Gamma_{3})}_{i_0xyz}\hat{\mathcal{P}}^{(\Gamma_2)}_{xyz}$, 
$\hat{\mathcal{P}}^{(\Gamma_2)}_{i_0xyz}\hat{\mathcal{P}}^{(\Gamma_2)}_{xyz}$, and $
\hat{\mathcal{P}}^{(\Gamma_{3'})}_{i_0xyz}\hat{\mathcal{P}}^{(\Gamma_2)}_{xyz}$.\\

\section{$GL(n,\mathbb{R})$ structure of an array of retarded correlators}
\label{GLstruct}

Closely related to permutations, is another structure of crucial importance that correlators carry. To see this, let us look at a correlator as a scalar-valued function of operators, $C(\mathcal{O})=\text{tr}(\mathcal{O})$. By the linearity of the trace, for scalars $a_1,a_2$, we have  $C(a_1\mathcal{O}_1+a_2\mathcal{O}_2)=a_1C(\mathcal{O}_1)+a_2C(\mathcal{O}_2)$. Similarly, we can define correlators as scalar-valued functions with multiple operator arguments, $C(A^0,\dots,A^n)=\text{tr}(A^0\dots A^n)$, that are, in-turn, linear in each argument, i.e., $C(A^0,\dots,a_{i1}A^i_1+a_{i2}A^i_2,\dots,A^n)=a_{i_1}C(A^0,\dots,A^i_1,\dots,A^n)+a_{i2}C(A^0,\dots,A^i_2,\dots,A^n)$ for all $i$. This is important, because it means that correlators carry finite-dimensional linear representations of the general linear group, and thereby, its subgroups. This is what we shall elaborate on now and discuss the intimate connection to permutations, and thereby the different classes of responses arrived at via the permutation decomposition introduced in Section \ref{PermDecomp}.\\

\subsection{Tensor product representations and retarded correlators}
\label{RetTensor}

Let us look at the $n+1$-point retarded correlators of a collection of $m+1$ observables, $\tilde{C}^r_{A^{i_0}A^{i_1}\dots A^{i_n}}(t_1,t_2,\dots,t_n)$, where $i_0,\dots,i_n\in\{0,\dots,m\}$, and think of them as \textit{real}-valued functions with $n+1$ operator arguments,

\begin{equation}
\label{eq:RetOp}
\begin{split}
&\tilde{C}^r_{t_1\dots t_n}(A^{i_0},A^{i_1},\dots,A^{i_n})
\\
&\qquad\qquad\qquad\equiv\tilde{C}^r_{A^{i_0}A^{i_1}\dots A^{i_n}}(t_1,t_2,\dots,t_n).
\end{split}
\end{equation}

Now consider an element $g_{m+1}$ of the general linear group $GL(m+1,\mathbb{R})$ of real, invertible $(m+1)\times(m+1)$ matrices. We then have a natural action on observables as, for example,

\begin{equation}
\begin{split}
A^i\to (g_{m+1})_{i0}A^0+(g_{m+1})_{i1}A^1+\dots+(g_{m+1})_{im}A^m,
\end{split}
\end{equation}

and the retarded correlator; when looked at as a \textit{real}-valued function with $n+1$ operator arguments as in \eqref{eq:RetOp} above; is linear under such a change of each of its arguments. Note that the reason we restricted to \textit{real}-valued matrices is because the operators we are defining the action on are observables, hence Hermitian, and we want this action to map observables to other observables.\\

The action on observables, in-turn, defines a representation of $GL(m+1,\mathbb{R})$ on $n+1$-point retarded correlators as follows. Define the objects

\begin{equation}
\begin{split}
&P^{m+1,n+1}_{t_1\dots t_n}(A^0,A^1,\dots,A^m)_{i_0i_1\dots i_n}
\\
&\qquad\qquad=\tilde{C}^r_{t_1\dots t_n}(A^{i_0},A^{i_1},\dots,A^{i_n})
\\
&\qquad\qquad\equiv\tilde{C}^r_{A^{i_0}A^{i_1}\dots A^{i_n}}(t_1,t_2,\dots,t_n),
\end{split}
\end{equation}

which can be considered as real numbers in an $\underbrace{(m+1)\times\dots\times(m+1)}_{n+1}$ array. We then have a linear $GL(m+1,\mathbb{R})$ representation on these objects

\begin{widetext}
\begin{equation}
\label{eq:GLmCorr}
\begin{split}
&\left\{\left(\pi(g_{m+1})P^{m+1,n+1}_{t_1\dots t_n}\right)(A^0,A^1,\dots,A^m)
\right\}_{i_0i_1\dots i_n}
\\
&\qquad\qquad=P^{m+1,n+1}_{t_1\dots t_n}\bigg(\sum_{j_0=0}^{m}(g^{-1}_{m+1})_{0j_0}A^{j_0},\sum_{j_1=0}^{m}(g^{-1}_{m+1})_{1j_1}A^{j_1},\dots,\sum_{j_m=0}^{m}(g^{-1}_{m+1})_{mj_m}A^{j_m}\bigg)_{i_0i_1\dots i_n}
\\
&\qquad\qquad=\tilde{C}^r_{t_1\dots t_n}\bigg(\sum_{j_0=0}^{m}(g^{-1}_{m+1})_{i_0j_0}A^{j_0},\sum_{j_1=0}^{m}(g^{-1}_{m+1})_{i_1j_1}A^{j_1},\dots,\sum_{j_n=0}^{m}(g^{-1}_{m+1})_{i_nj_n}A^{j_n}\bigg)
\\
&\qquad\qquad=\sum_{j_0=0}^m\sum_{j_1=0}^m\dots\sum_{j_n=0}^m(g^{-1}_{m+1})_{i_0j_0}(g^{-1}_{m+1})_{i_1j_1}\dots(g^{-1}_{m+1})_{i_nj_n}
P^{m+1,n+1}_{t_1\dots t_n}(A^0,A^1,\dots,A^m)_{j_0j_1\dots j_n},
\end{split}
\end{equation}
\end{widetext}

where we used the linearity in each argument of the retarded correlator to get the final equality. Consequently, we can consider these objects as components of rank-$(n+1)$ tensors. Thus, there is a natural correspondence between elements of an $(n+1)$-fold tensor product $T_{n+1}\mathsf{V}^{m+1}=\mathsf{V}^{m+1}\otimes\dots\otimes\mathsf{V}^{m+1}$ of an $(m+1)$-dimensional vector space $\mathsf{V}^{m+1}$, and $n+1$-point retarded correlators of $m+1$ observables. Note that, in general, the fact that retarded correlators, and, by extension, the response coefficients they determine, can be looked at as components of (most commonly rotation group) tensors is trivially assumed, but our perspective will be rather beneficial when looking at the connection with permutations.\\
Let us now look at two examples. First, we take $m=n=1$, meaning that our objects become elements of a $2\times 2$ matrix

\begin{equation}
\label{eq:P22Matrix}
P^{2,2}_{t_1}(A^0,A^1)=
\begin{pmatrix}
\tilde{C}^r_{A^{0}A^{0}}(t_1)&\tilde{C}^r_{A^{0}A^{1}}(t_1)\\
\tilde{C}^r_{A^{1}A^{0}}(t_1)&\tilde{C}^r_{A^{1}A^{1}}(t_1)
\end{pmatrix},
\end{equation}

and the $GL(2,\mathbb{R})$ representation can be written as

\begin{equation}
\label{eq:P22GL2}
\begin{split}
&(\pi(g_2)\,P^{2,2}_{t_1})(A^0,A^1)
\\
&=P^{2,2}_{t_1}\bigg((g^{-1}_2)_{00}A^0+(g^{-1}_2)_{01}A^1,(g^{-1}_2)_{10}A^0+(g^{-1}_2)_{11}A^1\bigg)
\\
&=g^{-1}_2P^{2,2}_{t_1}(A^0,A^1)(g_2^{-1})^T.
\end{split}
\end{equation}

Now we take $m=2$ and $n=1$, and our objects can be seen as elements of a $3\times 3$ matrix

\begin{equation}
\label{eq:P32Matrix}
P^{3,2}_{t_1}(A^0,A^1,A^2)=
\begin{pmatrix}
\tilde{C}^r_{A^{0}A^{0}}(t_1)&\tilde{C}^r_{A^{0}A^{1}}(t_1)&\tilde{C}^r_{A^{0}A^{2}}(t_1)\\
\tilde{C}^r_{A^{1}A^{0}}(t_1)&\tilde{C}^r_{A^{1}A^{1}}(t_1)&\tilde{C}^r_{A^{1}A^{2}}(t_1)\\
\tilde{C}^r_{A^{2}A^{0}}(t_1)&\tilde{C}^r_{A^{2}A^{1}}(t_1)&\tilde{C}^r_{A^{2}A^{2}}(t_1)\\
\end{pmatrix},
\end{equation}

with the $GL(3,\mathbb{R})$ representation being

\begin{equation}
\label{eq:P32GL3}
\begin{split}
&(\pi(g_3)\,P^{3,2}_{t_1})(A^0,A^1,A^2)
\\
&\qquad\qquad=g^{-1}_3P^{3,2}_{t_1}(A^0,A^1,A^2)(g_3^{-1})^T.
\end{split}
\end{equation}

We thus see that retarded correlators can be arranged into arrays that transform in tensor product representations of the general linear group.\\

\subsection{Schur-Weyl duality}

Tensor product representations are \textit{reducible} and it would be conducive to decompose them into direct sums of representations that are \textit{irreducible}. While such a decomposition is standard textbook material, we want to emphasize the fundamental interplay with the irreducible representations of permutation groups, which, while ubiquitously discussed in the mathematics literature (see for example the introductory textbook \cite{FultonHarris}), and present in the general physics literature \cite{TungGroups,SternbergPhysics}, is practically absent in an explicit form from the quantum transport and non-linear optics literature.\\

The fundamental observation is that the natural action of the general linear group on a tensor product commutes with the natural action of permutations. Indeed, let $\mathsf{V}^{m+1}$ be an $m+1$-dimensional vector space and $GL(\mathsf{V}^{m+1})$ be the general linear group on $\mathsf{V}^{m+1}$. Suppose $\textbf{v}_0\otimes\textbf{v}_1\otimes\dots\otimes\textbf{v}_n\in T_{n+1}\mathsf{V}^{m+1}$, where $T_{n+1}\mathsf{V}^{m+1}=\mathsf{V}^{m+1}\otimes\dots\otimes\mathsf{V}^{m+1}$ is the $n+1$-fold tensor product of $\mathsf{V}^{m+1}$ and $\textbf{v}_{i}\in\mathsf{V}^{m+1}$ for all $i\in\{0,\dots,m\}$. On the one hand, we have for $g_{m+1}\in GL(\mathsf{V}^{m+1})$ the natural action via $T_{n+1}g_{m+1}\equiv g_{m+1}\otimes\dots\otimes g_{m+1}$ as

\begin{equation}
\begin{split}
&(T_{n+1}g_{m+1})(\textbf{v}_0\otimes\textbf{v}_1\otimes\dots\otimes\textbf{v}_n)
\\
&=(g_{m+1}\textbf{v}_0)\otimes (g_{m+1}\textbf{v}_1)\otimes\dots\otimes (g_{m+1}\textbf{v}_n).
\end{split}
\end{equation}

On the other hand, we can also permute the vectors in the tensor product, meaning that we have a permutation action $\Pi\in P(n+1)$ as

\begin{equation}
\Pi(\textbf{v}_0\otimes\textbf{v}_1\otimes\dots\otimes\textbf{v}_n)=\textbf{v}_{\Pi^{-1}(0)}\otimes\textbf{v}_{\Pi^{-1}(1)}\otimes\dots\otimes\textbf{v}_{\Pi^{-1}(n)}.
\end{equation}

It is straightforward to see that

\begin{equation}
\begin{split}
&\Pi(T_{n+1}g_{m+1})(\textbf{v}_0\otimes\textbf{v}_1\otimes\dots\otimes\textbf{v}_n)
\\
&=(T_{n+1}g_{m+1})\Pi(\textbf{v}_0\otimes\textbf{v}_1\otimes\dots\otimes\textbf{v}_n),
\end{split}
\end{equation}

i.e., the two actions commute. This is important, because it means that we can find a basis of the tensor product space in which it decomposes into spaces that are invariant under $GL(\mathsf{V}^{m+1})$ and $P(m+1)$ \textit{simultaneously}. In fact, there is a one-to-one correspondence between the irreducible representations of the two groups occuring in the direct sum decomposition and they uniquely determine each other. This deep and fundamental connection between the representations of the two groups is generally known as Schur-Weyl duality and we refer to \cite{FultonHarris,TungGroups,SternbergPhysics} for detailed proofs. Here we simply state the general result. Suppose we have the decomposition $T_{n+1}\mathsf{V}^{m+1}=\mathsf{W}_0\oplus\mathsf{W}_1\oplus\dots\oplus\mathsf{W}_r$, where the $\mathsf{W}_i$, $i\in\{0,\dots,r\}$, transform under a $\Gamma_i$ irreducible representation of $P(n+1)$. If the dimension of this representation is $d_i$, then all $\mathsf{W}_i$ can be further decomposed into $d_i$ copies of spaces $\mathsf{U}_i$ that are in turn irreducible representations of $GL(\mathsf{V}^{m+1})$, i.e., we have

\begin{equation}
\label{eq:SchurWeyDecomp}
\begin{split}
T_{n+1}\mathsf{V}^{m+1}&=\underbrace{\mathsf{V}^{m+1}\otimes\dots\otimes\mathsf{V}^{m+1}}_{n+1}
\\
&=\underbrace{\mathsf{U}_0\oplus\dots\oplus \mathsf{U}_0}_{d_0}\oplus\underbrace{\mathsf{U}_1\oplus\dots\oplus \mathsf{U}_1}_{d_1}\oplus\dots
\\
&\quad\oplus\underbrace{\mathsf{U}_r\oplus\dots\oplus \mathsf{U}_r}_{d_r}.
\end{split}
\end{equation}

We show two examples of the duality, as applied to retarded correlators, explicitly. First, we look at $P^{2,2}_{t_1}(A^0,A^1)$, the array of 2-point retarded correlators of two observables. The tensor product representation $T_{2}g_{2}$ corresponds to the $\pi(g_{2})$ representation on $P^{2,2}_{t_1}(A^0,A^1)$ (see \eqref{eq:GLmCorr} with $n=m=1$). The relevant permutation group is $P(2)$ which has two 1-dimensional irreducible representations, the symmetric $\Gamma_1^+$ and antisymmetric $\Gamma_1^-$. Let us decompose our tensor into these as

\begin{equation}
P^{2,2;\Gamma_1^{\pm}}_{t_1}(A^0,A^1)=\frac{P^{2,2}_{t_1}(A^0,A^1)\pm P^{2,2}_{t_1}(A^1,A^0)}{2}.
\end{equation}

Looking first at the antisymmetric part, by virtue of \eqref{eq:P22GL2} we have 

\begin{equation}
\left(\pi(g_2)P^{2,2;\Gamma_1^{-}}_{t_1}\right)(A^0,A^1)=\frac{1}{\text{det}(g_2)}P^{2,2;\Gamma_1^{-}}_{t_1}(A^0,A^1),
\end{equation}

meaning that $P^{2,2;\Gamma_1^{-}}_{t_1}(A^0,A^1)$ transforms under a 1-dimensional irreducible representation of $GL(2,\mathbb{R})$. For the symmetric part, \eqref{eq:P22GL2} shows that $\left(\left(\pi(g_2)P^{2,2;\Gamma_1^{+}}_{t_1}\right)(A^0,A^1)\right)^T=\left(\pi(g_2)P^{2,2;\Gamma_1^{+}}_{t_1}\right)(A^0,A^1)$, in other words, as expected, the $GL(2,\mathbb{R})$ action preserves the symmetry. Since a symmetric $2\times 2$ matrix has 3 independent components, $P^{2,2;\Gamma_1^{-}}_{t_1}(A^0,A^1)$ transforms under a 3-dimensional irreducible representation of $GL(2,\mathbb{R})$. We can obtain the explicit $3\times 3$ representation matrix easily by defining the 3 component vector $(\tilde{C}^r_{A^{0}A^{0}}(t_1),\tilde{C}^{r,\Gamma_1^+}_{A^{0}A^{1}}(t_1),\tilde{C}^r_{A^{1}A^{1}}(t_1))^T$ from the elements of the symmetrized form of the $2\times 2$ matrix \eqref{eq:P22Matrix}, and realizing that the transformation rule \eqref{eq:P22GL2} corresponds to

\begin{widetext}
\begin{equation}
\begin{split}
&\begin{pmatrix}\tilde{C}^r_{A^{0}A^{0}}(t_1)\\
\tilde{C}^{r,\Gamma_1^+}_{A^{0}A^{1}}(t_1)\\
\tilde{C}^r_{A^{1}A^{1}}(t_1)
\end{pmatrix}
\to \frac{1}{\text{det}^2(g_2)}
\begin{pmatrix}
g^2_{11}&-2g_{11}g_{01}&g^2_{01}\\
-g_{11}g_{10}&g_{00}g_{11}+g_{01}g_{10}&-g_{01}g_{00}\\
g^2_{10}&-2g_{10}g_{00}&g^2_{00}
\end{pmatrix}
\begin{pmatrix}\tilde{C}^r_{A^{0}A^{0}}(t_1)\\
\tilde{C}^{r,\Gamma_1^+}_{A^{0}A^{1}}(t_1)\\
\tilde{C}^r_{A^{1}A^{1}}(t_1)
\end{pmatrix}.
\end{split}
\end{equation}
\end{widetext}

Note that for $\text{det}(g_2)=1$, this is nothing but the adjoint representation of $SL(2,\mathbb{R})$ on its own Lie algebra $\mathfrak{sl}(2,\mathbb{R})$, and the 3-component vector of correlators can be looked at as components of an element of $\mathfrak{sl}(2,\mathbb{R})$ in a Cartan-Weyl basis \cite{FuchsLie} of the latter . In this simplest example of Schur-Weyl duality, we thus see how a rank-2 tensor is decomposed into a direct sum of two spaces that are each irreducible representations of both $GL(2,\mathbb{R})$ and $P(2)$. Similar arguments can be made for the case of $P^{3,2}_{t_1}(A^0,A^1,A^2)$, the array of 2-point retarded correlators of three observables. Glancing at its explicit $3\times 3$ matrix form, we can define its symmetric and antisymmetric parts as the corresponding symmetric and antisymmetric $3\times 3$ matrices. A symmetric $3\times 3$ matrix has 6 independent components, whereas an antisymmetric one has 3, hence we have 6 and 3-dimensional irreducible representations of $GL(3,\mathbb{R})$ dual to, respectively, 1-dimensional symmetric and antisymmetric irreducible representations of $P(2)$. Similarly, we can consider an array of 2-point retarded correlators of $m+1$ observables and find a duality between relevant irreducible representations of $GL(m+1,\mathbb{R})$ and $P(2)$.\\

In all the cases discussed so far, however, the permutation group $P(2)$ had only 1-dimensional irreducible representations. We now briefly show the simplest example of the decomposition for the $P(3)$ case in which we also have a 2-dimensional one, and hence, multiple copies of the same irreducible representation of the relevant general linear group in the decomposition (cf. \eqref{eq:SchurWeyDecomp}). Consider the $3\times 3\times 3$ array $P^{3,3}_{t_1}(A^0,A^1,A^2)$ of 3-point retarded correlators of three observables, meaning that the relevant general linear group is $GL(3,\mathbb{R})$. The array has 27 elements $\tilde{C}^r_{A^{i_0}A^{i_1}A^{i_2}}(t_1,t_2)$, where $i_0,i_1,i_2\in\{0,1,2\}$, and, for the purposes of this demonstration, we label the elements as $(i_0i_1i_2)$. We project these elements onto the three irreducible representations $\Gamma_1^{\pm},\,\Gamma_2$ of $P(3)$ and write $(i_0i_1i_2)^{\Gamma_1^{\pm}}, (i_0i_1i_2)^{\Gamma_2}$. The totally symmetric irreducible representation $\Gamma_1^+$ yields 10 independent terms
\begin{equation}
\begin{split}
&((000),(001)^{\Gamma_1^+},(011)^{\Gamma_1^+},(111),
\\
&\qquad(002)^{\Gamma_1^+},(012)^{\Gamma_1^+},(112)^{\Gamma_1^+},
\\
&\qquad\qquad(022)^{\Gamma_1^+},(122)^{\Gamma_1^+},
\\
&\qquad\qquad\qquad(222)),
\end{split}
\end{equation}

which transform according to a 10-dimensional irreducible representation of $GL(3,\mathbb{R})$. For the totally antisymmetric $\Gamma_1^-$, all three observables must differ, hence we only have a single independent term $(012)^{\Gamma_1^-}$ which transforms according to a 1-dimensional irreducible representation of $GL(3,\mathbb{R})$. We are left with the most involved $\Gamma_2$ case. From the explicit expression \eqref{eq:S3Gamma2Proj} of the projector, we can notice that at least two of the three numbers in $(i_0i_1i_2)$ must differ for the result to be non-vanishing, furthermore, it is clear that $(i_0i_1i_2)^{\Gamma_2}+(i_2i_0i_1)^{\Gamma_2}+(i_1i_2i_0)^{\Gamma_2}=0$. If all three numbers in $(i_0i_1i_2)$ differ, i.e., $(012),(120)$ etc., we can have six possible terms, out of which only four are independent when projected to $\Gamma_2$, since we have the two cycle relations $(012)^{\Gamma_2}+(201)^{\Gamma_2}+(120)^{\Gamma_2}=0$ and $(021)^{\Gamma_2}+(102)^{\Gamma_2}+(210)^{\Gamma_2}=0$. If only two out of the three numbers in $(i_0i_1i_2)$ differ, i.e., $(001),(211)$ etc., we can have 18 terms, but only 12 of them are independent, since we have the cycle relations $(i_0i_0i_1)^{\Gamma_2}+(i_0i_1i_0)^{\Gamma_2}+(i_1i_0i_0)^{\Gamma_2}=0$ that lead to 6 independent conditions. Thus, in total we have $4+12=16$ independent terms which correspond to two copies of an 8-dimensional irreducible representation of $GL(3,\mathbb{R})$. We can show a realization of the two copies explicitly by choosing a basis for $\Gamma_2$. We choose the natural one dictated by the consideration that retarded correlators have intrinsic permutation symmetry, just as we did for the permutation decomposition in Section \ref{PermDecomp}. The projectors are then given by \eqref{eq:Gamma2Proj}, and we have terms $(i_0i_1i_2)^{\Gamma_2^{\pm}}\equiv\hat{\mathcal{P}}^{\Gamma_{1}^{\pm}}_{i_1i_2}(i_0i_1i_2)^{\Gamma_2}$ that are symmetric or antisymmetric in the last two numbers. We can then group the terms as two 8-component vectors

\begin{equation}
\begin{split}
&(\qquad\qquad(100)^{\Gamma_2^+},\qquad\qquad\quad(011)^{\Gamma_2^+},\\\\
&\quad(200)^{\Gamma_2^+},\qquad(012)^{\Gamma_2^+},(201)^{\Gamma_2^+},\qquad(211)^{\Gamma_2^+},\\\\
&\qquad\qquad\,\,(022)^{\Gamma_2^+},\qquad\qquad\quad\,(122)^{\Gamma_2^+}\qquad\qquad\qquad),
\end{split}
\end{equation}

and

\begin{equation}
\begin{split}
&(\qquad\qquad(001)^{\Gamma_2^-},\qquad\qquad\quad(101)^{\Gamma_2^-},\\\\
&\quad(002)^{\Gamma_2^-},\qquad(012)^{\Gamma_2^-},(201)^{\Gamma_2^-},\qquad(112)^{\Gamma_2^-},\\\\
&\qquad\qquad\,\,(202)^{\Gamma_2^-},\qquad\qquad\quad(212)^{\Gamma_2^-}\qquad\qquad\qquad),
\end{split}
\end{equation}

that each transform according to an 8-dimensional irreducible representation of $GL(3,\mathbb{R})$. However, from a purely group theoretical perspective, we have no means of explicitly distinguishing between the copies---a basis for the 2-dimensional representation $\Gamma_2$ has to be chosen.\\

\subsection{Schur-Weyl duality and nonlinear responses of crystals}
\label{SchWcryst}

The key application of Schur-Weyl duality in the context of quantum transport is in finding the crystal structures that can support the nonlinear responses classified according to their behaviour under irreducible representations of the permutation group via the permutation decomposition. The possible responses supported by a crystal are restricted by the Neumann principle to those that are invariant under the symmetry group describing the crystal structure \cite{BurnsGlazer}. Specifically, the point group of a three-dimensional crystal is a finite group that is a subgroup of the continuous, full orthogonal group $O(3)$ of $3\times 3$ orthogonal matrices, which, in turn is a subgroup of $GL(3,\mathbb{R})$. This means that the \textit{irreducible} representations of $GL(3,\mathbb{R})$ are, in fact, \textit{reducible} representations of $O(3)$, which can then be decomposed into irreducible representations of $O(3)$ itself. The irreducible representations of $O(3)$ are reducible representations of the crystal point groups, and hence can further be decomposed into irreducible representation of the crystal point groups themselves. This means that Schur-Weyl duality allows us to associate irreducible representations of permutation groups and hence distinct responses to a collection of irreducible representations of crystal point groups. Note that the correspondence is \textit{not} one-to-one since the permutation group irreducible representations are associated uniquely only to the irreducible representations of $GL(3,\mathbb{R})$ itself, of which the crystal point groups are subgroups. Following the standard argument \cite{DresselhausSymm}, should the final collection of crystal point group irreducible representations contain the trivial representation---which is an $O(3)$ singlet---,Neumann's principle would allow the corresponding response to be non-vanishing, and the number of trivial representation copies would provide the number of independent such responses.\\

The explicit decomposition of the first few relevant irreducible representations of $O(3)$ into those of the 32 point groups can be found, for example, in the standard reference \cite{KosterGroup}. Similarly, the non-vanishing independent components of tensors upto rank 4 allowed by the 32 point groups can be found in standard textbooks \cite{Birss,BoydNLO}. However, a general classification of non-vanishing terms according to the irreducible representations of permutations groups via Schur-Weyl duality, is, to the best of our knowledge, not widely available. We remedy this, by performing it for rank-3 and rank-4 tensors, and the results are shown in Tables \ref{tab:P3ClassificationPolar}, \ref{tab:P3ClassificationAxial}, \ref{tab:P4ClassificationPolar}, and \ref{tab:P4ClassificationAxial} in Appendix \ref{PointGroupClass}. The tables are prepared as follows. We first restrict ourselves to the special linear group $SL(3,\mathbb{R})$ of elements in $GL(3,\mathbb{R})$ with unit determinant, since, despite the restriction, irreducible representations of the latter remain irreducible representations of the former \cite{TungGroups}. Denoting by $\textbf{3}$ the defining 3-dimensional irreducible representation of $SL(3,\mathbb{R})$ we can use the standard technique with Young tableaux \cite{TungGroups,SternbergPhysics,FuchsLie}, to decompose tensor products $\textbf{3}\otimes\dots\otimes\textbf{3}$ into direct sums of irreducible representations. We have the results

\begin{equation}
\label{eq:TensorDecomp}
\begin{split}
&\textbf{3}\otimes\textbf{3}=\textbf{6}\oplus\textbf{3},
\\
&\textbf{3}\otimes\textbf{3}\otimes\textbf{3}=\textbf{10}\oplus 2\cdot\textbf{8}\oplus\textbf{1}
\\
&\textbf{3}\otimes\textbf{3}\otimes\textbf{3}\otimes\textbf{3}=\textbf{15}'\oplus 3\cdot\textbf{15}\oplus2\cdot\textbf{6}\oplus3\cdot\textbf{3},
\\
&\dots,
\end{split}
\end{equation}

where the numbers in bold refer to the irreducible representation of the given dimension. Then, using the facts that; irreducible representations of $SL(3,\mathbb{R})$ are also those of $SU(3)$, since, roughly speaking, the complexification of their Lie algebras is the same \cite{FultonHarris,FuchsLie}; and that the odd-dimensional irreducible representations of $SU(2)$ are also (linear) irreducible representations of $SO(3)$ \cite{TungGroups}, we can use the branching rules of $SU(3)$ into $SU(2)$ (see for example ref. \cite{LieAlgebrasBig}) to find the decompositions into irreducible representations of $SO(3)$. We have

\begin{equation}
\label{eq:SU3Decomp}
\begin{split}
&\textbf{6}_{SU(3)}=(\textbf{5}\oplus\textbf{1})_{SO(3)},
\\
&\textbf{8}_{SU(3)}=(\textbf{5}\oplus\textbf{3})_{SO(3)},
\\
&\textbf{10}_{SU(3)}=(\textbf{7}\oplus\textbf{3})_{SO(3)},
\\
&\textbf{15}'_{SU(3)}=(\textbf{9}\oplus\textbf{5}\oplus\textbf{1})_{SO(3)},
\\
&\textbf{15}_{SU(3)}=(\textbf{7}\oplus\textbf{5}\oplus\textbf{3})_{SO(3)}.
\end{split}
\end{equation}

It is known that irreducible representations of $SO(3)$ are also irreducible representations of $O(3)$ \cite{Louck1972}, however, due to the presence of inversion with negative determinant in $O(3)$, we can have two irreducible representations of the latter corresponding to each one of the former, meaning that we have to distinguish between those that are even and odd under inversion. Under inversion, polar (axial) tensors of even (odd) rank are even, whereas polar (axial) tensors of odd (even) rank are odd. Thus, in the former (latter) case we have to look at those irreducible representations of $O(3)$ that are even (odd) under inversion, and we label them as $(\dots)^{+(-)}$. After deciding which tensor we want to look at, we can use the full rotation group compatibility tables from \cite{KosterGroup} to find out how the $O(3)$ irreducible representations such as $\textbf{5}^{\pm},\textbf{7}^{\pm}$ etc., decompose into irreducible representations of the different point groups and jot down the number of times the trivial representation appears in the result. By Neumann's principle, this is the number of independent responses the given crystal can support. Note that \cite{KosterGroup} uses the notation $D^{\pm}_l$, where $l=0,1,\dots$, for the odd-dimensional $O(3)$ representations, which corresponds to our $(\textbf{2}l+\textbf{1})^{\pm}$, i.e., $\textbf{1}^{\pm},\textbf{3}^{\pm},\dots$.\\

As a concrete example, consider the case of a rank 3 polar tensor for the point group $6mm(C_{6v})$. It is rank 3, so we need $\textbf{3}\otimes\textbf{3}\otimes\textbf{3}$, whose decomposition into $SU(3)$ irreducible representations is given in \eqref{eq:TensorDecomp} and contains $\textbf{10},\textbf{8},\textbf{1}$. By Schur-Weyl duality we can map each term in the decomposition to one of the permutation group irreducible representations, $\textbf{10}\to\Gamma_1^+,\,2\cdot\textbf{8}\to\Gamma_2,\,\textbf{1}\to\Gamma_1^-$. Then, we can decompose each $SU(3)$ irreducible representation into $SO(3)$ ones according to \eqref{eq:SU3Decomp}, and will have copies of $\textbf{7},\textbf{5},\textbf{3},\textbf{1}$ . At this point, we extend the representations to those of $O(3)$ and since we are interested in an odd rank polar tensor, we shall need $\textbf{7}^-,\textbf{5}^-,\textbf{3}^-,\textbf{1}^-$. Next, we look at the tables in \cite{KosterGroup}, check how each of these $O(3)$ representations decompose into those of $6mm(C_{6v})$, and find that only $\textbf{7}^-$ and $\textbf{3}^-$ contain one copy each of the trivial representation. Since the $\textbf{7}^-$ comes from the $\textbf{10}$ of $SU(3)$, the corresponding response transforms under the $\Gamma_1^+$ of $P(3)$ and is totally symmetric. On the other hand, $\textbf{3}^-$ can come from both $\textbf{10}$ and $\textbf{8}$, hence we have one of the corresponding responses transforming under $\Gamma_1^+$, whereas two others under $\Gamma_2$ (since we have two copies of the $\textbf{8}$). Thus, in total, we have 4 independent responses, with 2 of them being totally symmetric and 2 being mixed symmetric. As discussed in Section \ref{PermDecomp}, the $\Gamma_1^+$ responses can be interpreted as being `in-line longitudinal', whereas the $\Gamma_2$ as being either the latters' transverse counterpart or `in-plane longitudinal'. In the static case, we only have the `in-line longitudinal' $\Gamma_1^+$ and its transverse $\Gamma_2$, and can consider the latter as corresponding to, for example, the second order Hall response. In the non-static case, on the other hand, we can also interpret $\Gamma_2$ as `in-plane longitudinal' with an example of such a response being the second order injection current (see formula \eqref{eq:Cond2FlucDissFinal}). The symmetrized responses themselves can be obtained by projecting the non-vanishing terms detailed in textbooks (see e.g. refs. \cite{Birss,BoydNLO}) onto the $P(3)$ irreducible representations. For our case of $6mm(C_{6v})$ we have \cite{BoydNLO}: $xzx=yzy,\,xxz=yyz,\,zxx=zyy,\,zzz$; meaning that the two independent, totally symmetric combinations will be $zzz,\,(xzx+xxz+zxx)/3=(yzy+yyz+zyy)/3$, whereas, using the projection \eqref{eq:S3Gamma2Proj}, two independent $\Gamma_2$ combinations, among others, can be $(2xzx-xxz-zxx)/3,\,(2xxz-xzx-zxx)/3$. As discussed earlier, from a purely group theoretical point of view, the choice of basis for the 2-dimensional $\Gamma_2$ is ambiguous, and we have to be guided by the physical setup at hand, i.e., the driving field configuration and the frequency dependence. This then leads to the natural basis choice of the permutation decomposition \eqref{eq:3PermDecomp}. In our particular case, suppose we look at the second harmonic response to a field in the $x$ direction. Then, the response tensor can be $zxx$, and the permutation decomposition gives 1 independent transverse response $(2zxx-xzx-xxz)/3$.\\

In Appendix \ref{PointGroupClass}, we provide the number of independent coefficients classified in this way for all 32 point groups for rank 3 polar (axial) tensors in Table \ref{tab:P3ClassificationPolar} (\ref{tab:P3ClassificationAxial}), whereas for rank 4 polar (axial) tensors in Table \ref{tab:P4ClassificationPolar} (\ref{tab:P4ClassificationAxial}). The header of all tables contains the $P(n)$ irreducible representations and their dual $SU(3)$ irreducible representations together with the $O(3)$ irreducible representations resulting from the decompositions of the $SU(3)$ irreducible representations. The body of the table contains, for each point group, the number of independent responses classified according to the scheme discussed above. Note that for polar tensors of even rank and axial tensors of odd rank we need only specify the Laue groups not containing improper rotations \cite{Birss}.\\

We discuss some physical examples pertaining to the tables. Rank 3 and 4 polar tensors can represent, respectively, second and third order electric susceptibilities and charge conductivities, quantities of interest in non-linear optics and quantum transport. Similarly, rank 3 axial tensors can describe second order spin responses, and first order spin current responses to an electric field, whereas rank 4 axial tensors, among other effects, second order spin current responses. Let us first look at the third order charge conductivity, with the corresponding table being Table \ref{tab:P4ClassificationPolar}. Looking at the cubic point groups, we see that $432,\bar{4}3m$ and $m\bar{3}m$ do not have any responses corresponding to $\Gamma_3$, which, as discussed in Section \ref{PermDecomp}, gives the response transverse to the `in-line longitudinal' response, i.e., can be looked at as providing a third order Hall effect, and can potentially remain non-vanishing even in the static limit. Thus, crystals described by these groups, an important example being the class of zinc-blende semiconductors, cannot support such a third order Hall effect. Should we look beyond the static limit, however, they can support an effect described by $\Gamma_2$. In particular, for a self-focusing $\sigma_{i_0i_1i_2i_3}(\omega,-\omega,\omega)$ response, the two independent $\Gamma_2$ terms can be explicitly isolated from the $\Gamma_1^+$'s via the special beam configuration discussed in Section \ref{PermDecomp4Point}. Moving on to an example of a third order spin response to an electric field, we have to take a peek at Table \ref{tab:P4ClassificationAxial}. For the cubic $\bar{4}3m$, it is clear that the only response supported corresponds to $\Gamma_3$, a transverse one, hence, at third order, an electric field in direction, say $y$, can only give rise to spin polarization transverse to $y$, consequently, it can only contribute to the higher-order version of the `damping-like' component of the spin-orbit torque. For static electric fields, the same analysis applies to $6mm$; note that for an oscillating electric field, $6mm$ can also support a $\Gamma_{3'}$ response. Second order spin current responses $\sigma_{(s,i_0)i_1i_2}$ are also described by the same type of tensors, thus, with an electric field in, say, the $y$ direction, for the same group $\bar{4}3m$, the spin current response in the $y$ direction is required to be polarized transverse to $y$. In the same manner, we could discuss a large number of nonlinear physical effects and look at their possible manifestations in different crystals, however, in this brief note, we simply wanted to highlight the usefulness of the permutation decomposition and the corresponding point group classification scheme.\\

\section{Discussion}

\subsection{Weak 4-point fluctuation-response relations}
In Section \ref{FlucDissReact}, we presented a detailed discussion of the 2-point fluctuation-dissipation and 3-point fluctuation-reaction relation, moreover, we introduced the weak 3-point fluctuation-dissipation relation. The key insight was that these relations connect not fully time-ordered correlators with combinations of fully time-ordered correlators. The former can be thought of as `fluctuations', since not all of their time arguments are constrained, whereas the latter as responses, since full time-ordering is necessary to implement causality---there can be no response before a driving is applied. In the 2-point case, we can only have no time-ordering and full time-ordering, and obtain the standard fluctuation-dissipation relation. In the 3-point case, however, we can have no time-ordering leading to a fluctuation-reaction relation, and partial time-ordering rendering us with a weak fluctuation-dissipation relation, with the latter relation following from the former. What about the case of 4-point correlators? It turns out, that it is impossible to have a relation between such non-time-ordered stripped correlators and fully time-ordered ones. This can be checked directly, via expression \eqref{eq:Ret4Pl}. Forming $C^{r\pm a}$ we cannot find combinations that allow us to factor out a sum of step functions that covers all time-orderings and thereby isolate any stripped correlators. However, weak relations between partially time-ordered stripped correlators and fully time-ordered ones could, in fact, exist. We explore the details in an upcoming paper on third order responses \cite{Bonbien3rdOrder}.

\subsection{Totally symmetric nonlinear responses, invariant theory, and elliptic curves}
Consider a general homogeneous polynomial of degree $n+1$ in $m+1$ variables $x_0,\dots,x_m$,

\begin{equation}
p_{n+1}(x_0,\dots,x_m)=\sum_{i_0=1}^{m}\dots\sum_{i_n=0}^{m}A_{i_0\dots i_n}x_{i_0}\cdots x_{i_n},
\end{equation}

where the coefficents $A_{i_0\dots i_n}\in\mathbb{C}$ are complex numbers, and, since any two of the scalar variables $x_i,x_j$ can be exchanged, they can be considered totally symmetric in their indices. Now suppose we apply a transformation $g_{m+1}\in SL(m+1,\mathbb{R})$ to the variables as $x_{i_0}\to x_{j_0}(g_{m+1})_{j_0i_0}$, where summation over $j_0\in\{0,\dots,m\}$ is implied. Then, it is straightforward to see that the coefficients transform as components of a tensor, i.e., 

\begin{equation}
A_{i_0\dots i_n}\to\sum_{j_0=0}^m\dots\sum_{j_n=0}^m (g^{-1}_{m+1})_{i_0j_0}\dots(g^{-1}_{m+1})_{i_nj_n}A_{j_0\dots j_n}.
\end{equation}

In other words, it is always possible to represent components of totally symmetric tensors as coefficents of homogeneous polynomials. An important and widely used practical example is the case of a quadratic in 3 variables, i.e., the ternary quadratic

\begin{equation}
\begin{split}
p_2(x,y,z)=&A_{00}x^2+A_{11}y^2+A_{22}z^2
\\
&+2A_{12}xy+2A_{23}yz+2A_{13}xz,
\end{split}
\end{equation}

in which the $A_{ij}\in\mathbb{R}$ can be considered as elements of a symmetric $3\times3$ matrix. The resulting 2-dimensional surface $p_2(x,y,z)=1$ is generally referred to as the `representation quadric' and is used to visualize rank-2 (real) symmetric tensors \cite{LovettTensor}.\\

Now consider a nonlinear response, $\tilde{P}_{i_0i_1\dots i_n}(\omega_1,\dots,\omega_n)$. This is a collection of retarded correlators (cf. Eq. \eqref{eq:respn}), which, as we showed in Section \ref{RetTensor}, can be considered as components of a tensor. Hence, we can think of the totally symmetric nonlinear responses, i.e., the projections to the $\Gamma_1^+$ of $P(n+1)$, as coefficients of degree $n+1$ homogeneous polynomials. Natural examples of such nonlinear responses would be the longitudinal $n$-th harmonic responses $\tilde{P}^{\Gamma_1^+}_{i_0i_1\dots i_n}(\omega,\dots,\omega)$. The question we ask, is whether it is possible to associate simple geometric objects to some of these responses? We show two cases in which there is a link to a rather exotic object.

In order to do this, we go back to the behaviour of the polynomials and their coefficients under transformations. As we saw, a change of the variables leads to a change of the coefficients. There are then two rather natural questions we can ask. Are there any polynomials, referred to as \textit{invariants}, of the coefficients themselves that are left invariant under a transformation, and, if yes, how many independent ones are there? Do there exist polynomials in the coefficients and perhaps the variables, referred to as \textit{covariants}, that are left unchanged under a transformation, and, should they exist, can we find them? The field of study looking at such questions is known as `invariant theory' and was central to the mathematics of the 19th century \cite{HilbertInvariant}. As a simple example, in the case of a degree 2 polynomial in 2 variables, $p_2(x,y)=A_{xx}x^2+2A_{xy}xy+A_{yy}y^2$, a so-called binary quadratic, we have only one independent invariant $(A_{xy})^2-A_{xx}A_{yy}$ \cite{HilbertInvariant}, which, as we can see, is proportional to the determinant of the symmetric $2\times 2$ matrix formed by the coefficients. As for the covariants, we can note that invariants are always covariants, and the polynomial itself is always a covariant. For the binary quadratic, these are the only independent covariants \cite{HilbertInvariant}.\\

For our application to nonlinear responses, we first look at the case of a homogeneous polynomial of degree 4 in 2 variables, $x,y$, in other words, a binary quartic

\begin{equation}
\begin{split}
&p_4(x,y;\omega_1,\omega_2,\omega_3)
\\
&=\tilde{P}_{xxxx}(\omega_1,\omega_2,\omega_3)x^4+4\tilde{P}^{\Gamma_1^+}_{xxxy}(\omega_1,\omega_2,\omega_3)x^3y
\\
&\quad+6\tilde{P}^{\Gamma_1^+}_{xxyy}(\omega_1,\omega_2,\omega_3)x^2y^2+4\tilde{P}^{\Gamma_1^+}_{xyyy}(\omega_1,\omega_2,\omega_3)xy^3
\\
&\quad+\tilde{P}_{yyyy}(\omega_1,\omega_2,\omega_3)y^4,
\end{split}
\end{equation}

with the coefficients describing some longitudinal nonlinear response in 2-dimensions, for example the longitudinal third harmonic charge conductivity $\sigma^{\Gamma_1^+}_{i_0i_1i_2i_3}(\omega,\omega,\omega)$, where $i_0,i_1,i_2,i_3\in\{x,y\}$, of a 2-dimensional material. The binary quartic has two independent invariants which are well-known to be \cite{HilbertInvariant}

\begin{equation}
\label{eq:BinaryQuarticInv}
\begin{split}
&I_2=\tilde{P}_{xxxx}\tilde{P}_{yyyy}-4\tilde{P}^{\Gamma_1^+}_{xxxy}\tilde{P}^{\Gamma_1^+}_{xyyy}+3(\tilde{P}^{\Gamma_1^+}_{xxyy})^2,
\\
&I_3=\tilde{P}_{xxxx}\tilde{P}^{\Gamma_1^+}_{xxyy}\tilde{P}_{yyyy}-\tilde{P}_{xxxx}(\tilde{P}^{\Gamma_1^+}_{xyyy})^2-(\tilde{P}^{\Gamma_1^+}_{xxxy})^2\tilde{P}_{yyyy}
\\
&\qquad+2\tilde{P}^{\Gamma_1^+}_{xxxy}\tilde{P}^{\Gamma_1^+}_{xxyy}\tilde{P}^{\Gamma_1^+}_{xyyy}-(\tilde{P}^{\Gamma_1^+}_{xxyy})^3,
\end{split}
\end{equation}

where we dropped the frequency arguments for brevity. Furthermore, we also have three independent covariants, one of them being the polynomial $p_4$ itself, and two others that can be written as

\begin{equation}
\begin{split}
&H=\frac{1}{144}\left(\left(\frac{\partial^2 p_4}{\partial x\partial y}\right)^2-\frac{\partial^2p_4}{\partial x^2}\frac{\partial^2p_4}{\partial y^2}\right),
\\
&J=\frac{1}{8}\left(\frac{\partial p_4}{\partial x}\frac{\partial H}{\partial y}-\frac{\partial p_4}{\partial y}\frac{\partial H}{\partial x}\right),
\end{split}
\end{equation}

where we can recognize $H$ as the determinant of the Hessian, and $J$ as the Jacobian determinant of $p_4$ and $H$. Remarkably, the two invariants $I_2,I_3$ and three covariants $p_4,H,J$ form a rather simple relation known as a syzygy \cite{CubicSyzygy}

\begin{equation}
J^2=4H^3-I_2Hp^2_4-I_3p_4^3.
\end{equation} 

Defining the double cover of the polynomial, $z^2=p_4(x,y)$, we can rewrite the syzygy as

\begin{equation}
\left(\frac{J}{z^3}\right)^2=4\left(\frac{H}{z^2}\right)^3-I_2\left(\frac{H}{z^2}\right)
-I_3,
\end{equation}

and recognize the Weierstrass form of a cubic that can be parameterized using the $\wp$-function as \cite{EllipticFunctions}

\begin{equation}
\label{eq:Weierstrass}
\wp'^2(z)=4\wp^3(z)-g_2\wp(z)-g_3.
\end{equation}

We see that the invariants $I_2,I_3$ can be interpreted as the elliptic invariants $g_2,g_3$. The Weierstrass $\wp$-function is an elliptic function, i.e., it is doubly periodic with domain $\mathbb{C}/\Lambda$ where $\Lambda$ is a lattice on the complex plane. The two periods are encoded by $g_2$ and $g_3$, meaning that they define a lattice and their variation leads to different lattices. In our case, $I_2,I_3$ are given in terms of complex response coefficients which are measurable, and frequency dependent. As we change the frequency, $I_2$ and $I_3$ vary, meaning that they define different lattices. Thus, we can, in fact, associate a family of lattices over the complex numbers to our nonlinear responses. The $\wp$-function is also closely related to elliptic curves over the complex numbers, meaning that we can also think of attaching a family of elliptic curves to our nonlinear responses. Note that the number of independent response coefficients are restricted by crystal symmetry, hence, in most cases, the 6 coefficients of the binary quartic are closely related, thereby simplifying the corresponding invariants $I_2,I_3$.\\

There is another case in which families of elliptic curves can be associated to responses. This time we look at a homogeneous polynomial of degree 3 in 3 variables---a ternary cubic. The 10 coefficients are given by 3-index totally symmetric responses $\tilde{P}^{\Gamma_1^+}_{i_0i_1i_2}(\omega_1,\omega_2)$, where $i_0,i_1,i_2\in\{x,y,z\}$, but we drop the frequency arguments for brevity. We have

\begin{equation}
\begin{split}
&p_3(x,y,z;\omega_1,\omega_2)
\\
&=\tilde{P}_{xxx}x^3+\tilde{P}_{yyy}y^3+\tilde{P}_{zzz}z^3
\\
&\quad+ 3\tilde{P}^{\Gamma_1^+}_{xxy}x^2y+3\tilde{P}^{\Gamma_1^+}_{xxz}x^2z+3\tilde{P}^{\Gamma_1^+}_{xyy}xy^2
\\
&\quad+3\tilde{P}^{\Gamma_1^+}_{xzz}xz^2
+3\tilde{P}^{\Gamma_1^+}_{yyz}y^2z+3\tilde{P}^{\Gamma_1^+}_{yzz}yz^2
\\
&\quad+6\tilde{P}^{\Gamma_1^+}_{xyz}xyz.
\end{split}
\end{equation}

Examples of such response coefficients would be the longitudinal second harmonic charge conductivities $\sigma^{\Gamma_1^+}_{i_0i_1i_2}(\omega,\omega)$. Now let us interpret the variables $x,y,z$ as homogeneous coordinates on the projective plane and define the cubic plane curve $p_3(x,y,z;\omega_1,\omega_2)=0$. It is clear that as $\omega_1,\omega_2$ vary, the coefficients give us different cubic plane curves and it is possible to asssociate a family of cubic plane curves to the response coefficients. However, we can go further. The ternary cubic has two independent invariants, labelled $S$ and $T$. They were calculated in the 19th century and are known as Aronhold invariants, but they are rather lengthy polynomial expressions of the coefficients; hence we simply refer to \cite{SturmfelsInv,CubicSyzygy} for their explicit form. We also have four covariants, $p_4,H,\Theta,J$, where $H$ is related to the determinant of the Hessian, for the construction of $\Theta$ we refer to \cite{CubicSyzygy}, and $J$ is related to the Jacobian determinant of $p_4,\,H$ and $\Theta$. Yet again, we have a syzygy between the two invariants and four covariants, which, for a cubic plane curve $p_4=0$ reads \cite{CubicSyzygy}

\begin{equation}
\left(\frac{J}{H^3}\right)^2=4\left(\frac{\Theta}{H^2}\right)^3+108S\frac{\Theta}{H^2}-27T.
\end{equation}

With $g_2=-108S$ and $g_3=27T$ we get the Weierstrass cubic \eqref{eq:Weierstrass} and can once again interpret the invariants $S,T$ as determining a lattice in the complex plane. Since $S$ and $T$ can be measured, we can also associate families of lattices and thereby elliptic curves to totally symmetric responses determined by 3-index quantities such as longitudinal second harmonic responses.\\

While our presentation was more of a brief sketch, such ideas could provide a fruitful visualization of the described responses and, perhaps, offer a deeper probe into the material itself by providing an alternative way of interpreting the measured data. We shall explore this in future work.\\

\section{Conclusion}

The purpose of this paper was to delve into the heart of the Kubo formalism---the retarded correlator, and present a detailed analysis of its structure. The primary theme was to decompose the retarded correlator itself, or collections of retarded correlators into terms transforming in the irreducible representations of certain groups and thereby provide a means to extract information on the possible and otherwise distinct physical effects. We considered three such groups: the time-reversal group $\mathcal{G}_{\mathfrak{K}}\times\mathcal{G}_{\mathfrak{K}^{\textbf{m}}}$ classifying the responses based on them being reactive or dissipative and present or absent in the absence of the time-reversal symmetry of the equilibrium Hamiltonian $\mathcal{H}_0$, the permutation group $P(n)$ yielding insights into what responses are possible for given driving field and observational configurations, and, finally, the general linear group, closely linked to the permutation group via Schur-Weyl duality, allowing us to provide a general scheme for discerning whether crystals support a given effect or not.\\

We presented a simple derivation of the 3-point fluctuation-reaction relation and showed that it can be used to express a specific reactive direct current charge conductivity response transforming under the $\Gamma_2$ representation of $P(3)$ and noted that in clean semiconductors it corresponds to the injection current, a contribution to the bulk photovoltaic effect. We showed that the 3-point fluctuation-\textit{reaction} relation gives rise to a weak 3-point fluctuation-\textit{dissipation} relation, the latter being, in essence, a connection between partially time-ordered correlators and combinations of fully time-ordered correlators. We then found that this weak relation describes a number of second order dissipative responses, moreover, that it defines an assortment of new frequency sum rules.\\

We introduced the permutation decomposition of retarded correlators into irreducible representations of permutation groups and showed that the terms in the decomposition can be related to a number of physical effects such as higher order Hall responses. We also discussed how, in some cases, we can use driving field configurations to restrict the system response to specific terms in the permutation decomposition.\\

We arranged the retarded correlators into arrays that transform under a reducible representation of the general linear group. Referring to Schur-Weyl duality between irreducible representations of general linear groups and irreducible representations of permutation groups, we showed how it is possible to decide which terms in our permutation decomposition can be supported by different crystals described via the 32 point groups, and performed the full point group classification for rank 3 and 4 polar and axial tensors.\\

We believe that the results and narrative of this paper provide a baseline for a robust framework and new perspective for nonlinear quantum transport. It is made abundantly clear how a general and systematic analysis of the main mathematical structures can lead very far and to findings that are surprisingly concrete.\\

In paper III \cite{Bonbien2021c}, we apply the described  decompositions to second order charge conductivity responses, and show that they provide a highly useful scheme for separating different physical effects.\\

\begin{acknowledgments}
This research was supported by the King Abdullah University of Science and Technology (KAUST). A. M. acknowledges support from the Excellence Initiative of Aix-Marseille Université—A*Midex, a French ‘Investissements d’Avenir’ program.
\end{acknowledgments}

\appendix
\section{Point group classification of rank 3 and 4, polar and axial tensors according to permutation group irreducible representations}
\label{PointGroupClass}

\begin{table}[h]
\begin{center}
\caption{Point group classification of rank 3 polar tensors according to irreducible representations of $P(3)$}
\label{tab:P3ClassificationPolar}
\begin{tabular}{c||c|c||c|c||c}
\hline\hline
$P(3)$&\multicolumn{2}{c||}{$\Gamma_{1}^{+}$}&\multicolumn{2}{|c||}{$\Gamma_2$}&$\Gamma_1^-$\\\hline
$SU(3)$&\multicolumn{2}{c||}{$\textbf{10}$}&\multicolumn{2}{|c||}{$2\cdot\textbf{8}$}&$\textbf{1}$\\\hline
$O(3)$&$\textbf{7}^-$&$\textbf{3}^-$&$2\cdot\textbf{5}^-$&$2\cdot\textbf{3}^-$&$\textbf{1}^-$\\\hline\hline
$1$&7&3&10&6&1\\\hline
$2$&3&1&6&2&1\\\hline
$m$&4&2&4&4&-\\\hline
$222$&1&-&4&-&1\\\hline
$mm2$&2&1&2&2&-\\\hline
$4$&1&1&2&2&1\\\hline
$\bar{4}$&2&-&4&-&-\\\hline
$422$&-&-&2&-&1\\\hline
$4mm$&1&1&-&2&-\\\hline
$\bar{4}2m$&1&-&2&-&-\\\hline
$3$&3&1&2&2&1\\\hline
$32$&1&-&2&-&1\\\hline
$3m$&2&1&-&2&-\\\hline
$6$&1&1&2&2&1\\\hline
$\bar{6}$&2&-&-&-&-\\\hline
$622$&-&-&2&-&1\\\hline
$6mm$&1&1&-&2&-\\\hline
$\bar{6}m2$&1&-&-&-&-\\\hline
$23$&1&-&-&-&1\\\hline
$432$&-&-&-&-&1\\\hline
$\bar{4}3m$&1&-&-&-&-\\\hline
\end{tabular}
\end{center}
\end{table}

\begin{table}[h]
\begin{center}
\caption{Point group classification of rank 4 axial tensors according to the irreducible representations of $P(4)$}
\label{tab:P4ClassificationAxial}
\begin{tabular}{c||c|c|c||c|c|c||c|c||c}
\hline\hline
$P(4)$&\multicolumn{3}{c||}{$\Gamma_{1}^{+}$}&\multicolumn{3}{c||}{$\Gamma_3$}&\multicolumn{2}{c||}{$\Gamma_2$}&$\Gamma_{3'}$\\\hline
$SU(3)$&\multicolumn{3}{c||}{$\textbf{15}'$}&\multicolumn{3}{c||}{$3\cdot \textbf{15}$}&\multicolumn{2}{c||}{$2\cdot\textbf{6}$}&$3\cdot\textbf{3}$\\\hline
$O(3)$&$\textbf{9}^-$&$\textbf{5}^-$&$\textbf{1}^-$&$3\cdot\textbf{7}^-$&$3\cdot\textbf{5}^-$&$3\cdot\textbf{3}^-$&$2\cdot\textbf{5}^-$&$2\cdot\textbf{1}^-$&$3\cdot\textbf{3}^-$\\\hline\hline
$1$&9&5&1&21&15&9&10&2&9\\\hline
$2$&5&3&1&9&9&3&6&2&3\\\hline
$m$&4&2&-&12&6&6&4&-&6\\\hline
$222$&3&2&1&3&6&-&4&2&-\\\hline
$mm2$&2&1&-&6&3&3&2&-&3\\\hline
$4$&3&1&1&3&3&3&2&2&3\\\hline
$\bar{4}$&2&2&-&6&6&-&4&-&-\\\hline
$422$&2&1&1&-&3&-&2&2&-\\\hline
$4mm$&1&-&-&3&-&3&-&-&3\\\hline
$\bar{4}2m$&1&1&-&3&3&-&2&-&-\\\hline
$3$&3&1&1&9&3&3&2&2&3\\\hline
$32$&2&1&1&3&3&-&2&2&-\\\hline
$3m$&1&-&-&6&-&3&-&-&3\\\hline
$6$&1&1&1&3&3&3&2&2&3\\\hline
$\bar{6}$&2&-&-&6&-&-&-&-&-\\\hline
$622$&1&1&1&-&3&-&2&2&-\\\hline
$6mm$&-&-&-&3&-&3&-&-&3\\\hline
$\bar{6}m2$&1&-&-&3&-&-&-&-&-\\\hline
$23$&1&-&1&3&-&-&-&2&-\\\hline
$432$&1&-&1&-&-&-&-&2&-\\\hline
$\bar{4}3m$&-&-&-&3&-&-&-&-&-\\\hline
\end{tabular}
\end{center}
\end{table}

\begin{table}[h]
\begin{center}
\caption{Point group classification of rank 3 axial tensors according to the irreducible representations of $P(3)$}
\label{tab:P3ClassificationAxial}
\begin{tabular}{c||c|c||c|c||c}
\hline\hline
$P(3)$&\multicolumn{2}{c||}{$\Gamma_{1}^{+}$}&\multicolumn{2}{|c||}{$\Gamma_2$}&$\Gamma_1^-$\\\hline
$SU(3)$&\multicolumn{2}{c||}{$\textbf{10}$}&\multicolumn{2}{|c||}{$2\cdot\textbf{8}$}&$\textbf{1}$\\\hline
$O(3)$&$\textbf{7}^+$&$\textbf{3}^+$&$2\cdot\textbf{5}^+$&$2\cdot\textbf{3}^+$&$\textbf{1}^+$\\\hline\hline
$1$&7&3&10&6&1\\\hline
$2$&3&1&6&2&1\\\hline
$222$&1&-&4&-&1\\\hline
$4$&1&1&2&2&1\\\hline
$422$&-&-&2&-&1\\\hline
$3$&3&1&2&2&1\\\hline
$32$&1&-&2&-&1\\\hline
$6$&1&1&2&2&1\\\hline
$622$&-&-&2&-&1\\\hline
$23$&1&-&-&-&1\\\hline
$432$&-&-&-&-&1\\\hline
\end{tabular}
\end{center}
\end{table}

\begin{table}[h]
\begin{center}
\caption{Point group classification of rank 4 polar tensors according to the irreducible representations of $P(4)$}
\label{tab:P4ClassificationPolar}
\begin{tabular}{c||c|c|c||c|c|c||c|c||c}
\hline\hline
$P(4)$&\multicolumn{3}{c||}{$\Gamma_{1}^{+}$}&\multicolumn{3}{c||}{$\Gamma_3$}&\multicolumn{2}{c||}{$\Gamma_2$}&$\Gamma_{3'}$\\\hline
$SU(3)$&\multicolumn{3}{c||}{$\textbf{15}'$}&\multicolumn{3}{c||}{$3\cdot \textbf{15}$}&\multicolumn{2}{c||}{$2\cdot\textbf{6}$}&$3\cdot\textbf{3}$\\\hline
$O(3)$&$\textbf{9}^+$&$\textbf{5}^+$&$\textbf{1}^+$&$3\cdot\textbf{7}^+$&$3\cdot\textbf{5}^+$&$3\cdot\textbf{3}^+$&$2\cdot\textbf{5}^+$&$2\cdot\textbf{1}^+$&$3\cdot\textbf{3}^+$\\\hline\hline
$1$&9&5&1&21&15&9&10&2&9\\\hline
$2$&5&3&1&9&9&3&6&2&3\\\hline
$222$&3&2&1&3&6&-&4&2&-\\\hline
$4$&3&1&1&3&3&3&2&2&3\\\hline
$422$&2&1&1&-&3&-&2&2&-\\\hline
$3$&3&1&1&9&3&3&2&2&3\\\hline
$32$&2&1&1&3&3&-&2&2&-\\\hline
$6$&1&1&1&3&3&3&2&2&3\\\hline
$622$&1&1&1&-&3&-&2&2&-\\\hline
$23$&1&-&1&3&-&-&-&2&-\\\hline
$432$&1&-&1&-&-&-&-&2&-\\\hline
\end{tabular}
\end{center}
\end{table}

\section{Character tables of $\mathsf{Z}_2$, $\mathsf{Z}_2\times\mathsf{Z}_2$, $P(3)$ and $P(4)$}
\label{GroupTools}

For convenience, we reproduce the character tables of the groups used in the main text. Details can be found in standard texts such as \cite{DresselhausSymm}.

\begin{table}[h]
\caption{Character table of $\mathsf{Z}_2$.}
\label{tab:Z2}
\begin{center}
\begin{tabular}{|c||c|c|}
\hline
$\mathsf{Z}_2$&$E$ & $A$\\\hline\hline
$\Gamma_1^{+}$&1&1\\\hline
$\Gamma_1^{-}$&1&-1\\\hline
\end{tabular}
\end{center}
\end{table}

\begin{table}[h]
\caption{Character table of $\mathsf{Z}_2\times \mathsf{Z}_2$.}
\label{tab:Z2Z2}
\begin{center}
\begin{tabular}{|c||c|c|c|c|}
\hline
$\mathsf{Z}_2\times Z_2$&$E$ & $A$ &$B$&$A*B$\\\hline\hline
$++$&1&1&1&1\\\hline
$+-$&1&-1&1&-1\\\hline
$-+$&1&-1&-1&1\\\hline
$--$&1&1&-1&-1\\\hline
\end{tabular}
\end{center}
\end{table}

\begin{table}[h]
\caption{Character table of $P(3)$.}
\label{tab:P3}
\begin{center}
\begin{tabular}{|c||c|c|c|}
\hline
$P(3)$&$(1^3)$&$3(2,1)$&$2(3)$\\\hline\hline
$\Gamma_1^{+}$&1&1&1\\\hline
$\Gamma_1^{-}$&1&-1&1\\\hline
$\Gamma_2$&2&0&-1\\\hline
\end{tabular}
\end{center}
\end{table}

\begin{table}[h]
\caption{Character table of $P(4)$.}
\label{tab:P4}
\begin{center}
\begin{tabular}{|c||c|c|c|c|c|}
\hline
$P(4)$&$(1^4)$&$8(3,1)$&$3(2^2)$&$6(2,1^2)$&6(4)\\\hline\hline
$\Gamma_1^{+}$&1&1&1&1&1\\\hline
$\Gamma_1^{-}$&1&1&1&-1&-1\\\hline
$\Gamma_2$&2&-1&2&0&0\\\hline
$\Gamma_3$&3&0&-1&1&-1\\\hline
$\Gamma_{3'}$&3&0&-1&-1&1\\\hline
\end{tabular}
\end{center}
\end{table}

For the permutation groups $P(3)$ and $P(4)$ the conjugation classes are denoted using the cycle notation. For example $8(3,1)$ means that the conjugation class consists of 8 permutations that act on a 4-tuple $1234$ by leaving one object fixed and permuting the remaining 3 e.g.$1234\to 1432$ etc. Another example is $3(2^2)$ which contains 3 permutations that act by exchanging two at a time e.g. $1234\to 2143$. See \cite{DresselhausSymm,SternbergPhysics} for details.\\

We can project a function $f$ to an irreducible representation $\Gamma_i$ of the group $\mathcal{G}$ using the following formula \cite{DresselhausSymm}:

\begin{equation}
\label{eq:ProjIrrep}
\hat{\mathcal{P}}^{(\Gamma_i)}f=\frac{l_i}{|\mathcal{G}|}\sum_{g\in \mathcal{G}}\chi^{\Gamma_i}(g)\pi(g)f,
\end{equation}

where $l_i$ is the dimension of $\Gamma_i$ (the number in the first column of the character tables), $|\mathcal{G}|$ is the number of elements in $\mathcal{G}$, $\chi^{\Gamma_i}(g)$ is the character of the element $g$ in the representation $\Gamma_i$, and $\pi(g)$ is the action of the element $g$ on the function $f$.

\section{The 3-point fluctuation-reaction relation}
\label{FluctDiss}

\subsection{Proof of Eq. \eqref{eq:3FlucDiss}, the general relation}

First, recall the $n+1$-point retarded and advanced correlators from equations  \eqref{eq:Retn} and \eqref{eq:Advn}, which we write for $n=2$:

\begin{equation}
\label{eq:3retApp}
\begin{split}
-2\hbar^2C^{r}_{A^0A^1A^2}(t_0,t_1,t_2)=&\theta_{t_0t_1}\theta_{t_1t_2}C_{[A^0A^1A^2]}(t_0,t_1,t_2)
\\
&+\theta_{t_0t_2}\theta_{t_2t_1}C_{[A^0A^2A^1]}(t_0,t_2,t_1),
\end{split}
\end{equation}
\begin{equation}
\label{eq:3advApp}
\begin{split}
-2\hbar^2C^{a}_{A^0A^1A^2}(t_0,t_1,t_2)=&\theta_{t_2t_1}\theta_{t_1t_0}C_{[A^0A^1A^2]}(t_0,t_1,t_2)
\\
&\theta_{t_1t_2}\theta_{t_2t_0}C_{[A^0A^2A^1]}(t_0,t_2,t_1),
\end{split}
\end{equation}

where $C_{[A^0A^1A^2]}(t_0,t_1,t_2)$ is the 3-point stripped correlator

\begin{equation}
C_{[A^0A^1A^2]}(t_0,t_1,t_2)=\text{tr}(\rho_0[[A^0_{t_0},A^1_{t_1}],A^2_{t_2}]),
\end{equation}

and, keeping brevity in sight, we introduced the shorthand notation $A^i_{t_i}\equiv A^i_{\mathcal{H}_0}(t_i)$. Next, we have the identity 

\begin{equation}
\label{eq:ThetaPermSumApp}
\sum_{\pi\in S_n}\theta_{t_{\pi(1)}t_{\pi(2)}}\theta_{t_{\pi(2)}t_{\pi(3)}}\cdots\theta_{t_{\pi(n-1)}t_{\pi(n)}}=1,
\end{equation}

where $\pi$ label the permutations of the set $\{0,\dots,n-1\}$. This identity is true since we are summing over all permutations of time labels and thus covering all possible time-orderings. Making use of it for $n=3$ we manipulate the nested commutator

\begin{widetext}
\begin{equation}
\nonumber
\begin{split}
[[A^0_{t_0},A^1_{t_1}],A^2_{t_2}]=&(\theta_{t_0t_1}\theta_{t_1t_2}+\theta_{t_0t_2}\theta_{t_2t_1}+\theta_{t_1t_0}\theta_{t_0t_2}+\theta_{t_1t_2}\theta_{t_2t_0}+\theta_{t_2t_0}\theta_{t_0t_1}+\theta_{t_2t_1}\theta_{t_1t_0})[[A^0_{t_0},A^1_{t_1}],A^2_{t_2}]
\\
\\
=&\theta_{t_0t_1}\theta_{t_1t_2}[[A^0_{t_0},A^1_{t_1}],A^2_{t_2}] + \theta_{t_0t_2}\theta_{t_2t_1}\underbrace{[[A^0_{t_0},A^1_{t_1}],A^2_{t_2}]}_{=[[A^0_{t_0},A^2_{t_2}],A^1_{t_1}]-[[A^1_{t_1},A^2_{t_2}],A^0_{t_0}]}
\\
&+\theta_{t_2t_1}\theta_{t_1t_0}[[A^0_{t_0},A^1_{t_1}],A^2_{t_2}] +\theta_{t_1t_2}\theta_{t_2t_0}\underbrace{[[A^0_{t_0},A^1_{t_1}],A^2_{t_2}]}_{=[[A^0_{t_0},A^2_{t_2}],A^1_{t_1}]-[[A^1_{t_1},A^2_{t_2}],A^0_{t_0}]}
\\
&+\theta_{t_1t_0}\theta_{t_0t_2}\underbrace{[[A^0_{t_0},A^1_{t_1}],A^2_{t_2}]}_{=-[[A^1_{t_1},A^0_{t_0}],A^2_{t_2}]} + \theta_{t_2t_0}\theta_{t_0t_1}\underbrace{[[A^0_{t_0},A^1_{t_1}],A^2_{t_2}]}_{=-[[A^1_{t_1},A^0_{t_0}],A^2_{t_2}]}
\\
\\
=&\theta_{t_0t_1}\theta_{t_1t_2}[[A^0_{t_0},A^1_{t_1}],A^2_{t_2}] + \theta_{t_0t_2}\theta_{t_2t_1}[[A^0_{t_0},A^2_{t_2}],A^1_{t_1}]\qquad(\to r)
\\
&+\theta_{t_2t_1}\theta_{t_1t_0}[[A^0_{t_0},A^1_{t_1}],A^2_{t_2}] +\theta_{t_1t_2}\theta_{t_2t_0}[[A^0_{t_0},A^2_{t_2}],A^1_{t_1}]\qquad(\to a)
\\
&-\theta_{t_1t_0}\theta_{t_0t_2}[[A^1_{t_1},A^0_{t_0}],A^2_{t_2}]-\theta_{t_1t_2}\theta_{t_2t_0}[[A^1_{t_1},A^2_{t_2}],A^0_{t_0}]\qquad(\to r)
\\
&-\theta_{t_2t_0}\theta_{t_0t_1}[[A^1_{t_1},A^0_{t_0}],A^2_{t_2}]-\theta_{t_0t_2}\theta_{t_2t_1}[[A^1_{t_1},A^2_{t_2}],A^0_{t_0}]\qquad(\to a),
\end{split}
\end{equation}

where we made use of the Jacobi identity. Taking a look at the retarded and advanced correlators in \eqref{eq:3retApp} and \eqref{eq:3advApp}, we recognize them appearing after the last equality above, as indicated by $(\to r,a)$. Averaging, we have

\begin{equation}
\label{eq:3FlucDissApp}
C_{[A^0A^1A^2]}(t_0,t_1,t_2)=2\hbar^2(C^{r+a}_{A^1A^0A^2}(t_1,t_0,t_2)-C^{r+a}_{A^0A^1A^2}(t_0,t_1,t_2)),
\end{equation}

which is equation \eqref{eq:3FlucDiss} in the main text.
\end{widetext}

\subsection{Proof of Eq. \eqref{eq:Cond2FlucDiss2}, the application to second order charge conductivity}
\label{FluctDissCondProof}

For clarity, throughout the proof we shall use the following short-hands: $012(\omega_1,\omega_2)\equiv\tilde{C}^{r+a}_{J_{i_0}J_{i_1}J_{i_2}}(\omega_1,\omega_2)$ and $\hat{\mathcal{P}}^{(\Gamma_1^{-})}_{12}\equiv\hat{\mathcal{P}}^{(\Gamma_1^{-})}_{i_1i_2}$ with which the relevant conductivity expression \eqref{eq:Cond2FlucDiss1} becomes $2\omega^2\sigma^{\mathcal{T}^{\text{full}}_+,\mathcal{K}^{\textbf{m}}_+}_{i_0i_1i_2}(\omega,-\omega)\equiv \hat{\mathcal{P}}^{(\Gamma_1^{-})}_{12}(012(\omega,-\omega)-012(\omega,0)-012(0,-\omega))$; furthermore, we shall write $[012](\omega_1,\omega_2)\equiv \tilde{C}_{[J_{i_0}J_{i_1}J_{i_2}]}(\omega_1,\omega_2)$ for the stripped correlator. We shall also make ample use of intrinsic permutation symmetry, $012(\omega_1,\omega_2)=021(\omega_2,\omega_1)$, valid for the retarded and advanced correlators.\\

The manifestations of the fluctuation-reaction relation that we shall be needing are \eqref{eq:3FlucDissFreqCurr1} and \eqref{eq:3FlucDissFreqCurr2}, which read

\begin{align}
\label{eq:FlucReacCondApp1a}
&012(\omega,-\omega)-102(0,-\omega)=-\frac{1}{2\hbar^2}[012](\omega,-\omega),
\\
\label{eq:FlucReacCondApp1b}
&012(\omega,0)-120(0,-\omega)=0.
\end{align}

We also need further relations. Exchanging 0 and 2 in \eqref{eq:FlucReacCondApp1b}, and then using the result in \eqref{eq:FlucReacCondApp1a} allows us to write 

\begin{equation}
\label{eq:FlucReacCondApp1c}
012(\omega,-\omega)-210(\omega,0)=-\frac{1}{2\hbar^2}[012](\omega,-\omega).
\end{equation}

Useful cyclic permutations of \eqref{eq:FlucReacCondApp1a}, 
\eqref{eq:FlucReacCondApp1b} and \eqref{eq:FlucReacCondApp1c} are

\begin{align}
\label{eq:FlucReacCondApp1d}
&201(\omega,-\omega)-021(0,-\omega)=-\frac{1}{2\hbar^2}[201](\omega,-\omega),
\\
\label{eq:FlucReacCondApp1e}
&120(\omega,-\omega)-021(\omega,0)=-\frac{1}{2\hbar^2}[120](\omega,-\omega),
\\
\label{eq:FlucReacCondApp1f}
&201(\omega,0)-012(0,-\omega)=0.
\end{align}

Let us now project the conductivity $2\omega^2\sigma^{\mathcal{T}^{\text{full}}_+,\mathcal{K}^{\textbf{m}}_+}_{i_0i_1i_2}(\omega,-\omega)$ onto $\Gamma_2$ according to \eqref{eq:S3Gamma2Proj}. Then, we have

\begin{equation}
\begin{split}
&\hat{\mathcal{P}}^{(\Gamma_2)}_{012}\hat{\mathcal{P}}^{(\Gamma_1^{-})}_{12}(012(\omega,-\omega)-012(\omega,0)-012(0,-\omega))
\\
&=\frac13\hat{\mathcal{P}}^{(\Gamma_1^{-})}_{12}(2\times 012(\omega,-\omega)-201(\omega,-\omega)-120(\omega,-\omega)
\\
&\qquad\qquad\quad -2\times 012(\omega,0)+201(\omega,0)+120(\omega,0)
\\
&\qquad\qquad\quad -2\times 012(0,-\omega)+201(0,-\omega)+120(0,-\omega))
\\
&=\frac13\hat{\mathcal{P}}^{(\Gamma_1^{-})}_{12}(012(\omega,-\omega)-210(\omega,0)
\\
&\qquad\qquad\quad+012(\omega,-\omega)-102(0,-\omega)
\\
&\qquad\qquad\quad -120(\omega,-\omega)+021(\omega,0)
\\
&\qquad\qquad\quad-201(\omega,-\omega)+021(0,-\omega)
\\
&\qquad\qquad\quad +201(\omega,0)-012(0,-\omega)
\\
&\qquad\qquad\quad-012(\omega,0)+120(0,-\omega))
\\
&=-\frac{1}{2\hbar^2}\hat{\mathcal{P}}^{(\Gamma_2)}_{012}\hat{\mathcal{P}}^{(\Gamma_1^{-})}_{12}[012](\omega,-\omega),
\end{split}
\end{equation}

where we used relations \eqref{eq:FlucReacCondApp1a}-\eqref{eq:FlucReacCondApp1f} to obtain the final result. This means that the relevant conductivity can be written as

\begin{equation}
2\omega^2\sigma^{\mathcal{T}^{\text{full}}_+,\mathcal{K}^{\textbf{m}}_+,\Gamma_2^-}_{i_0i_1i_2}(\omega,-\omega)\equiv-\frac{1}{2\hbar^2}\hat{\mathcal{P}}^{(\Gamma_2)}_{012}\hat{\mathcal{P}}^{(\Gamma_1^{-})}_{12}[012](\omega,-\omega),
\end{equation}

which is Eq. \eqref{eq:Cond2FlucDiss2} in the main text.

\section{The retarded correlator as a combination of plain correlators}
\label{RetPlain}
The retarded correlator is defined as a symmetrized version of the time-ordered stripped correlator

\begin{equation}
\nonumber
\begin{split}
&C^r_{A^0A^1\dots A^n}(t_0,t_1,\dots,t_n)
\\
&=\frac{(-i)^n}{\hbar^n}\hat{\mathcal{P}}^{(\Gamma_1^+)}_{A^{1}_{t_{1}}\dots A^{n}_{t_{n}}}\theta_{t_0t_1}\cdots\theta_{t_{n-1}t_n}C_{[A^0A^1\dots A^n]}(t_0,t_1,\dots,t_n).
\end{split}
\end{equation}

The stripped correlator is the expectation value of nested commutators as in \eqref{eq:StrN}

\begin{equation}
\label{eq:strippedApp}
\begin{split}
&C_{[A^0A^1\dots A^n]}(t_0,t_1,\dots,t_n)
\\
&= \text{tr}\left(\rho_0\left[\left[\dots\left[A^0_{\mathcal{H}_0}(t_0),A^1_{\mathcal{H}_0}(t_1)\right],\dots\right],A^n_{\mathcal{H}_0}(t_n)\right]\right).
\end{split}
\end{equation}

We define the 'plain' correlator as the expectation value

\begin{equation}
\label{eq:planApp}
\begin{split}
&C_{A^0A^1\dots A^n}(t_0,t_1,\dots,t_n)
\\
&\qquad = \text{tr}\left(\rho_0A^0_{\mathcal{H}_0}(t_0)A^1_{\mathcal{H}_0}(t_1)\cdots A^n_{\mathcal{H}_0}(t_n)\right).
\end{split}
\end{equation}

When expanded out, the $n+1$-point stripped correlator has $2^n$ plain terms, which, when symmetrized, lead to $2^nn!$ plain terms for the retarded correlator. This is 2, 8 and 48 terms for the retarded 2, 3 and 4-point correlators, respectively. Calculating such a large number of plain terms is impractical, but the symmetrization provides a means to reduce the number of such terms we need to look at. This follows from the fact that different permutations of the elements in a nested commutator contain equivalent plain terms, which can then be combined upon summation. We perform this explicitly for the retarded 2, 3 and 4-point correlator and express these correlators as the symmetrized combination of plain correlators.\\

The 2-point retarded correlator is particularly simple. We have

\begin{equation}
\label{eq:Ret2Pl}
\begin{split}
C^r_{A^0A^1}(t_0,t_1)&=\frac{-i}{\hbar}\theta_{t_0t_1}(C_{A^0A^1}(t_0,t_1)-C_{A^1A^0}(t_1,t_0))
\\
&=\frac{2}{\hbar}\mathfrak{Re}(-i\theta_{t_0t_1}C_{A^0A^1}(t_0,t_1)),
\end{split}
\end{equation}

where $\mathfrak{Re}(\dots)$ refers to the real part with respect to the complex structure. More specifically, we have that the complex conjugate of the plain correlator

\begin{equation}
\nonumber
\begin{split}
&\bigg((-i)^n C_{A^0A^1\dots A^n}(t_0,t_1,\dots,t_n)\bigg)^{*}
\\
&\qquad =i^n\text{tr}\left(\bigg(\rho_0A^0_{\mathcal{H}_0}(t_0)A^1_{\mathcal{H}_0}(t_1)\cdots A^n_{\mathcal{H}_0}(t_n)\bigg)^{\dagger}\right)
\\
&\qquad = i^n\text{tr}\left(\rho_0A^n_{\mathcal{H}_0}(t_n)A^{n-1}_{\mathcal{H}_0}(t_{n-1})\cdots A^0_{\mathcal{H}_0}(t_0)\right)
\\
&\qquad = i^nC_{A^nA^{n-1}\dots A^0}(t_n,t_{n-1},\dots,t_0),
\end{split}
\end{equation}

reverses the ordering of the observables and time-arguments. The real part is twice the sum of the plain correlator and its conjugate, meaning that it corresponds to the sum of the plain correlator and its 'reverse'.\\

In the next two subsections we shall streamline our notation and refer to the plain correlators as $C_{A^0A^1\dots A^n}(t_0,t_1,\dots,t_n)\equiv 01\dots n $ and similarly for the stripped correlator $C_{[A^0A^1\dots A^n]}(t_0,t_1,\dots,t_n)\equiv [01\dots n]$. The step functions will be denoted as $\theta_{t_0t_1}\equiv \theta_{01}$. 

\subsection{The 3-point retarded correlator}
We have 
\begin{equation}
\label{eq:3pointDef}
-2\hbar^2C^r_{A^0A^1A^2}(t_0,t_1,t_2)=\theta_{01}\theta_{12}[012]+\theta_{02}\theta_{21}[021].
\end{equation}

Expanding the nested commutators of the stripped correlators

\begin{align}
\nonumber
&[012]=012+210-102-201,
\\
\nonumber
&[021]=021+120-102-201.
\end{align}

Combining the terms we can write \eqref{eq:3pointDef} as
\begin{equation}
\label{eq:Ret3Deriv}
\begin{split}
&\theta_{01}\theta_{12}(012+210)+\theta_{02}\theta_{21}(021+120)
\\
&-(\theta_{01}\theta_{12}+\theta_{02}\theta_{21})(102+201).
\end{split}
\end{equation}

The sum of step functions can be written as $\theta_{01}\theta_{12}+\theta_{02}\theta_{21}=\theta_{01}\theta_{12}\theta_{02}+\theta_{02}\theta_{21}\theta_{01}=\theta_{01}\theta_{02}(\theta_{12}+\theta_{21})=\theta_{01}\theta_{02}$. Thus \eqref{eq:Ret3Deriv} becomes

\begin{equation}
\begin{split}
2\hat{\mathcal{P}}^{(\Gamma_1^+)}_{12}(\theta_{01}\theta_{12}(012+210)-\theta_{01}\theta_{02}102).
\end{split}
\end{equation}

Comparing with the left hand side of \eqref{eq:3pointDef} we obtain
\begin{equation}
\label{eq:Ret3Pl}
\begin{split}
&\hbar^2C^r_{A^0A^1A^2}(t_0,t_1,t_2)
\\
&=-\hat{\mathcal{P}}^{(\Gamma_1^+)}_{12}(\theta_{01}\theta_{12}(012+210)-\theta_{01}\theta_{02}102)
\\
&=2\mathfrak{Re}\hat{\mathcal{P}}^{(\Gamma_1^+)}_{12}(-\theta_{01}\theta_{12}012+\frac{1}{2}\theta_{01}\theta_{02}102).
\end{split}
\end{equation}

\subsection{The retarded 4-point correlator}
The retarded 4-point correlator is defined as

\begin{equation}
\label{eq:4pointDef}
\begin{split}
&-i3!\hbar^3C^r_{A^0A^1A^2A^3}(t_0,t_1,t_2,t_3)
\\
&=\theta_{01}\theta_{12}\theta_{23}[0123]+\theta_{01}\theta_{13}\theta_{32}[0132]+\theta_{03}\theta_{32}\theta_{21}[0321]
\\
&\quad+\theta_{02}\theta_{21}\theta_{13}[0213]+\theta_{02}\theta_{23}\theta_{31}[0231]+\theta_{03}\theta_{31}\theta_{12}[0312].
\end{split}
\end{equation}

Expanding the nested commutators of the stripped correlators
\begin{widetext}
\begin{align}
\nonumber
[0123]=0123-3210-1023+3201-2013+3102-3012+2103,
\\
\nonumber
[0132]=0132-2310-1032+2301-2013+3102-3012+2103,
\\
\nonumber
[0321]=0321-1230-1032+2301-2031+1302-3021+1203,
\\
\nonumber
[0213]=0213-3120-1023+3201-2013+3102-3021+1203,
\\
\nonumber
[0231]=0231-1320-1023+3201-2031+1302-3021+1203,
\\
\nonumber
[0312]=0312-2130-1032+2301-2031+1302-3012+2103.
\end{align}
\end{widetext}

Noticing the same terms in different commutators, we combine them and write \eqref{eq:4pointDef} as

\begin{align}
\label{eq:Ret4Deriv}
\nonumber
3!&\hat{\mathcal{P}}^{(\Gamma_1^+)}_{123}(\theta_{01}\theta_{12}\theta_{23}(0123-3210)
\\
&-(\theta_{01}\theta_{12}\theta_{23}+\theta_{02}\theta_{21}\theta_{13}+\theta_{02}\theta_{23}\theta_{31})(1023-3201)).
\end{align}

The sum of step functions on the second line can be found as $\theta_{01}\theta_{12}\theta_{23}+\theta_{02}\theta_{21}\theta_{13}+\theta_{02}\theta_{23}\theta_{31}=\theta_{01}\theta_{02}(\theta_{12}\theta_{23}+\theta_{21}\theta_{13}+\theta_{23}\theta_{31})=\theta_{01}\theta_{02}(\theta_{13}\theta_{23}(\theta_{12}+\theta_{21})+\theta_{23}\theta_{31})=\theta_{01}\theta_{02}(\theta_{13}\theta_{23}+\theta_{23}\theta_{31})=\theta_{01}\theta_{02}\theta_{23}(\theta_{13}+\theta_{31})=\theta_{01}\theta_{02}\theta_{23}$, where we repeatedly used the facts that step functions satisfy $\theta_{ij}\theta_{jk}=\theta_{ij}\theta_{jk}\theta_{ik}$ and $\theta_{ij}+\theta_{ji}=1$. We can thus write \eqref{eq:Ret4Deriv} as

\begin{align}
\nonumber
3!\hat{\mathcal{P}}^{(\Gamma_1^+)}_{123}(\theta_{01}\theta_{12}\theta_{23}(0123-3210)-\theta_{01}\theta_{02}\theta_{23}(1023-3201)).
\end{align}

Comparing with the left hand side of \eqref{eq:4pointDef} we are left with

\begin{equation}
\label{eq:Ret4Pl}
\begin{split}
&\hbar^3C^r_{A^0A^1A^2A^3}(t_0,t_1,t_2,t_3)
\\
&=\hat{\mathcal{P}}^{(\Gamma_1^+)}_{123}(\theta_{01}\theta_{12}\theta_{23}(i0123-i3210)
\\
&\qquad\qquad\quad-\theta_{01}\theta_{02}\theta_{23}(i1023-i3201))
\\
&=2\mathfrak{Re}\hat{\mathcal{P}}^{(\Gamma_1^+)}_{123}(i\theta_{01}\theta_{12}\theta_{23}0123-i\theta_{01}\theta_{02}\theta_{23}1023).
\end{split}
\end{equation}

This is a considerable simplification, since we need a detailed study of only two kinds of plain correlators in order to analyze the retarded 3 and 4-point correlators, furthermore it establishes a platform allowing explicit calculations, that we shall elaborate on next.\\

\subsection{Time-translation invariance}
\label{TimeTrans}
The cyclicity of the trace operation can now be used to show that the correlators depend only on the differences of time arguments. Indeed, we have for the plain correlator

\begin{equation}
\nonumber
\begin{split}
C_{A^0A^1\dots A^n}&(t_0,t_1,\dots,t_n)
\\
&=C_{A^0A^1\dots A^n}(0,t_1-t_0,\dots,t_n-t_0).
\end{split}
\end{equation}

Following the standard convention, we can define

\begin{equation}
\nonumber
\begin{split}
\tilde{C}_{A^0A^1\dots A^n}&(t_0-t_1,\dots,t_0-t_n)
\\
&\equiv C_{A^0A^1\dots A^n}(0,t_1-t_0,\dots,t_n-t_0).
\end{split}
\end{equation}

Redefining the time labels as $t_0-t_i\to t_i$ for $i\in\{1,\dots,n\}$ leads to

\begin{equation}
\tilde{C}_{A^0A^1\dots A^n}(t_1,\dots,t_n)\equiv C_{A^0A^1\dots A^n}(0,-t_1,\dots,-t_n).
\end{equation}

This procedure is straightforward for the stripped and plain correlators, moreover, it also extends to the retarded correlator since the step functions are also functions of time-differences.\\

Using this convention, we can rewrite equations \eqref{eq:Ret2Pl}, \eqref{eq:Ret3Pl}, and \eqref{eq:Ret4Pl} expressing the retarded correlators with plain correlators as

\begin{equation}
\label{eq:Ret2Pl1n}
\begin{split}
\tilde{C}^r_{A^0A^1}(t_1)&=\frac{-i}{\hbar}\theta_{t_1}(\tilde{C}_{A^0A^1}(t_1)-\tilde{C}_{A^1A^0}(-t_1))
\\
&=\frac{2}{\hbar}\mathfrak{Re}(-i\theta_{t_1}\tilde{C}_{A^0A^1}(t_1)),
\end{split}
\end{equation}

\begin{equation}
\label{eq:Ret3Pl1n}
\begin{split}
&\tilde{C}^r_{A^0A^1A^2}(t_1,t_2)=
\\
&=\frac{2}{\hbar^2}\mathfrak{Re}\hat{\mathcal{P}}^{(\Gamma_1^+)}_{A^{1}_{t_{1}}A^{2}_{t_{2}}}\bigg(-\theta_{t_1}\theta_{t_2t_1}\tilde{C}_{A^0A^1A^2}(t_1,t_2)
\\
&\qquad\qquad+\frac12(\theta_{t_1}\theta_{t_2t_1}+\theta_{t_2}\theta_{t_1t_2})\tilde{C}_{A^1A^0A^2}(-t_1,t_2-t_1)\bigg),
\\
&=\frac{2}{\hbar^2}\mathfrak{Re}\hat{\mathcal{P}}^{(\Gamma_1^+)}_{A^{1}_{t_{1}}A^{2}_{t_{2}}}\bigg(-\theta_{t_1}\theta_{t_2t_1}\tilde{C}_{A^0A^1A^2}(t_1,t_2)
\\
&\qquad\qquad\qquad\quad+\frac12\theta_{t_1}\theta_{t_2}\tilde{C}_{A^1A^0A^2}(-t_1,t_2-t_1)\bigg),
\end{split}
\end{equation}
and
\begin{equation}
\label{eq:Ret4Pl1n}
\begin{split}
&\tilde{C}^r_{A^0A^1A^2A^3}(t_1,t_2,t_3)=
\\
&=\frac{2}{\hbar^3}\mathfrak{Re}\hat{\mathcal{P}}^{(\Gamma_1^+)}_{A^{1}_{t_{1}}A^{2}_{t_{2}}A^{3}_{t_{3}}}\bigg(
i\theta_{t_1}\theta_{t_2t_1}\theta_{t_3t_2}\tilde{C}_{A^0A^1A^2A^3}(t_1,t_2,t_3)
\\
&\qquad\qquad\qquad-i\theta_{t_1}\theta_{t_2}\theta_{t_3t_2}\tilde{C}_{A^1A^0A^2A^3}(-t_1,t_2-t_1,t_3-t_1)
\bigg).
\end{split}
\end{equation}

Note that we have given two different expressions for the 3-point retarded correlator. Both of these are useful.\\

\subsection{Correlators in the frequency-domain}
\label{FreqDom}

The formulae for the retarded correlators just derived are all time-domain expressions. While conceptually clear, in order to make contact with experiment, we ought to transfer our results to the frequency domain. This is performed by taking the Fourier transform

\begin{equation}
\begin{split}
&\tilde{C}_{A^0\dots A^n}(\omega_1,\dots,\omega_n)
\\
&=\int dt_1\cdots\int dt_n \tilde{C}_{A^0\dots A^n}(t_1,\dots,t_n)e^{i\omega_1 t_1}\cdots e^{i\omega_n t_n}.
\end{split}
\end{equation}

We now introduce some notation that will simplify our expressions considerably. When looking at \eqref{eq:Ret2Pl1n}, \eqref{eq:Ret3Pl1n} and \eqref{eq:Ret4Pl1n}, we have to take the Fourier transform of a quantity's \textit{real} part, but the result will not be real. Indeed, suppose $f(t)$ is a complex-valued function with real part $\mathfrak{Re}f(t)=(f(t)+f^*(t))/2$, its Fourier transform is $(f(\omega)+f^*(-\omega))/2$ which is not real. However, we would like to keep the notation concise, so we introduce an operator $\mathcal{K}^*_{\omega}$ that acts on frequency-dependent, complex-valued functions and its action is to complex-conjugate and frequency-invert as $\mathcal{K}^*_{\omega}f(\omega)=f^*(-\omega)$. It is clear that the Fourier transform of the real part is an eigenfunction of this operator, since $\mathcal{K}^*_{\omega}(f(\omega)+f^*(-\omega))/2=(f(\omega)+f^*(-\omega))/2$. From the representation theory point of view, the pair $\{\mathfrak{I},\mathfrak{K}^*_{\omega}\}$, with $\mathfrak{I}$ the identity, form an abstract group $\mathcal{G}_{\mathfrak{K}^*_{\omega}}$ that is a realization of the cyclic group $\mathsf{Z}_2$ and the Fourier transform transforms in the representation of this group, with the Fourier transform of the real part transforming in the positive irreducible representation. We can thus use the projector notation to write

\begin{equation}
\label{eq:FreqInv}
 \hat{\mathcal{P}}^{(+)}_{\mathcal{K}^*_{\omega}}f(\omega)=\frac{I+\mathcal{K}^*_{\omega}}{2}f(\omega)=\frac{f(\omega)+f^*(-\omega)}{2}.
\end{equation}

Consequently, the frequency-domain expressions for the retarded correlators \eqref{eq:Ret2Pl1n}, \eqref{eq:Ret3Pl1n} and \eqref{eq:Ret4Pl1n} can be arrived at by simply taking the Fourier transform of the expression behind the real part and then applying the projector \eqref{eq:FreqInv} to the result.

\section{Spectral representation of stripped and retarded correlators}
\label{SpectRep}

The formal expressions arrived at for the structure of the retarded correlator in Appendix \ref{RetPlain} lend themselves well to theoretical discussions, but are difficult to handle when moving on to practical calculations. For example, averaging over the full many-body density matrix is impractical if not impossible, so it would be convenient to derive formulae that can easily be adapted to integrals over single particle statistical operators \cite{Lax1958}. Furthermore it would be conducive to find expressions that, apart from providing a firm handle on analytical calculations, can be ported to a computational platform in a straightforward manner. As recognized decades ago by many, moving to the spectral representation and introducing Green's functions, or, as we refer to them, Green's operators, accomplishes all of these objectives.\\

\subsection{Green's operators}
\label{Green}
Green's operators are important correlators intimately connected to the spectral properties of operators. In the following, however, we will not be writing them in a basis to obtain functions, but keep them in operator form and refer to them as 'Green's operators'. We shall now briefly summarize their basic properties. This is standard textbook material \cite{HaugQuantumKinetics,RadiJishi}, but we present it here for completeness.\\

The spectral resolvent of the equilibrium Hamiltonian $\mathcal{H}_0$ is

\begin{equation}
\nonumber
G(z)=\frac{1}{zI-\mathcal{H}_0},
\end{equation}

where $z$ is a complex number and $I$ refers to the identity operator, which we will refrain from writing out explicitly any longer since this should not cause any misunderstandings. The spectrum of $\mathcal{H}_0$ is then defined as the set of complex numbers over which $G(z)$ is not bounded. Since $\mathcal{H}_0$ is Hermitian, this is a subset of the real line, so non-analyticity is only present along the real line. Thus, $G(z)$ is analytic in both the upper and lower half planes and we consider approaching---analytically continuing to---the real line from these two directions. Approaching from the upper(lower) half plane we obtain the retarded(advanced) resolvents or Green's operators

\begin{equation}
\label{eq:GrGaEps}
G^r(\varepsilon) = \lim_{\eta\to 0^+}\frac{1}{\varepsilon-\mathcal{H}_0+i\eta},\, G^a(\varepsilon) = \lim_{\eta\to 0^+}\frac{1}{\varepsilon-\mathcal{H}_0-i\eta}.
\end{equation}
 
It is then possible to look at $G^r(\varepsilon),\,G^a(\varepsilon)$ as---sufficiently regularized---Fourier transforms of certain time-domain operators. Define

\begin{equation}
\label{eq:GrGaT}
G^r(t) = -\frac{i}{\hbar}\theta(t)e^{-i\mathcal{H}_0t/\hbar},\,G^a(t) = \frac{i}{\hbar}\theta(-t)e^{-i\mathcal{H}_0t/\hbar},
\end{equation}

which indeed satisfy

\begin{equation}
G^{r(a)}(\varepsilon) = \int dt \,G^{r(a)}(t)e^{i\varepsilon t/\hbar}.
\end{equation}

We have the useful identities following from the definitions

\begin{align}
\label{eq:GrmGat}
&G^r(t)-G^a(t) = -\frac{i}{\hbar}e^{-i\mathcal{H}_0 t/\hbar},
\\
\label{eq:GrmGa}
&G^r(\varepsilon)-G^a(\varepsilon) = -i2\pi\delta(\varepsilon-\mathcal{H}_0),
\end{align}

furthermore it is apparent from \eqref{eq:GrGaEps} and \eqref{eq:GrGaT} that Green's operators satisfy $(G^{r}(\varepsilon))^{\dagger}=G^a(\varepsilon)$ and $(G^{r}(t))^{\dagger}=G^a(-t)$.\\

In order to keep notation under control, we introduce the short-hands,

\begin{equation}
\label{eq:GreenShorthand}
\begin{split}
&G^{r(a)}_{t}\equiv G^{r(a)}(t),\, G^{r(a)}_{\omega}\equiv G^{r(a)}_{\varepsilon+\hbar\omega}\equiv G^{r(a)}(\varepsilon+\hbar\omega),
\\
&G^{r\pm a}_{\omega}\equiv G^{r}(\varepsilon+\hbar\omega)\pm G^a(\varepsilon+\hbar\omega),
\end{split}
\end{equation}

that we shall use depending on the situation at hand.\\

Thus, it follows from \eqref{eq:GrGaT} that attaching step functions to interaction picture evolution operators $e^{-i\mathcal{H}_0t/\hbar}$ result in Green's operators. Furthermore, from equations \eqref{eq:Ret2Pl1n}, \eqref{eq:Ret3Pl1n} and \eqref{eq:Ret4Pl1n} we see that the retarded correlators consist of step functions attached to plain correlators, with the time-dependence of the latter being given by interaction picture evolution operators. As a consequence, we can rewrite the retarded correlators using Green's operators and obtain the spectral representation.\\

\subsection{Spectral representation of the 2-point stripped correlator}
\label{stripped2specApp}

The 2-point stripped correlator $C_{[A^0A^1]}(t_0,t_1)$ can be written in terms of plain correlators (defined in \eqref{eq:planApp}) as

\begin{equation}
C_{[A^0A^1]}(t_0,t_1)=C_{A^0A^1}(t_0,t_1)-C_{A^1A^0}(t_1,t_0),
\end{equation}

and, by virtue of time-translation invariance (see Appendix \ref{RetPlain}.\ref{TimeTrans}), this is equivalent to

\begin{equation}
\tilde{C}_{[A^0A^1]}(t_1)=\tilde{C}_{A^0A^1}(t_1)-\tilde{C}_{A^1A^0}(-t_1).
\end{equation}

We want to find the spectral representation of the Fourier transform

\begin{equation}
\tilde{C}_{[A^0A^1]}(\omega_1)=\tilde{C}_{A^0A^1}(\omega_1)-\tilde{C}_{A^1A^0}(-\omega_1).
\end{equation}

In order to do this, we look at the plain correlators. The explicit expression for the 2-point plain correlator $\tilde{C}_{A^0A^1}(t_1)\equiv C_{A^0A^1}(0,-t_1)$ is given by the $n=1$ case of \eqref{eq:planApp} as

\begin{equation}
\begin{split}
\tilde{C}_{A^0A^1}(t_1)&\equiv C_{A^0A^1}(0,-t_1)
\\
&=\text{tr}(\rho_0A^0A^1_{\mathcal{H}_0}(-t_1))
\\
&=\text{tr}(\rho_0A^0e^{-i\mathcal{H}_0t_1/\hbar}A^1e^{i\mathcal{H}_0t_1/\hbar}).
\end{split}
\end{equation}

Writing the time evolution factors as

\begin{equation}
\label{eq:timevolspecApp}
\begin{split}
e^{\pm i\mathcal{H}_0t_1/\hbar}=&\int d\varepsilon \delta(\varepsilon-\mathcal{H}_0)e^{\pm i\varepsilon t_1/\hbar}
\\
=&\frac{i}{2\pi}\int d\varepsilon (G^r(\varepsilon)-G^a(\varepsilon)) e^{\pm i\varepsilon t_1/\hbar}
\\
\equiv&\frac{i}{2\pi}\int d\varepsilon G^{r-a}_{\varepsilon} e^{\pm i\varepsilon t_1/\hbar},
\end{split}
\end{equation} 

where we used \eqref{eq:GrmGa}, we have for the Fourier transform

\begin{equation}
\begin{split}
&\tilde{C}_{A^0A^1}(\omega_1)=\int dt_1 \tilde{C}_{A^0A^1}(t_1) e^{i\omega_1t_1}
\\
&=\int dt_1\text{tr}(\rho_0A^0e^{-i\mathcal{H}_0t_1/\hbar}A^1e^{i\mathcal{H}_0t_1/\hbar})e^{i\omega_1t_1}
\\
&=\int d\varepsilon\int d\varepsilon'\text{tr}(\rho_0A^0\delta(\varepsilon'-\mathcal{H}_0)A^1\delta(\varepsilon-\mathcal{H}_0))
\\
&\qquad\times\int dt_1 e^{i\left(\omega_1-\frac{\varepsilon'-\varepsilon}{\hbar}\right)t_1}
\\
&=2\pi\hbar\int d\varepsilon\text{tr}(\rho_0A^0\delta(\varepsilon+\hbar\omega_1-\mathcal{H}_0)A^1\delta(\varepsilon-\mathcal{H}_0)),
\end{split}
\end{equation}

where we used the identity

\begin{equation}
\int dt_1 e^{i(\omega_1-(\varepsilon-\varepsilon')/\hbar)t_1}=2\pi\hbar\delta(\varepsilon-\varepsilon'-\hbar\omega_1).
\end{equation}

We can thus write the spectral representation of the 2-point stripped correlator as

\begin{equation}
\begin{split}
\tilde{C}_{[A^0A^1]}(\omega_1)=&\tilde{C}_{A^0A^1}(\omega_1)-\tilde{C}_{A^0A^1}(-\omega_1)
\\
=&2\pi\hbar\int d\varepsilon\text{tr}(\rho_0(A^0\delta(\varepsilon+\hbar\omega_1-\mathcal{H}_0)A^1
\\
&-A^1\delta(\varepsilon-\hbar\omega_1-\mathcal{H}_0)A^0)\delta(\varepsilon-\mathcal{H}_0)).
\end{split}
\end{equation}

Using the fact that the equilibrium density matrix is a function of $\mathcal{H}_0$, i.e., $\rho_0\equiv \rho_0(\mathcal{H}_0)$, we can use the cyclic property of the trace and the standard trick of writing $\rho_0(\mathcal{H}_0)\delta(\varepsilon-\mathcal{H}_0)=\rho_0(\varepsilon)\delta(\varepsilon-\mathcal{H}_0)$ to `pull out' the scalar-valued function $\rho_0(\varepsilon)$ from the trace

\begin{equation}
\begin{split}
\tilde{C}_{[A^0A^1]}(\omega_1)=&2\pi\hbar\int d\varepsilon\rho_0(\varepsilon)\text{tr}((A^0\delta(\varepsilon+\hbar\omega_1-\mathcal{H}_0)A^1
\\
&-A^1\delta(\varepsilon-\hbar\omega_1-\mathcal{H}_0)A^0)\delta(\varepsilon-\mathcal{H}_0)).
\end{split}
\end{equation}

At this point, we shift the integration variable in the second term within the trace as $\varepsilon\to \varepsilon+\hbar\omega_1$ and once again use the cyclic property to arrive at

\begin{equation}
\begin{split}
\tilde{C}_{[A^0A^1]}(\omega_1)=-&2\pi\hbar\int d\varepsilon(\rho_0(\varepsilon+\omega_1)-\rho_0(\varepsilon))
\\
&\times\text{tr}((A^0\delta(\varepsilon+\hbar\omega_1-\mathcal{H}_0)A^1\delta(\varepsilon-\mathcal{H}_0))
\\
\equiv-&2\pi\hbar\int d\varepsilon(\rho_0(\varepsilon+\omega_1)-\rho_0(\varepsilon))
\\
&\times\text{tr}(A^0G^{r-a}_{\omega_1}A^1G^{r-a}).
\end{split}
\end{equation}

\subsection{Spectral representation of the 3-point stripped correlator}
\label{stripped3specApp}

The 3-point stripped correlator (the $n=2$ case of \eqref{eq:strippedApp}) in terms of plain correlators is

\begin{equation}
\begin{split}
&C_{[A^0A^1A^2]}(t_0,t_1,t_2)
\\
&=C_{A^0A^1A^2}(t_0,t_1,t_2)+C_{A^2A^1A^0}(t_2,t_1,t_0)
\\
&\quad-C_{A^2A^0A^1}(t_0,t_1,t_2)-C_{A^1A^0A^2}(t_2,t_1,t_0).
\end{split}
\end{equation}

Time-translation invariance allows us to write

\begin{equation}
\begin{split}
&\tilde{C}_{[A^0A^1A^2]}(t_1,t_2)
\\
&=\tilde{C}_{A^0A^1A^2}(t_1,t_2)+\tilde{C}_{A^2A^1A^0}(t_1-t_2,-t_2)
\\
&\quad-\tilde{C}_{A^2A^0A^1}(-t_2,t_1-t_2)-\tilde{C}_{A^1A^0A^2}(-t_1,t_2-t_1).
\end{split}
\end{equation}

The Fourier transform is

\begin{equation}
\label{eq:stripped3FourierApp}
\begin{split}
&\tilde{C}_{[A^0A^1A^2]}(\omega_1,\omega_2)
\\
&=\tilde{C}_{A^0A^1A^2}(\omega_1,\omega_2)+\tilde{C}_{A^2A^1A^0}(\omega_1,-\omega_1-\omega_2)
\\
&\quad-\tilde{C}_{A^2A^0A^1}(-\omega_1-\omega_2,\omega_2)-\tilde{C}_{A^1A^0A^2}(-\omega_1-\omega_2,\omega_1).
\end{split}
\end{equation}

The explicit expression for the 3-point plain correlator $\tilde{C}_{A^0A^1A^2}(t_1,t_2)\equiv C_{A^0A^1A^2}(0,-t_1,-t_2)$ is given by the $n=2$ case of \eqref{eq:planApp} as

\begin{equation}
\begin{split}
&\tilde{C}_{A^0A^1A^2}(t_1,t_2)
\\
&=\text{tr}(\rho_0A^0A^1_{\mathcal{H}_0}(-t_1)A^2_{\mathcal{H}_0}(-t_2))
\\
&=\text{tr}(\rho_0A^0e^{-i\mathcal{H}_0t_1/\hbar}A^1e^{i\mathcal{H}_0(t_1-t_2)/\hbar}A^2e^{i\mathcal{H}_0t_2/\hbar}).
\end{split}
\end{equation}

Writing the time-evolution factors using delta functions, or, equivalently, $G^{r-a}_{\varepsilon}$ terms, according to \eqref{eq:timevolspecApp}, we obtain the Fourier transform

\begin{widetext}
\begin{equation}
\begin{split}
&\tilde{C}_{A^0A^1A^2}(\omega_1,\omega_2)=\int dt_1\int dt_2 \tilde{C}_{A^0A^1A^2}(t_1,t_2) e^{i\omega_1t_1}e^{i\omega_2t_2}
\\
&=\int d\varepsilon\int d\varepsilon'\int d\varepsilon''\text{tr}(\rho_0A^0\delta(\varepsilon''-\mathcal{H}_0)A^1\delta(\varepsilon'-\mathcal{H}_0))A^2\delta(\varepsilon-\mathcal{H}_0))
\int dt_1 e^{i\left(\omega_1-\frac{\varepsilon''-\varepsilon'}{\hbar}\right)t_1}\int dt_2 e^{i\left(\omega_2-\frac{\varepsilon'-\varepsilon}{\hbar}\right)t_2}
\\
&=4\pi^2\hbar^2\int d\varepsilon\text{tr}(\rho_0A^0\delta(\varepsilon+\hbar\omega_1+\hbar\omega_2-\mathcal{H}_0)A^1\delta(\varepsilon+\hbar\omega_2-\mathcal{H}_0)A^2\delta(\varepsilon-\mathcal{H}_0))
\\
&\equiv-\frac{i\hbar^2}{2\pi}\int d\varepsilon\text{tr}(\rho_0A^0G^{r-a}_{\omega_1+\omega_2}A^1G^{r-a}_{\omega_2}A^2G^{r-a}).
\end{split}
\end{equation}

We can then combine the corresponding terms according \eqref{eq:stripped3FourierApp}, and, just as for the 2-point case above, use the delta function to pull out the density matrix from the trace, and shift arguments to find

\begin{equation}
\begin{split}
\tilde{C}_{[A^0A^1A^2]}(\omega_1,\omega_2)=&-4\pi^2\hbar^2\int d\varepsilon(\rho_0(\varepsilon+\hbar\omega_2)-\rho_0(\varepsilon))\text{tr}(A^0\delta(\varepsilon+\hbar\omega_1+\hbar\omega_2-\mathcal{H}_0)A^1\delta(\varepsilon+\hbar\omega_2-\mathcal{H}_0)A^2\delta(\varepsilon-\mathcal{H}_0))
\\
&+4\pi^2\hbar^2\int d\varepsilon(\rho_0(\varepsilon)-\rho_0(\varepsilon-\hbar\omega_2))\text{tr}(A^1\delta(\varepsilon-\hbar\omega_1-\hbar\omega_2-\mathcal{H}_0)A^0\delta(\varepsilon-\mathcal{H}_0)A^2\delta(\varepsilon-\hbar\omega_2-\mathcal{H}_0))
\\
\equiv&\frac{i\hbar^2}{2\pi}\int d\varepsilon(\rho_0(\varepsilon+\hbar\omega_2)-\rho_0(\varepsilon))\text{tr}(A^0G^{r-a}_{\omega_1+\omega_2}A^1G^{r-a}_{\omega_2}A^2G^{r-a})
\\
&-\frac{i\hbar^2}{2\pi}\int d\varepsilon(\rho_0(\varepsilon)-\rho_0(\varepsilon-\hbar\omega_2))\text{tr}(A^1G^{r-a}_{-\omega_1-\omega_2}A^0G^{r-a}A^2G^{r-a}_{-\omega_2}).
\end{split}
\end{equation}

\end{widetext}

\subsection{Spectral representation of the 2-point retarded correlator}

The 2-point plain correlator needed in \eqref{eq:Ret2Pl1n} gives

\begin{equation}
\label{eq:Ret2SpecDeriv}
\begin{split}
&-\frac{i}{\hbar}\theta_{t_1}\tilde{C}_{A^0A^1}(t_1)
\\
&=\text{tr}\left(\rho_0A^0\left(\frac{-i}{\hbar}\theta_{t_1}e^{-i\mathcal{H}_0t_1/\hbar}\right)A^1e^{i\mathcal{H}_0t_1/\hbar}\right)
\\
&=i\hbar\text{tr}\left(\rho_0A^0G^r_{t_1}A^1(G^r_{-t_1}-G^a_{-t_1})\right)
\\
&=i\hbar\int \frac{d\varepsilon}{2\pi\hbar} \int \frac{d\varepsilon'}{2\pi\hbar}\text{tr}(\rho_0A^0G^r_{\varepsilon}A^1(G^r_{\varepsilon'}-G^a_{\varepsilon'}))
\\
&\qquad\qquad\qquad\qquad\qquad\times e^{-i(\varepsilon-\varepsilon')t_1/\hbar},
\end{split}
\end{equation}

where we used \eqref{eq:GrmGat}. Taking the Fourier transform results in

\begin{equation}
\begin{split}
&\int dt_1\left( -\frac{i}{\hbar}\theta_{t_1}\tilde{C}_{A^0A^1}(t_1)\right)e^{i\omega_1 t_1}
\\
&=\frac{i}{2\pi}\int d\varepsilon \text{tr}(\rho_0A^0G^r_{\varepsilon+\hbar\omega_1}A^1(G^r_{\varepsilon}-G^a_{\varepsilon})),
\\
&\equiv\frac{i}{2\pi}\int d\varepsilon \text{tr}(\rho_0A^0G^r_{\omega_1}A^1 G^{r-a}).
\end{split}
\end{equation}

With the projector notation introduced in \eqref{eq:FreqInv}, the spectral representation of the 2-point retarded correlator \eqref{eq:Ret2Pl1n} in the frequency domain becomes

\begin{equation}
\label{eq:Ret2SpecF}
\tilde{C}^r_{A^0A^1}(\omega_1)=\frac{1}{\pi}\hat{\mathcal{P}}^{(+)}_{\mathcal{K}^*_{\omega}}\,i\int d\varepsilon\,\text{tr}\left(\rho_0A^0G^r_{\omega_1}A^1G^{r-a}\right).
\end{equation}

\subsection{Spectral representation of the 3-point retarded correlator}

The 3-point retarded correlator \eqref{eq:Ret3Pl1n} contains two different kinds of plain correlators and is expressed in two different ways depending on what combinations of step functions are attached to the second plain correlator.\\

The first correlator becomes

\begin{widetext}
\begin{equation}
\begin{split}
-\frac{1}{\hbar^2}\theta_{t_1}\theta_{t_2t_1}\tilde{C}_{A^0A^1A^2}(t_1,t_2)
&=\text{tr}\left(\rho_0A^0\left(\frac{-i}{\hbar}\theta_{t_1}e^{-i\mathcal{H}_0t_1/\hbar}\right)A^1\left(\frac{-i}{\hbar}\theta_{t_2t_1}e^{-i\mathcal{H}_0(t_2-t_1)/\hbar}\right)A^2e^{i\mathcal{H}_0t_2/\hbar}\right)
\\
&=i\hbar\text{tr}(\rho_0 A^0G^r_{t_1}A^1G^r_{t_2-t_1}A^2(G^r_{-t_2}-G^a_{-t_2}))
\\
&=i\hbar\int \frac{d\varepsilon}{2\pi\hbar} \int \frac{d\varepsilon'}{2\pi\hbar}\int \frac{d\varepsilon''}{2\pi\hbar}\text{tr}(\rho_0A^0G^r_{\varepsilon}A^1G^r_{\varepsilon'}A^2(G^r_{\varepsilon''}-G^a_{\varepsilon''}))e^{-i(\varepsilon-\varepsilon')t_1/\hbar}e^{-i(\varepsilon'-\varepsilon'')t_2/\hbar}.
\end{split}
\end{equation}

Moving to the frequency domain, we obtain

\begin{equation}
\label{eq:Ret3G1}
\begin{split}
&\int dt_1\int dt_2\left(-\frac{1}{\hbar^2}\theta_{t_1}\theta_{t_2t_1}\tilde{C}_{A^0A^1A^2}(t_1,t_2)\right)e^{i\omega_1t_1}e^{i\omega_2t_2}
\\
&=\frac{i}{2\pi}\int d\varepsilon\text{tr}(\rho_0A^0G^r_{\omega_1+\omega_2}A^1G^r_{\omega_2}A^2G^{r-a}).
\end{split}
\end{equation}

The other plain correlator in \eqref{eq:Ret3Pl1n} has two different combinations of step functions attached to it. We have for the first type

\begin{equation}
\begin{split}
\frac{1}{\hbar^2}\theta_{t_1}\theta_{t_2t_1}\tilde{C}_{A^1A^0A^2}(-t_1,t_2-t_1)&=-\text{tr}\left(\rho_0A^1\left(\frac{i}{\hbar}\theta_{t_1}e^{i\mathcal{H}_0t_1/\hbar}\right)A^0 e^{-i\mathcal{H}_0t_2/\hbar}A^2\left(\frac{i}{\hbar}\theta_{t_2t_1}e^{i\mathcal{H}_0(t_2-t_1)/\hbar}\right)\right)
\\
&=-i\hbar\text{tr}(\rho_0A^1 G^a_{-t_1} A^0 (G^r_{t_2}-G^a_{t_2}) A^2 G^a_{t_1-t_2}).
\end{split}
\end{equation}

Similarly,

\begin{equation}
\frac{1}{\hbar^2}\theta_{t_2}\theta_{t_1t_2}\tilde{C}_{A^1A^0A^2}(-t_1,t_2-t_1)=-i\hbar \text{tr}(\rho_0A^1(G^r_{-t_1}-G^a_{-t_1})A^0 G^r_{t_2}A^2 G^r_{t_1-t_2}).
\end{equation}

The frequency-domain expression of the spectral representation then becomes

\begin{equation}
\label{eq:Ret3G21}
\begin{split}
&\int dt_1\int dt_2 \left(\frac{1}{\hbar^2}(\theta_{t_1}\theta_{t_2t_1}+\theta_{t_2}\theta_{t_1t_2})\tilde{C}_{A^1A^0A^2}(-t_1,t_2-t_1)\right)e^{i\omega_1t_1}e^{i\omega_2t_2}=
\\
&=-\frac{i}{2\pi}\int d\varepsilon(\text{tr}(\rho_0 A^1G^{r-a}A^0 G^r_{\omega_1+\omega_2}A^2G^r_{\omega_1})+\text{tr}(\rho_0 A^1G^a_{-\omega_1-\omega_2}A^0G^{r-a}A^2G^a_{-\omega_2})).
\end{split}
\end{equation}

We also have the second type 

\begin{equation}
\frac{1}{\hbar^2}\theta_{t_1}\theta_{t_2}\tilde{C}_{A^1A^0A^2}(-t_1,t_2-t_1)=i\hbar \text{tr}(\rho_0A^1G^a_{-t_1}A^0 G^r_{t_2}A^2(G^r_{t_1-t_2}-G^a_{t_1-t_2}),
\end{equation}

which yields

\begin{equation}
\label{eq:Ret3G22}
\begin{split}
&\int dt_1\int dt_2\left(\frac{1}{\hbar^2}\theta_{t_1}\theta_{t_2}\tilde{C}_{A^1A^0A^2}(-t_1,t_2-t_1)\right)e^{i\omega_1t_1}e^{i\omega_2t_2}
\\
&=\frac{i}{2\pi}\int d\varepsilon\text{tr}(\rho_0 A^1G^a_{-\omega_1}A^0G^r_{\omega_2}A^2G^{r-a}).
\end{split}
\end{equation}

Combining \eqref{eq:Ret3G1} with first \eqref{eq:Ret3G21} and then \eqref{eq:Ret3G22} and attaching the projectors, the two expressions for the full 3-point retarded correlator \eqref{eq:Ret3Pl1n} become

\begin{align}
\label{eq:Ret3SpecF}
C^r_{A^0A^1A^2}(\omega_1,\omega_2)=
&\frac{1}{\pi}\hat{\mathcal{P}}^{(+)}_{\mathcal{K}^*_{\omega}}\hat{\mathcal{P}}^{(\Gamma_1^+)}_{A^{1}_{\omega_{1}}A^{2}_{\omega_{2}}}i\int d\varepsilon\,\text{tr}\bigg(\rho_0\bigg[A^0G^r_{\omega_1+\omega_2}A^1G^r_{\omega_2},A^2G^{r-a}\bigg]\bigg)
\\\nonumber
\\\nonumber
=&\frac{1}{\pi}\hat{\mathcal{P}}^{(+)}_{\mathcal{K}^*_{\omega}}\hat{\mathcal{P}}^{(\Gamma_1^+)}_{A^{1}_{\omega_{1}}A^{2}_{\omega_{2}}}i\int d\varepsilon\,\text{tr}\bigg(\rho_0\bigg(A^0G^r_{\omega_1+\omega_2}A^1G^r_{\omega_2}A^2
+\frac12A^1G^a_{-\omega_1}A^0G^r_{\omega_2}A^2\bigg)G^{r-a}\bigg),
\end{align}

where $[\dots,\dots]$ is the usual commutator and we defined the projector $\hat{\mathcal{P}}^{(\Gamma_1^+)}_{A^{1}_{\omega_{1}}\dots A^{n}_{\omega_{n}}}$ onto the $\Gamma_1^+$ irreducible representation of the permutation group $P^{A^1_{\omega_1}\dots A^n_{\omega_n}}(n)$ that acts by permuting the observables and frequency labels simultaneously.

\subsection{Spectral representation of the 4-point retarded correlator}
Pairing the step functions with the interaction picture evolution operators, recognizing Green's operators and taking their Fourier transforms followed by moving to the frequency domain in the same manner as we did for the 2 and 3-point correlators above, enables us to write \eqref{eq:Ret4Pl1n} as 

\begin{equation}
\label{eq:Ret4SpecF}
\begin{split}
C^r_{A^0A^1A^2A^3}(\omega_1,\omega_2,\omega_3)
=\frac{1}{\pi}\hat{\mathcal{P}}^{(+)}_{\mathcal{K}^*_{\omega}}\hat{\mathcal{P}}^{(\Gamma_1^+)}_{A^{1}_{\omega_{1}}A^{2}_{\omega_{2}}A^{3}_{\omega_{3}}}i
\int d\varepsilon\,\text{tr}\bigg(\rho_0\bigg(&A^0G^r_{\omega_1+\omega_2+\omega_3}A^1G^r_{\omega_2+\omega_3}A^2G^r_{\omega_3}
\\
&+A^1G^a_{-\omega_1}A^0G^r_{\omega_2+\omega_3}A^2G^r_{\omega_3}\bigg)A^3G^{r-a}\bigg).
\end{split}
\end{equation}

\end{widetext}

\subsection{Spectral representation of advanced correlators}

Up to this point, the center stage has been given to the retarded correlator $C^r$ since it expresses the total physical response of a system. However, based on our discussion in section \ref{tstruct}, the advanced correlator $C^a$ plays an equally significant role when we are concerned with discerning the $C^r$'s behaviour under the action of the time-reversal group. Indeed, formula \eqref{eq:RDTRSTRB} describes the reactive(dissipative) part of the response as proportional to $C^{r+a}$($C^{r-a}$). Even when we are not interested in the reactive/dissipative distinction, only the behaviour under magnetic-inversion, we are left with formula \eqref{eq:RetMagInv} which contains both $C^r$ and a time-inverted $C^a$. Consequently, it is  important to obtain the spectral representation of $C^a$ in addition to $C^r$ arrived at above. The recipe for this is very simple.\\
Consider the 2-point advanced correlator arrived at by taking $n=1$ in \eqref{eq:Advn} and making use of time-translation invariance

\begin{equation}
\begin{split}
&C^a_{A^0A^1}(t_1)=2\mathfrak{Re}(\frac{i}{\hbar}\theta_{-t_1}C_{A^0A^1}(0,-t_1))
\\
&=2\mathfrak{Re}\left(\text{tr}\left(\rho_0A^0\left(\frac{i}{\hbar}\theta_{-t_1}e^{-i\mathcal{H}_0t_1/\hbar}\right)A^1e^{i\mathcal{H}_0t_1/\hbar}\right)\right)
\\
&=2\mathfrak{Re}(i\hbar\text{tr}\left(\rho_0A^0G^a(t_1)A^1(G^r(-t_1)-G^a(-t_1))\right)).
\end{split}
\end{equation}

Contrasting this with the retarded case \eqref{eq:Ret2SpecDeriv}, we note that the only difference is that in the latter, we had $G^r(t_1)$ as a result of the step-function attachment, whereas in the present case it is $G^a(t_1)$ that takes the stage. This means that we can obtain the spectral representation of $C^a_{A^0A^1}(\omega_1)$ by simply exchanging $G^r_{\omega_1}$ for $G^a_{\omega_1}$ in \eqref{eq:Ret2SpecF} and vice-versa. Note, however, that $G^{r}(\varepsilon)-G^a(\varepsilon)$ should be left as is since this factor is not a result of step-function attachments. In a similar manner, the spectral representations of the advanced 3 and 4-point correlators can be found from the retarded expressions \eqref{eq:Ret3SpecF} and \eqref{eq:Ret4SpecF} by performing the exchange $G^r\leftrightarrow G^a$ while leaving the $G^{r-a}\equiv G^{r}(\varepsilon)-G^a(\varepsilon)$ factor unchanged.\\

\nocite{*}

\bibliography{2ndord1}

\end{document}